\documentclass[iop,apj,twocolumn,tighten,twocolappendix,numberedappendix]{
emulateapj}

\usepackage{mathptmx,subfigure,graphicx,lpic,amssymb,amsmath,rotating,epsfig}
\def\aj{{AJ}}%
\def\apj{{ApJ}}%
\def\apjl{{ApJ}}%
\def\apjs{{ApJS}}%
\def\ao{{Appl.~Opt.}}%
\def\aap{{A\&A}}%
\def\mnras{{MNRAS}}%
\def\pasp{{PASP}}%
\def\pasj{{PASJ}}%
\def\nat{{Nature}}%
\def\procspie{{Proc.~SPIE}}%
\def\na{{NewA}}
\begin{document}
\title[]{The Contribution of Halos with Different Mass Ratios to the Overall
Growth of Cluster-Sized Halos}

\author
{Doron Lemze\altaffilmark{1}, Marc Postman\altaffilmark{2}, Shy Genel\altaffilmark{3}, Holland C. Ford\altaffilmark{1}, Italo Balestra\altaffilmark{4,5}, Megan Donahue\altaffilmark{6}, Daniel Kelson\altaffilmark{7}, Mario Nonino\altaffilmark{4}, Amata Mercurio\altaffilmark{5}, Andrea Biviano\altaffilmark{4}, Piero Rosati\altaffilmark{8}, Keiichi Umetsu\altaffilmark{9}, David Sand\altaffilmark{10}, Anton Koekemoer\altaffilmark{2}, Massimo Meneghetti\altaffilmark{11,12}, Peter Melchior\altaffilmark{13,14}, Andrew B. Newman\altaffilmark{15}, Waqas A. Bhatti\altaffilmark{16}, G. Mark Voit\altaffilmark{7}, Elinor Medezinski\altaffilmark{1}, Adi Zitrin\altaffilmark{17}, Wei Zheng\altaffilmark{1}, Tom Broadhurst\altaffilmark{18,19},
Matthias Bartelmann\altaffilmark{17}, Narciso Benitez\altaffilmark{20}, Rychard Bouwens\altaffilmark{21}, Larry Bradley\altaffilmark{2}, Dan Coe\altaffilmark{2}, Genevieve Graves\altaffilmark{16,22}, Claudio Grillo\altaffilmark{23}, Leopoldo Infante\altaffilmark{24}, Yolanda Jimenez-Teja\altaffilmark{20}, Stephanie Jouvel\altaffilmark{25}, Ofer Lahav\altaffilmark{26}, Dan Maoz\altaffilmark{27}, Julian Merten\altaffilmark{28}, Alberto Molino\altaffilmark{20}, John Moustakas\altaffilmark{29}, Leonidas Moustakas\altaffilmark{28}, Sara Ogaz\altaffilmark{2}, Marco Scodeggio\altaffilmark{30}, and Stella Seitz\altaffilmark{31,32}
}
\altaffiltext{1}{Department of Physics \& Astronomy, Johns Hopkins University, 3400 North Charles Street, Baltimore, MD 21218, USA}
\altaffiltext{2}{Space Telescope Science Institute, 3700 San Martin Drive, Baltimore, MD 21208, USA}
\altaffiltext{3}{Harvard-Smithsonian Center for Astrophysics, 60 Garden Street Cambridge, MA 02138 USA}
\altaffiltext{4}{INAF/Osservatorio Astronomico di Trieste, via G.B. Tiepolo 11, 34143 Trieste, Italy}
\altaffiltext{5}{INAF/Osservatorio Astronomico di Capodimonte, Salita Moiariello 16, 80131 Napoli, Italy}
\altaffiltext{6}{Department of Physics and Astronomy, Michigan State University, East Lansing, MI 48824-2320, USA}
\altaffiltext{7}{Carnegie Institute for Science, Carnegie Observatories, Pasadena, CA, USA}
\altaffiltext{8}{European Southern Observatory, Karl-Schwarzschild Strasse 2, 85748 Garching, Germany}
\altaffiltext{9}{Institute of Astronomy and Astrophysics, Academia Sinica, P.O. Box 23-141, Taipei 10617, Taiwan}
\altaffiltext{10}{Department of Physics, Texas Tech University, Box 41051, Lubbock, TX 79409-1051, USA} 
\altaffiltext{11}{INAF, Osservatorio Astronomico di Bologna, via Ranzani 1, 40127 Bologna, Italy}
\altaffiltext{12}{INFN, Sezione di Bologna, viale Berti Pichat 6/2, 40127, Bologna, Italy}
\altaffiltext{13}{Center for Cosmology and Astro-Particle Physics, The Ohio State University, 191 W. Woodruff Ave., Columbus, OH 43210, USA}
\altaffiltext{14}{Department of Physics, The Ohio State University, 191 W. Woodruff Ave., Columbus, OH 43210, USA}
\altaffiltext{15}{Cahill Center for Astronomy and Astrophysics, California Institute of Technology, MS 249-17, Pasadena, CA 91125, USA}
\altaffiltext{16}{Department of Astrophysical Sciences, Princeton University, Princeton, NJ 08544, USA}
\altaffiltext{17}{Institut f\"ur Theoretische Astrophysik, Zentrum f\"ur Astronomie, Universit\"at Heidelberg, Philosophenweg 12, D-69120 Heidelberg, Germany}
\altaffiltext{18}{Department of Theoretical Physics, University of Basque Country UPV/EHU, Bilbao, Spain}
\altaffiltext{19}{IKERBASQUE, Basque Foundation for Science, Bilbao, Spain}
\altaffiltext{20}{Instituto de Astrofisica de Andalucia (CSIC), Glorieta de la Astronomia s/n, E18008 Granada, Spain}
\altaffiltext{21}{Leiden Observatory, Leiden University, NL-2333 Leiden, The Netherlands}
\altaffiltext{22}{Department of Astronomy, University of California, Berkeley, CA, USA}
\altaffiltext{23}{Dark Cosmology Centre, Niels Bohr Institute, University of Copenhagen, DK-2100 Copenhagen, Denmark}
\altaffiltext{24}{Institute of Astrophysics and Center for Astroengineering, Pontificia Universidad Cat\'olica de Chile, Santiago, Chile}
\altaffiltext{25}{Institut de Cincies de l'Espai (IEE-CSIC), E-08193 Bellaterra (Barcelona), Spain}
\altaffiltext{26}{Department of Physics and Astronomy, University College London, London, UK}
\altaffiltext{27}{The School of Physics and Astronomy, Tel Aviv University, Tel Aviv, Israel}
\altaffiltext{28}{Jet Propulsion Laboratory, California Institute of Technology, Pasadena, CA 91109, USA}
\altaffiltext{29}{Department of Physics and Astronomy, Siena College, Loudonville, NY, USA}
\altaffiltext{30}{INAF-IASF Milano, Via Bassini 15, I-20133, Milano, Italy}
\altaffiltext{31}{Instituts f\"ur Astronomie und Astrophysik, Universit\"as-Sternwarte M\"unchen, D-81679 M\"unchen, Germany}
\altaffiltext{32}{Max-Planck-Institut f\"ur extraterrestrische Physik (MPE), D-85748 Garching, Germany}

%%%%%%%%%%%%%%

\label{firstpage}
\begin{abstract}

We provide a new observational test for a key prediction of the $\Lambda$CDM
cosmological model: the contributions of mergers with different halo-to-main-cluster mass
ratios to cluster-sized halo growth. We perform this test by dynamically
analyzing seven galaxy clusters, spanning the redshift range $0.13 < z_{\rm c} <
0.45$ and caustic mass range $0.4-1.5$ $10^{15} h_{0.73}^{-1}$ M$_{\odot}$,
with an average of $293$ spectroscopically-confirmed bound galaxies to each
cluster. The large radial coverage (a few virial radii), which covers the whole
infall region, with a high number of spectroscopically identified galaxies enables
this new study.
For each cluster, we identify bound galaxies. Out of these galaxies, we identify
infalling and accreted halos and estimate their masses and their dynamical
states. Using the estimated masses, we derive the contribution of different mass
ratios to cluster-sized halo growth. For mass ratios between $\sim 0.2$ and
$\sim 0.7$, we find a $\sim 1 \sigma$ agreement with $\Lambda$CDM expectations 
based on the Millennium simulations I and II. At low mass ratios, 
$\lesssim 0.2$, our derived contribution is underestimated since the detection efficiency 
decreases at low masses, $\sim 2 \times 10^{14}$ $h_{0.73}^{-1}$ M$_{\odot}$. At large 
mass ratios, $\gtrsim 0.7$, we do not detect halos probably because our sample, which 
was chosen to be quite X-ray relaxed, is biased against large mass ratios. Therefore, at large
mass ratios, the derived contribution is also underestimated.

\end{abstract}

\keywords{dark matter - galaxies: clusters: individual (Abell 611, Abell 963,
Abell 1423, Abell 2261, MACS J1206.2-0848, RX J2129.7+0005, 
CL2130.4-0000) - galaxies: kinematics and dynamics}

\section{Introduction}
\label{Introduction}

$\Lambda$CDM makes clear predictions for the growth of mass in halos due to
accretion of other halos with different masses (Fakhouri \& Ma 2008, hereafter
FM08; Berrier et al.\ 2009; Fakhouri, Ma, \& Boylan-Kolchin 2010, hereafter
FMB10; Genel et al.\ 2010, hereafter G10). For example, G10, who used several
N-body simulations to construct merger trees while taking special care of halo
fragmentation, suggested that mergers with mass ratios larger than $1/3$
($1/10$) contribute $\approx 20\%$ ($\approx 30\%$) of the total halo mass
growth. 

Confirming these predictions observationally is an important test for the
cosmological model. Using spectroscopically identified galaxies that cover the
whole cluster infall region is ideal for this purpose. The 3D spectroscopic data
(2D spatial information plus the redshift data) enable identification of bound
galaxies (den Hartog \& Katgert 1996, hereafter HK96; Diaferio 1999, hereafter
D99). Then, out of all the bound galaxies, one can identify the cluster and the
infalling and accreted satellite halos. The mass ratios between the cluster and
these identified satellite halos can reach low values, e.g. $\sim
1/100$ (Adami et al.\ 2005; this work), which enable a wide range for comparing
the estimated growth of cluster mass due to mergers with different
mass ratios and the theoretical predicted one. 

In a previous work, Adami et al.\ (2005) identified $17$
groups within the Coma cluster. They grossly estimated these groups' masses
and survival times in the cluster. Then they estimated that at least $\sim
10-30\%$ of the cluster mass was accreted since $z \sim 0.2-0.3$ by halos with
mass ratios larger than about $1/100$, in agreement with the theoretical
prediction. However, since the simulations-based predictions are made by averaging over a
large number of clusters, we need to compare these predictions with estimations
made from a cluster sample. We also need to observe the halos while they infall
and accrete in order to follow the cluster mass accretion process itself. 

In this work, we first identify and exclude unbound galaxies from a sample of seven
galaxy clusters. We measure the clusters' masses using different mass
estimators, and derive various mass uncertainties. Then, out of all the bound
galaxies, we also define infalling and accreted\footnote{In some cases, the
identified accreted satellite halos are fully within the cluster virial radius
and can be considered as substructure. However, in this work we consider them as
accreted satellites.} satellite halos and estimate their masses and dynamical
states. We use the measured masses of both the satellites and clusters to
estimate the growth of cluster-sized halos via the accretion of smaller halos.
Finally, we compare our observational constraints of cluster growth to those
from simulations. 

The Cluster Lensing and Supernova survey with Hubble (CLASH, Postman et al.\
2012, hereafter P12) is a large Hubble program imaging 25 galaxy clusters.
This cluster survey is also covered by supporting observations from a large
number of space and ground based telescopes, enabling an unprecedented
multi-wavelength study of clusters. Extensive ground-based spectroscopy for
galaxies in the environs of the CLASH clusters was either available from the
Sloan Digital Sky Survey (SDSS; Stoughton et al.\ 2002) and the Hectospec
Cluster Survey (HeCS; Rines et al.\ 2013) or was initiated specifically in
support of the CLASH program using the VLT/VIMOS and Magellan/IMACS instruments.
The large radial coverage (a few virial radii), which covers the whole infall
region, with a high number of
spectroscopically identified galaxies enables this new study. In this paper, we
study the dynamics of a subset of the CLASH clusters: Abell 611 (hereafter
A611), Abell 1423 (hereafter A1423), Abell 2261 (hereafter A2261), MACS
J1206.2-0848 (hereafter MACSJ1206), and RX J2129.7+0005 (hereafter RXJ2129). In
addition, we also analyze CL2130.4-0000 (hereafter CL2130), which is in the
foreground of RXJ2129, and Abell 963 (hereafter A963), which was originally in
the CLASH sample but was replaced by A1423 when it was found that A963 could not
be scheduled for HST observations due to a lack of usable guide stars. 

The paper is organized as follows. In \textsection~\ref{data}, we present a
short description of the data we used and their reduction process. In
\textsection~\ref{Methods and tests}, we describe the methods and tests
used to identify non-cluster members using the spectroscopic data, estimate
halos' masses, identify halos, and estimate
halos' substructure levels and relaxation states. In \textsection~\ref{Expected
fraction of cluster mass accretion}, we calculate the simulations-based expected
fraction of cluster mass accretion. In \textsection~\ref{Results}, we show our
results, where in \textsection~\ref{Clusters mass profiles} we present the
derived mass profiles, in \textsection~\ref{Substructure level} we estimate the
clusters' dynamical state, in \textsection~\ref{Accreted halos} we show our
findings regarding the growth of cluster-sized halos, and in
\textsection~\ref{Substructure-Relaxation connection} we show our estimations
for identified satellites' substructure level and relaxation state and the
correlation between them. In \textsection~\ref{Discussion}, we discuss
our results and summarize them in \textsection~\ref{Summary}.

Unless explicitly mentioned, the cosmology used throughout this paper
is WMAP7 (Komatsu et al.\ 2011), i.e. $\Omega_{\rm m} = 0.272$,
$\Omega_{\rm \Lambda} = 0.728$, and $h \equiv 0.73 h_{\rm 0.73} = 0.73$ where
$H_0 = 100 h$ km s$^{-1}$ Mpc$^{-1}$. Errors represent
a confidence level of $68.3\%$ ($1\sigma$). 

\section{Cluster Sample and Observational Data}
\label{data}

The sample selected for the dynamical analyses here consists of seven clusters
with extensive spectroscopic redshift measurements. These clusters are the
first ones we acquired (out of about the 20 X-ray selected) in the CLASH
project with a high number (a few hundred) of cluster members and infalling
galaxies. The clusters locations and redshifts are given in
\textsection~\ref{Results} (where we derive most of them). For more of their 
X-ray properties (except for CL2130), see P12 (table 4).

\subsection{Spectroscopy}
\label{Spectroscopy}

Spectra for galaxies in the environs of each cluster were drawn from a
combination of existing data and new observations. For A611, A963, A1423, A2261,
RXJ2129, and CL2130 the bulk of the redshifts were obtained using
the Hectospec instrument on the MMT (Fabricant et al.\ 2005). The sample
selection, observational parameters, and data reduction procedures for the
Hectospec data are described in detail in Rines et al.\ (2013), except for
$\sim 50\%$ of A611 redshifts, which were first used in Newman et al.\ (2013a,b). 
The Hectospec's circular 1$^{\circ}$ diameter field of view covers
the entire virial region and a significant fraction of the infall region of the
above clusters in a single pointing. The SDSS DR7 release (Abazajian et al.\
2009) was used to provide additional redshift information for $439$ galaxies. 

In tables~\ref{MMT-CL2130} and~\ref{MMT-A611}, we present the MMT/Hectospec
redshifts for CL2130 and A611, respectively. In these tables, columns 1 and 2 
list the galaxy's equatorial coordinates (in degrees) for epoch J2000.0. Columns 
3 and 4 list the heliocentric redshift and the redshift error, respectively. The 
fifth column contains the Tonry cross correlation coefficient (Tonry \& Davis 1979), 
$R_{\rm cross}$. Rines et al.\ (2013) assigned quality flags to each spectral fit. 
The flags are "Q" for high-quality redshifts, "?" for marginal cases, and "X" for 
poor fits. Although repeated observations of several targets with "?" flags show that 
these redshifts are generally reliable, we use only the high-quality redshifts in this 
paper. In tables~\ref{MMT-CL2130} and~\ref{MMT-A611}, we present only the high-quality 
redshifts (including the ones we have for stars). The MMT/Hectospec redshifts for A963, 
A1423, A2261, and RXJ2129 are published in Rines et al.\ (2013). 

\begin{table*}
\caption{CL2130 MMT/Hectospec Spectroscopic Redshifts\label{MMT-CL2130}}
\begin{center}
\begin{tabular}{ccccc}
\hline
  RA (J2000 deg)  &  DEC  (J2000 deg)  & $z$ & $\Delta z$ & $R_{\rm cross}$  \\
\hline

322.6146364 & -0.0816536 & 0.134668 & 0.000058 & 13.64  \\ 
322.7121945 & -0.2050228 & 0.135396 & 0.000156 & 7.07   \\
322.3012405 & -0.3572667 & 0.136382 & 0.000153 & 7.33   \\
322.2246194 & -0.0703775 & 0.184074 & 0.000120 & 8.04   \\
322.3168903 & -0.0935747 & 0.143829 & 0.000054 & 21.69  \\

 \hline 
\end{tabular}
\end{center}
\tablecomments{This table is available in its entirety in a machine-readable
form in the online journal. A portion is shown here for guidance regarding its
form and content.}
\end{table*}

\begin{table*}
\caption{A611 MMT/Hectospec Spectroscopic Redshifts\label{MMT-A611}}
\begin{center}
\begin{tabular}{ccccc}
\hline
  RA (J2000 deg)  &  DEC  (J2000 deg)  & $z$ & $\Delta z$ & $R_{\rm cross}$  \\
\hline

120.2032223 & 36.1133006 & 0.144207 & 0.000023 & 19.06  \\ 
120.0800936 & 36.4262502 & 0.257982 & 0.000108 & 13.33  \\
119.8793292 & 36.3154402 & 0.176505 & 0.000075 & 17.47  \\
120.1680427 & 36.1790189 & 0.503699 & 0.000115 & 11.16  \\
119.7670295 & 36.4042486 & 0.271796 & 0.000080 & 19.48  \\

 \hline 
\end{tabular}
\end{center}
\tablecomments{This table is available in its entirety in a machine-readable
form in the online journal. A portion is shown here for guidance regarding its
form and content.}
\end{table*}

The vast majority (2485) of the 2535 galaxy redshifts we use for MACSJ1206 were
obtained using the VIMOS instrument (Le Fevre et al.\ 2003) on the VLT in
multi-object spectroscopy mode. We consider only the VLT high reliable, $\gtrsim 80\%$, 
redshift estimates. The VIMOS data were acquired using 4 separate
pointings, always keeping one quadrant centered on the cluster core. This
strategy allows us to get spectra for fainter arcs in the core down to R
$\approx$ 25.5 mag and to a $\sim$80\% success limit of R $\approx$ 24.5 in the
non-overlapping regions. A broad spectral range, from 370 nm -- 970 nm, was
achieved by using the LR blue and the MR grisms, which yield spectral
resolutions of 180 and 580, respectively. For more details about the VIMOS
target selection, survey design, and data, see Rosati et al.\ (in prep.).

We specifically targeted galaxies within the core of the MACSJ1206 using the Inamori Magellan
Areal Camera and Spectrograph (Dressler et al.\ 2011) on the Baade 6.5 m telescope. We utilized the
Gladders Image-Slicing Multislit Option which reformats a $4'\times 4'$ field over a wider
area of the telescope focal plane, thus enabling a large increase in the multiplexing capability
of the instrument. We accumulated 210 min of exposure time in
six exposures with the 300 mm$^{-1}$ grism, with a dispersion of $\sim 1.34$\AA, and a resolution
of $\sim 5$\AA\ (FWHM). We obtained redshifts for 21 galaxies to $I_{\rm F814W}=22$ mag.

The remaining 29 redshifts in MACSJ1206, were taken from the literature (Jones et al.\ 2004, 
Lamareille et al.\ 2006, Jones et al.\ 2009, Ebeling et al.\ 2009).

For the target selection plane of the sky completeness, see Appendix~\ref{Survey
completeness}.

\subsection{X-ray profiles}

We derived cumulative mass profiles from the public ACCEPT (Cavagnolo et al.\
2009) intracluster medium projected temperature and deprojected density
profiles by assuming the hot gas is in hydrostatic equilibrium with a
spherically symmetric cluster gravitational potential. These temperature and
density profiles were derived from archival Chandra data by Cavagnolo et al.\
(2009). Small calibration differences may change the absolute mass estimates by
a small amount, but within the uncertainty. The temperature profile was
interpolated to the resolution of the density profiles (as in Cavagnolo et al.\
). Some regularization of the resulting pressure profiles was required for
stable estimates of errors on the pressure gradients, so we applied a minimal
requirement that the shape of the pressure profile followed the "universal"
pressure profile derived by Arnaud et al.\ (2010). We did not assume an NFW
(Navarro et al.\ 1997) profile or a density radial profile (such as a
beta law). Best estimates and one sigma uncertainties for the cumulative mass
within each radius were estimated using a simulated annealing technique
(Kirkpatrick, Gelatt \& Vecchi 1983). These profiles are consistent with mass
profiles derived using other techniques (e.g. Umetsu et al.\ 2012; Donahue et
al.\ in prep.).

\section{Methods and tests}
\label{Methods and tests}

We describe the methods we use to identify galaxies that are bound to the
cluster, define (from the bound galaxies) satellites in the plane of the sky,
derive the cluster and infalling and accreted satellites mass estimates and the
corresponding uncertainties in those estimates, and estimate the level of
substructure and relaxation state of each halo.

\subsection{Determining the cluster redshift}
\label{First velocity cut}

We begin by obtaining an estimate for the mean redshift of each cluster, $z_{\rm
c}$. To estimate $z_{\rm c}$, we first select galaxies that lie within a
projected radius of 10 $h_{0.73}^{-1}$ Mpc from the cluster center, which is
chosen to be the peak of the cluster's X-ray surface brightness distribution. We
adopt an initial guess for $z_{\rm c}$ by fitting a Gaussian to the largest peak
in the redshift histogram of the galaxies lying within the above projected
radius. Velocity offsets from the mean cluster redshift are defined for each
galaxy as $v=c (z-z_{\rm c})/(1+z_{\rm c})$ (Harrison \& Noonan 1979). 

We then make an initial cut based on this velocity offset, excluding all
galaxies with $|v| > 4000$ km s$^{-1}$. A histogram of the redshifts of the
galaxies with $|v| \le 4000$ km s$^{-1}$ is then generated, using a velocity bin
size of $150$ km s$^{-1}$ (except for MACSJ1206 where it
is $200$ km s$^{-1}$). This bin size is larger than any of the redshift
measurement uncertainties used in this work (except for a few in MACSJ1206
which are not taken into consideration here). The galaxy number uncertainty in
each bin was derived assuming Poisson statistics, i.e. $\Delta N = \sqrt{N}$.
Poisson uncertainty of zero counts was defined as 1. We then refine our estimate
of the mean cluster redshift by fitting a Gaussian, $G_1$, to the redshift
histogram. We then see how the goodness of fit changes by adding in additional
Gaussians to the fit: 
\begin{equation}
N(v) = G_1(v) + \sum_{i=2}^N G_i(v)
\end{equation}
where $G_i = A_i \exp \left(-\frac{(v-\overline{v}_i)^2} {2\sigma_i^2} \right)$,
where $A_i$, $\overline{v}_i$, and $\sigma_i$ are the Gaussian $i$
normalization, average, and dispersion, respectively, and they all are taken to
be free parameters. The main cluster peak corresponds to $i = 1$. If an
addition of a subsequent Gaussian decreases the $\chi_r^2$, it is added.

The cluster's redshift is then determined to be the average value of the
cluster's halo Gaussian. Later in the paper (see
table~\ref{Clusters centers and redshifts info table}), this value is compared
with the median redshift of the galaxies found using D99 interloper removal
method (see \textsection~\ref{D99 interlopers removal
method}). It is important to note that the results of the above
Gaussian/Gaussians fitting are only used for determining $z_{\rm c}$ and not for
any other dynamical or mass estimation application.

\subsection{Removal of non-cluster galaxies}
\label{Removal of non-cluster galaxies}

Due to projection, any cluster dynamical data sample inevitably contains
galaxies that are not bound to the cluster, i.e. interlopers. Removing them is
important for accurately estimating the dynamical cluster properties. For
removing them, a velocity-space diagram is constructed where the galaxies'
line-of-sight (measured with respect to the cluster's mean redshift, hereafter
LOS) velocities, $v$, are plotted versus their projected distances from the
cluster center, $R$. For a well defined cluster, the galaxies should be
distributed in a characteristic ``trumpet'' shape, the boundaries of which are
termed caustics (Kaiser 1987; Regos \& Geller 1989). Galaxies that are outside
the caustics are considered to be interlopers. We test two widely used
interlopers removal techniques that take into account the combined position and
velocity information to estimate to caustic location. The first method relies
on calculating the maximum LOS velocity that a galaxy may be observed to have
(HK96). The second relies on first
arranging the galaxies in a binary tree according to a hierarchical method to 
determine the velocity dispersion and mean projected distance of the 
members, and
then estimating membership by the escape velocity (Diaferio \& Geller 1997;
D99). Both methods assume spherical symmetry. For consistency with other works,
in the rest of this paper, the caustic notation refers only to the second
method. We briefly describe the two methods in Appendix~\ref{Removal of
non-cluster galaxies - techniques description}.

\subsection{Mass estimators}
\label{Mass proxies}

In this section, we briefly describe the three different dynamical based mass
estimators used in this work. The use of different mass estimators shows the
uncertainties due to the mass profile estimator picked, and increases the
reliability of our results. 
In Appendix~\ref{Mass profile biases}, we examine various possible mass profile 
biases, irrespective of the mass estimator used.

\subsubsection{Virial and projected mass estimators}
\label{Virial and projected mass proxies}

Two widely used mass estimators are the virial and projected mass estimators.
Both of them are derived from the collisionless Boltzmann equation assuming
the system is in steady state and spherical symmetry. Further assumptions are
that galaxies trace the dark matter (hereafter DM) distribution and that all galaxies have the same mass
(Bahcall \& Tremaine 1981, and reference within; Heisler, Tremaine, \& Bahcall
1985; Binney \& Termain 2008). 

The virial mass profile estimator is
\begin{equation}
M_{\rm v}(\leqslant r)\approx M_{\rm v}(\leqslant R) = \frac{3\pi N}{2G}
\frac{\sum_i^N (v_i - \overline{v})^2}{\sum_{i<j}^N 1/R_{ij}} \ ,
\end{equation} where $r$ is the distance to the cluster center, $R$ is the
projection of $r$, $N$ is the number of cluster galaxies inside $R$, $v$ is
the LOS velocity, and $R_{ij}$ are projected distances of galaxy pairs within a
cylinder of radius $R$ around the center.

The projected mass profile estimator\footnote{The projected mass profile
estimator is based on the projected mass $q$, where $q \equiv v^2
R/G$, but is not the projected mass (Bahcall \& Tremaine 1981).}, 
\begin{equation}
M_{\rm proj}(\leqslant r)\approx M_{\rm proj} (\leqslant R)= \frac{f_{\rm
PM}}{\pi G}\sum_i^N {R_i(v_i-\overline{v})^2}/N \ ,
\end{equation} is more robust in the presence of close pairs because it sums 
$R$ rather than $1/(1/R)$ (Bahcall \& Tremaine 1981). The disadvantage
of the projected mass estimator is that it requires to define a center. The
constant of proportionality $f_{\rm PM}$ depends on the distribution of orbits,
where $f_{\rm PM}=64/\pi,32/\pi,16/\pi$ for radial, isotropic, and circular
orbits, respectively (Heisler, Tremaine, \& Bahcall 1985; Rines 2003). Because
both radial and circular orbits are considered, we set $f_{\rm PM}=32$
throughout this paper.

Halos are not isolated systems since matter is continuously falling onto them.
This infalling matter was claimed to have a significant overall contribution to
the pressure at the halos' boundaries (Shaw et al. 2006; Davis et al. 2011).
Therefore, the usual formula of the virial theorem $2T+U = 0$ (Binney \&
Tremaine 2008) should be replaced by $2T+U = 3PV$ where $3PV$ is the surface
pressure term (Chandrasekhar 1961; The \& White 1986; Carlberg et al.\ 1996,
1997a; Shaw et al. 2006; Davis et al. 2011; Lemze et al.\ 2012). If the pressure
term is not taken into consideration, the mass is overestimated. In
particular we assume that mass follows the galaxy distribution and follow
Girardi et al.\ (1998), so the corrected virial/projected mass profile, $M_{\rm
Cv/Cproj}$, is: 
\begin{equation}
M_{\rm Cv/Cproj} =  M_{\rm v/proj} \left\{1 - 4\pi
b^3\frac{\rho(b)}{\int_0^b 4\pi r^2 \rho dr}\left[
\frac{\sigma_{\rm r}(b)}{\sigma(<b)} \right]^2 \right\} \ ,
\end{equation} where $\rho(r)$ is the cluster mass density, $\sigma(<b)$
refers to the integrated velocity dispersion within the boundary radius $b$,
and $\sigma_{\rm r}(b)$ is the radial velocity dispersion at $b$. Assuming the
velocity anisotropy is constant (as assumed for deriving the projected mass
estimator) and in the limiting cases of circular, isotropic, and radial orbits,
the maximum value of the term involving the velocity dispersions is $0$, $1/3$,
and $1$, respectively. We use $1/3$ to be consistent with the $f_{\rm PM}$ value
adopted for an isotropic orbit distribution.

For estimating $\rho$, here we assume an NFW mass profile and fit it to
the mass profile. Since the NFW best-fit parameters are derived by fitting the
uncorrected mass profile, we repeat the fitting using the corrected mass
profile. This process is iterated where in each step we take the parameters
of the most recent corrected mass profile. After a few iterations the mass
profile converges (when we adopt $10^{-6}$ tolerance).

\paragraph{Error analysis}

We assess the virial and projected mass uncertainty using the Jackknife
technique, which was introduced by Quenouille (1949) and Tukey (1958). This is
one of the simplest, wide spread, and quickest resampling method techniques. The
mass uncertainty is
\begin{equation}
 dM_{\rm v/proj}(\leqslant r)\approx dM(\leqslant R) =
\frac{N-1}{N}\sqrt{\sum_{i=1}^N[M_{-i}(\leqslant
R)-\overline{M}_{-i}(\leqslant R)]^2} \ ,
\end{equation} where $M_{-i}(\leqslant R)$ is the mass estimation when
we do not use the galaxy $i$ out of all galaxies within $R$, and
$N$ is the number of galaxies inside $R$ (Efron 1981, and reference
within).

\subsubsection {Caustic mass estimator}
\label{Mass profile from velocity caustics}

Diaferio \& Geller (1997) \& D99 showed that the 3D mass profile can be
estimated directly from the amplitude of the velocity caustics,
$A(R)$, (for how to determine $A(R)$, see Appendix~\ref{Building the redshit-phase diagram}),
\begin{equation}
M_{\rm caustics}(\leq r) = \frac{F_{\beta}}{G}\int_0^rA^2(R)dR\ ,
\label{M_from_A}
\end{equation} and its uncertainty
\begin{equation}
dM_{\rm caustics}(\leq r) = \frac{2F_{\beta}}{G}\int_0^rA(R) dA(R)dR
\label{M_from_A}
\end{equation} \footnote{By analyzing a sample of 3000 simulated clusters 
with masses of $M_{200}>10^{14}$ $h_{0.73}^{-1}$ M$_{\odot}$, Serra et al.\ (2011, hereafter 
S11) tested the deviation 
of this mass uncertainty recipe from the 1 $\sigma$ confidence level. They found that on average 
at $r_{200}$ the upper mass uncertainty is underestimated by about 15\% and the lower mass 
uncertainty is overestimated about 25\% (see right panel in their figure 16; Ana Laura Serra, 
private communication).}. We adopt $F_{\rm \beta} = 0.7$ (for more details, see
Appendix~\ref{F_beta}).

\subsection{Halo identification}
\label{Halos identification}

We are interested in estimating the contribution of mergers with different mass
ratios to cluster growth. The first step is identifying the satellites accreted
and falling halos into the cluster. In \textsection~\ref{Over density}, we
present our scheme to identify these halos. Because our estimation depends on
the identification scheme, in \textsection~\ref{FoF}, we present a different
scheme, the Friends of Friends (hereafter FoF), to identify these halos. The
latter scheme will be used to test the sensitivity of our results to the halo
finder (hereafter HF) used, and increase reliability. 

In our case, the FoF scheme identifies more small and elongated mass halos. In addition, in
some cases, one overdensity halo is identified as two closeby FoF halos.
Nevertheless, using the two different schemes results in estimations which are
in 1 $\sigma$ agreement. 

Since interlopers are already removed in a previous step (as described in 
\textsection~\ref{Removal of non-cluster galaxies}), we only identify the halos using celestial 
coordinates and neglect line of sight separation within the halos.

\subsubsection{2D Overdensity}
\label{Over density}

The method we present here is close in spirit to a combination of 
denMAX (Bertschinger \& Gelb 1991; Gelb \& Bertschinger 1994) and spherical
overdensity (Lacey \& Cole 1994), which are two HFs commonly used in
simulations. For each cluster, we make an initial velocity cut, as was
described in \textsection~\ref{First velocity cut}, build a binary tree, as
described in Appendix~\ref{D99 interlopers removal method}, and identify
galaxies that are bound to the cluster (cluster member galaxies and infalling
galaxies) using the D99 procedure. The axes of the galaxy surface density map 
are cut to include all the identified bound galaxies, and scaled with the same bin 
resolution, $B_{\rm res}$. We then smoothed
the galaxy surface density using a 2D Gaussian kernel with a fixed side size of
$1/10$ ($1/15$ for MACSJ1206) of the identified galaxies squared field of view.
The amount of smoothing, $\sigma_{\rm kernel}$, is taken to be equal in both
axes, and $\sim 30\%$ of the kernel size. This kernel size and level of
smoothing are optimized to include at least a few galaxies at the densest areas,
but not to erase significant ($3 \sigma$ above the average) features. 

For identifying accreted halos (and infalling halos), after the smoothing we
identify significant ($3 \sigma$ above the average) galaxy surface density
peaks. Now we need to estimate the halos' boundaries. As Kravtsov et al.\
(2004) mentioned, the virial radius is meaningless for the subhalos within a
larger host as their outer layers are tidally stripped, and the extent of the
halo is truncated. The definitions of the outer boundary of a subhalo and its
mass are thus somewhat ambiguous. The truncation radius is commonly estimated
at the point where $d \ln \rho(r)/d \ln r = -0.5$ since it is not expected that
the density profile of the CDM halos will be flatter than this slope (Kravtsov
et al.\ 2004). Kravtsov et al.\ (2004) note that this empirical definition of
the truncation radius roughly corresponds to the radius at which the density of
the gravitationally bound particles is equal to the background density of the
host halo, albeit with a large scatter. In our case, although many of the halos
are beyond the cluster virial radius, they are embedded in high density regions
with many galaxies in their surrounding. Estimating any specific characteristic
radius, such as the truncation or the halo's virial radius, via a density
profile is not possible because the data is not sufficient to estimate a density
profile for all halos. Therefore, we follow Kravtsov et al.\ suggestion and
limit the halos to surface densities which are $2 \sigma$ above the average.
Below this threshold, e.g. $\sim 1.5 \sigma$, the halos' sizes do not increase
by much, and halos within the cluster virial radius are bridged to the
cluster background.

If there are a few significant peaks inside one $2 \sigma$ region, we
estimate the smoothed galaxy surface density minimum between them, i.e.
$\Sigma_{\rm ridge}$. If $\Delta_{\rm peak} = \Sigma_{\rm peak}/\Sigma_{\rm
ridge}$, where $\Sigma_{\rm peak}$ is the peak of the smoothed surface
density, is smaller than some threshold (which is taken to be $1.1$) we
do not consider the peak with the lowest density to be significant. If
$\Delta_{\rm peak}$ is above the threshold, the $2 \sigma$ region is divided
by a perpendicular (to the line connecting the two peaks) line which passes
via the location of $\Sigma_{\rm ridge}$. 

For each cluster, the surface density region which is $2 \sigma$ above the
average and includes the cluster center (determined by the X-ray peak) is notated as the cluster core, and is
not considered to be one of the accreted halos. The core region can be about the
size of the cluster (as defined by the virial radius, as is the case for
RXJ2129).

The uncertainties in the galaxies surface density peaks for both the cluster and
identified peaks are taken to be $0.5 B_{\rm res} \sigma_{\rm kernel}$.
Different kernel sizes can of course shift the galaxies surface density
peaks but they will be in agreement within 1$\sigma$.

\subsubsection{2D Friends of Friends (FoF)}
\label{FoF}

We adopt the widely used standard FoF algorithm (Huchra \& Geller 1982; Davis
et al.\ 1985). Groups are defined by linking together all pairs of galaxies with
separations less than some linking length, $l_{\rm linking}$, which is taken to
be some fraction, $b_{\rm frac}$, of the averaged inter-galaxy spacing,
$\overline{r}_{\rm inter}$. The latter is estimated as the mean of all the
distances between each galaxy and its closest neighbor. Thus, $l_{\rm linking} =
b_{\rm frac} \overline{r}_{\rm inter}$. 
Except for $b_{\rm frac}$, the second free parameter is the minimum number of
galaxies in a group, which we take to be
$N_{\rm gal,min} = 6$ for consistency with the minimum number of galaxies in a
halo found by the overdensity HF (see table~\ref{Halos info}). Lower values for
$N_{\rm gal,min}$ increase the number of low mass halos when many of them have a
more filamentary shape than approximately round shape. Values of $b_{\rm frac}
\lesssim 0.45$ are too small to identify obvious high mass halos, while values
$b_{\rm frac} \ga 0.7$ are too large so halos are joined to the core. We
set $b_{\rm frac} = 0.54$ (values in the $0.51-0.55$ range give almost identical 
result) for consistency with the overdensity HF. At
$b_{\rm frac} = 0.5$ one of A963 halos, which is identified by overdensity
HF, is not identified by the FoF HF and at $b_{\rm frac} = 0.56$ one of A1423
halos, which is identified by overdensity HF, is fused with the core).

We consider only FoF halos that have most of their galaxies outside the cluster
core (identified by the overdensity HF).

\subsection{Substructure and relaxation tests}
\label{Substructure and relaxation tests}

The mass estimators (virial and projected), which are used the estimate the
halos' masses, assume the halos are in steady state (see \textsection~\ref{Mass
proxies}). Therefore, it is important to have an indication for the halos'
relaxation levels. In addition, we are interested to see if there is a strong
correlation between the infalling and accreted satellites substructure and
relaxation levels. In this section, we briefly describe the Dressler-Shectman
test, which is used to measure the level of substructure, and center
displacement tests, which are used to estimate the relaxation state.

\subsubsection{Dressler-Shectman test}
\label{Dressler-Shectman test}

In order to check for the presence of substructure in the three-dimensional
space, we compute the statistics devised by Dressler \& Shectman (1988,
hereafter DS). Pinkney et al.\ (1996, hereafter P96) examined various
substructure tests, and DS was found to be the most sensitive
three-dimensional test. The test works in the following way: for each galaxy
that is a cluster member, the $N_{\rm local}$ nearest neighbors are found,
and the local velocity mean and dispersion are computed from this sample of
$N_{\rm local}+1$ galaxies. The deviation of the local velocity mean and
dispersion from the cluster velocity mean and dispersion are calculated,
\begin{equation}
\delta_i^2=\frac{N_{\rm local}+1}{\sigma^2} \left[
\left(\overline{v}_{local,i}-\overline{v}\right)^2 +
\left(\sigma_{local,i}-\sigma  \right)^2  \right],
\end{equation} where $\overline{v}$ and $\sigma$ are the global dynamical
parameters and $\overline{v}_{local,i}$ and $\sigma_{local,i}$ are the local
mean velocity and velocity dispersion of galaxy $i$, determined using itself and
its $N_{\rm local}$ closest galaxies. Note, other forms of $\delta_i$ were also
suggested (e.g. Biviano et al.\ 2002). Dressler \& Shectman also defined
the cumulative deviation
\begin{equation}
 \Delta=\sum_{i=1}^{N_{\rm gal}}\delta_{i} \ ,
\end{equation} were $N_{\rm gal}$ in the number of cluster
members. 

It is necessary to calibrate the $\Delta$ statistics for each cluster (DS; Knebe
\& Muller 2000). The calibration is done by randomly shuffling among the
positions. If substructure exists, the shuffling erases any correlation between
redshifts and positions. We have done this procedure $10^4$ times for each
cluster. Then, we define $P(\Delta_{\rm s}>\Delta_{\rm obs})$ as the fraction
of the total number of Monte Carlo models of the cluster that have shuffled
values, $\Delta_{\rm s}$, larger than the cluster observed value, $\Delta_{\rm
obs}$. $P(\Delta_{\rm s}>\Delta_{\rm obs})\sim 1$ means that the cluster
contains no substructure, while $P(\Delta_{\rm s}>\Delta_{\rm obs})\sim 0$
indicates that the cluster contains statistically significant substructure.

Originally, Dressler \& Shectman proposed the computation of $\Delta$
using $N_{\rm local} = 10$ independently of the number of galaxy cluster
members. Bird (1994, and references therein) pointed out that using a constant
value for the number of nearest neighbors reduces the sensitivity of this test
to significant structures, and suggested using $N_{\rm local} =
\sqrt{N_{\rm gal}}$.

The test is sensitive to outliers (P96), and therefore we use it also to
identify which interloper removal method is more adequate to deal with
outliers (see \textsection~\ref{Discussion}).

\subsubsection{Relaxation tests}
\label{Relaxation tests}

The distinction between relaxed and unrelaxed clusters is made by estimating
the displacement between two different definitions for the cluster center.
These relaxation proxies have been widely used both with data (e.g. P12) and
simulations (e.g. Neto et al.\ 2007; Maccio et al.\ 2007; Lemze et al.\ 2012).
Sometimes they are also interpreted as a measure of the substructure
level (see Thomas et al.\ 2001).

We used two different displacements. The first test is the displacement
between the galaxy surface density peak, ${\bf R}_{\rm sd}$, and the potential center
(potential minimum), ${\bf R}_{\rm p}$, which is estimated as follows, 
\begin{equation}
\Phi({\bf R}_{\rm p}) = \min \left\{-G M_{\rm gal} \sum_{i = 1}^{N_{\rm gal}}
\frac{1}{|{\bf R}_{\rm i}-{\bf R}|}\right\} \ ,
\end{equation} where $M_{\rm gal}$ is taken to be $10^{12}$ $h_{0.73}^{-1}$
M$_{\odot}$, the value used to build the binary tree (see
Appendix~\ref{D99 interlopers removal method}).
The potential center is calculated using
galaxies within the halo's boundaries, which are taken to be the caustic
virial radius for the clusters and the 2$\sigma$ contours for the infalling
and accreted satellites (for more details, see \textsection~\ref{Accreted
halos}). The displacement is normalized with respect to a scale radius,
$r_{\rm scale}$, which is the caustic virial radius for the clusters and an
effective radius, $r_{\rm eff} = \sqrt{S}/\pi$, for the halos, where $S$ is the
area inside the 2$\sigma$ contour
\begin{equation}
 r_{\rm sp} = |{\bf R}_{\rm sd} - {\bf R}_{\rm p}|/r_{\rm scale} \, .
\end{equation}
The second test is the displacement between the center of mass , ${\bf R}_{\rm cm}$,
and the potential center
\begin{equation}
 r_{\rm cp} = |{\bf R}_{\rm cm} - {\bf R}_{\rm p}|/r_{\rm scale} \, .
\end{equation}
The center of mass is estimated by taking the galaxies projected locations, and
assuming the DM particles are distributed like the galaxies. 

In equilibrium, $r_{\rm sp}$ and $r_{\rm cp}$ are expected to vanish, while
high values, $\gtrsim 1$, indicate an unrelaxed halo. The threshold between
the two phases is quite arbitrary. Note also that due to projections these
criteria are less sensitive to LOS mergers.

\section{Expected fraction of cluster mass accretion}
\label{Expected fraction of cluster mass accretion}

Here we estimate the expected fraction of cluster mass accretion by following
the merger rates and mass assembly histories of DM halos in the two Millennium
simulations (FMB10; G10). 

The halo's mass as a function of redshift can be fitted using a two parameter
($\beta_{\rm MAH}$ and $\gamma_{\rm MAH}$) function
\begin{equation}
M(M_0,z) = M_0(1+z)^{\beta_{\rm MAH}(M_0)}\exp(-\gamma_{\rm MAH}(M_0)z) \ ,
\label{M(z)}
\end{equation} where $M_0 = M(z=0)$ and can be expressed as a function of the
observed cluster mass, $M_{\rm obs}$, and observed redshift, $z_{\rm obs}$,
\begin{equation}
 M_0 = \frac{M_{\rm obs}}{(1+z_{\rm obs})^{\beta_{\rm
MAH}(M_0)}\exp(-\gamma_{\rm MAH}(M_0)z_{\rm obs})} \ .
\label{M0}
\end{equation} This function is versatile enough to accurately capture the
main features of most mass accretion histories (MAHs) in the Millennium
simulation (McBride, Fakhouri, \& Ma 2009, see also Tasitsiomi et al.\ 2004).
The parameters in eq.~\ref{M(z)} are derived by fitting this function to the mass
history inferred from the mean mass growth rates of halos (see eq. 2
in FMB10)
\begin{equation}
\left< \dot{M} \right> = 46.1 \; M_{\odot} \; yr^{-1} \; \left(
\frac{M(z)}{h_{0.73}^{-1} 10^{12}M_{\odot}} \right)^{1.1} (1+1.11z)\epsilon(z) \
, \label{mean mass growth rate}
\end{equation} where $\epsilon(z)$ is the normalized Hubble function, i.e.
$\epsilon(z) = \sqrt{(\Omega_{\rm m}(1+z)^3+\Omega_{\rm \Lambda})}$. For
$\beta_{\rm MAH}$ and $\gamma_{\rm MAH}$ explicit dependence on $M_0$, see
Appendix~\ref{beta_MAH and gamma_MAH}.

FM08 introduced merger rates that fit well the ones found in the Millennium
simulations I and II (G10; FMB10). The mean merger rate of halos with the
descendant mass $M$ with other halos with mass ratio $\zeta$ at $z$ in units of
mergers per halo per unit redshift per unit $\zeta$ is
\begin{equation}
\frac{dN_{\rm m}}{d\zeta dz}(M_0,\zeta,z) =
A_{\rm m}\left(\frac{M(M_0,z)}{h_{0.73}^{-1} 10^{12}
M_{\odot}}\right)^{\alpha} \zeta^{\rm \beta}
\exp\left(\left(\frac{\zeta}{\tilde{\zeta}}\right)^{\gamma}\right)(1+z)^{\eta}
\ ,
\label{N_m}
\end{equation} when $A_{\rm m}=0.065$, $\alpha=0.15$, $\beta=-1.7$,
$\gamma=0.5$, $\tilde{\zeta}=0.4$, and $\eta=0.0993$. These (especially $\beta$,
$\gamma$, $\tilde{\zeta}$ and $A_{\rm m}$) values are obtained by taking special
care of halo fragmentation and ensuring that the mass contribution of each
merger to halo's growth is counted just once (G10). There is some uncertainty in
the value of $\alpha$. FMB10 found a lower value of $\alpha$ ($\alpha=0.133$).
We show the expected fraction of cluster mass accretion assuming $0.133$ as
well.

The Millennium simulation has a low number of massive cluster-sized halos and,
therefore, the statistics in this mass regime is limited. Wu et al.\ (2012)
analyzed a sample of 96 halos in the $10^{14.8\pm0.05}$ $h^{-1}$ M$_{\odot}$
mass range (about 4 times the number of halos with similar mass in the
Millennium simulation) from the Rhapsody cluster re-simulation project. They
found that the number of mergers per halo per unit redshift per unit $\zeta$ is
consistent in 1$\sigma$ with the one found in the Millennium simulation (see
their figure 6 right panel).

The mean mass accumulation per halo per unit redshift per unit $\zeta$ is
\begin{equation}
M_{\rm acc}(M_0,\zeta,z) = \frac{dN_{\rm m}}{d\zeta dz} M_{\rm small} \ ,
\end{equation} where $M_{\rm small}$ is the mass of the less massive progenitor
of each merger (G10). Since $M = M_{\rm mp}+M_{\rm small}$, where
$M_{\rm mp}$ is the main progenitor mass, and $\zeta \equiv M_{\rm small}/M_{\rm
mp}$, 
\begin{equation}
M_{\rm small}(M_0,\zeta,z) = M(M_0,z) \frac{\zeta}{1+\zeta} \ .
\end{equation}

Thus, the fraction of the cluster mass at $z=z_1$ accumulated at $z_1 \leqslant
z \leqslant z_2$ by progenitors with mass ratios $\zeta_1 \leqslant \zeta
\leqslant \zeta_2$ is
\begin{equation}
F(M_{\rm obs},z_{\rm obs}) = \frac{1}{M(z_1)}\int_{\zeta_1}^{\zeta_2}
d\zeta \int_{z_1}^{z_2} dz \; M_{\rm acc}(M_0,\zeta,z) \ .
\label{Estimated F}
\end{equation} 

We compare these predicted values to the ones based on observations. For the
comparison, we took $z_2 = z_{\rm obs}$ and $z_1 = z_{\rm f}$, which is the
redshift when all the bound matter falls onto the cluster (for more details, see
\textsection~\ref{Estimating z_f}).

\subsection{Correcting $\sigma_8$}
\label{Correcting sigma_8}

In the Millennium simulations I (Springel et al.\ 2005) and II (Boylan-Kolchin
et al.\ 2009), the linear mass density fluctuation amplitude in an 8
$h_{0.73}^{-1}$ Mpc sphere at redshift zero, $\sigma_8$, is $0.9$, while
the latest value is $0.82$ (Bennett et al.\ 2012; Hinshaw et al.\ 2012). A
lower value for $\sigma_8$ means slower structure formation, and therefore a
lower $\sigma_8$ is equivalent to a higher redshift in a high $\sigma_8$
universe. More specifically, we follow Angulo et al.\ (2012) and correct the
expectations derived from the Millennium simulations, which are based on the
values found by analyzing WMAP1 (the first year data from the WMAP satellite,
Spergel et al.\ 2003), to the ones based on the values found by analyzing WMAP7
(Komatsu et al.\ 2011), which are more updated and closer to the latest value.
Following Angulo et al.\ (2012), we denote the Millennium ($\Omega_{\rm m} =
0.25$ and $\sigma_8 = 0.9$) and WMAP7 ($\Omega_{\rm m} = 0.272$ and $\sigma_8 =
0.807$, Komatsu et al.\ 2011) cosmologies by $A$ and $B$, respectively. Angulo
\& White (2010) suggested scaling the Millennium simulations final redshift,
$z_{\rm A}^f$ (so that $z_{\rm A} \geqslant z_{\rm A}^f$), and box size scale
factor, $s$ (so that $L_{\rm A} = s L_{\rm B}$), to keep the linear fluctuation
amplitude the same in the two cosmologies, which is equivalent for having the
same halo mass function derived in Press-Schechter theory. For a zero final
redshift in the WMAP7 cosmology, $z_{\rm B}^f=0$, they found $z_{\rm A}^f=0.319$
and $s=1.072$ (Angulo et al.\ 2012).

To find the target redshifts, $z_{\rm A}$, we follow Angulo \& White (2010,
their eq. 5)
\begin{equation}
 D_A(z_{\rm A}) = D_B(z_{\rm B}) \frac{D_A(z_{\rm A}^f)}{D_B(z_{\rm B}^f)} \ ,
\end{equation} where $D$ is the linear growth factor, 
\begin{equation}
 D(z) = D_0 \epsilon(z) \int_0^{a(z)} \frac{1}{a^3\epsilon(a)^3}da \ ,
\end{equation} where $a=(1+z)^{-1}$ is the scale factor and $D_0$ is a constant
set by the normalization $D(0)=1$. 
This redshift transformation is applied to $z_{\rm obs}$, $z_1$, and $z_2$.

After scaling the positions with $s$, to keep the mass density the same in both
cosmologies, the particle mass is scaled as follows\footnote{The power of $h$
is $2$ and not $3$ because $m_{\rm p} \propto h^{-1}$.}
\begin{equation}
 m_{\rm p,A} = \frac{\Omega_{\rm m,A}h_{\rm A}^2 L_{\rm A}^3}{\Omega_{\rm
m,B}h_{\rm B}^2L_{\rm B}^3} m_{\rm p,B} \ ,
\end{equation} where in our case $h_{\rm A}=h_{\rm B}=0.73$
(Angulo \& White 2010). Since the halos at the two different cosmologies have
the same particle number, the difference in their mass is only due to the
scaling of the particle mass. Thus, 
\begin{equation}
 M_{\rm obs,A} = \frac{\Omega_{\rm m,A}}{\Omega_{\rm m,B}} s^3 M_{\rm obs,B}
\ .
\end{equation} In this case, the difference in $M_{\rm obs}$ due to the
different cosmologies is negligible, but $M_{\rm obs}$ was still scaled for
correctness.

\subsection{Estimating $z_{\rm f}$}
\label{Estimating z_f}

We are interested in comparing our simulations-based expectations of clusters'
mass growth with our estimated ones. Therefore, in our expectation calculations
(eq.~\ref{Estimated F}) we need to take the same redshift range as the
observed one. For each cluster, the observed redshift range of the clusters'
mass accumulation starts at about the cluster's redshift, $z_2=z_{\rm
obs}=z_{\rm c}$. For simplicity, we neglect the fact that accreted satellites
are partly or fully within the virial radius, and so $z_2$ is actually higher
than $z_{\rm c}$ (which increases the expected $F$). The observed redshift range
ends at the redshift when all the infalling matter reaches the cluster's virial
radius, $z_1=z_{\rm f}$. Below we explain how we estimated $z_{\rm f}$.

For each cluster, we estimate how much time it takes for matter to fall from
the furthest radius where galaxies are bound and falling onto the cluster,
$r_{\rm max}$ (for more details, see \textsection~\ref{Estimating r_max}), to the
clusters' virial radius, $r_{\rm caustics,vir}$. We do that by first separating
this radial range into $N_{\rm sections}$ smaller sections, corresponding
to $\Delta r = (r_{\rm
max}-r_{\rm caustics,vir}) / N_{\rm sections}$. We take $N_{\rm sections}=100$,
when larger values do not change the result significantly. The infalling time
in section $i$ is $\Delta t (r_i) = (-v_0(r_i) + \sqrt{v_0(r_i)^2+2a_{\rm
r}(r_i) \Delta r})/a_{\rm r}(r_i)$, where $a_{\rm r}(r) = GM_{\rm
caustics,vir}/r^2$ and $v_0$ is the velocity at the beginning of the section,
which propagates as $v_0(r_{i+1}) = a_{\rm r}(r_i) \Delta t(r_i)+v_0(r_i)$. Then
we sum the time of all sections, $t = \sum_i \Delta t_i$. Finally, we
convert the time into redshift, $t =
H_0^{-1}\int_{z_{\rm c}}^{z_{\rm f}}dz \; 1/(a(z) \epsilon(z))$, to yield
$z_{\rm f}$. Because the cluster's mass depends on the redshift (see
eq.~\ref{M(z)}), for each cluster we iterate this process. In each iteration, we
determined the cluster's final mass by the final redshifts inferred in the
previous iteration. The process stops when the final cluster's mass converges
(when we adopt $10^{-4}$ tolerance).

\subsection{Estimating $r_{\rm max}$}
\label{Estimating r_max}

We follow Rines et al.\ (2013) and estimate the furthest radius where
galaxies are bound and falling onto the cluster, $r_{\rm max}$, in a
conservative manner, 
\begin{equation}
 r_{\rm max} = \min(r_{\rm caustics,max},r_{\rm bound,max}) \ ,
\end{equation} where $r_{\rm caustics,max}$ is the maximum extent of the
caustics (the maximum radius where the caustics are above zero) and $r_{\rm
bound,max}$ is the maximum radius where all the galaxies are bound. The latter
is estimated to be the radius where $\bar{\rho}(<r_{\rm bound}) = \Delta_c
\rho_{\rm crit}$, where $\rho_{\rm crit}$ is the critical density of the
universe at the cluster redshift. The final overdensity to the critical density
at collapse, $\Delta_c$, can be derived from the critical mean
overdensity interior to the last shell that will collapse in the future,
$\delta_c$,
\begin{equation}
 \Delta_c = \Omega_{\rm m}(\delta_c+1) 
\end{equation} since the critical (final) cluster mean density contrast is 
$\delta_c = (\bar{\rho}-\rho_{\rm m})/\rho_{\rm m}$ and $\rho_{\rm m} \equiv
\Omega_{\rm m} \rho_{crit}$.

For calculating $\delta_c$, we use the expressions presented in
Lokas \& Hoffman (2001), who used the formalism of spherical tophat collapse,  
\begin{equation}
\delta_c(z) = \frac{1}{\Omega_{\rm m}(z)}u(z)-1 \ , 
\end{equation} (their eq.~$22$) where $u(z) = 1 + 5\Omega_{\rm \Lambda}(z)/4 +
3\Omega_\Lambda(z)(8+\Omega_{\rm \Lambda}(z))/(4v(z)) + 3v(z)/4$ (their
eq.~$23$) and $v(z) = (\Omega_{\rm \Lambda}(z)(8-\Omega_{\rm \Lambda}(z)^2 +
20\Omega_{\rm \Lambda}(z) + 8(1-\Omega_{\rm \Lambda}(z))^{3/2}))^{1/3}$
(their eq.~$24$), where $\Omega_{\rm m}(z)$ and $\Omega_{\rm \Lambda}(z)$ are
the ratio of the matter and dark energy densities to the critical density,
respectively.

Nagamine \& Loeb (2003) showed that although Lokas \& Hoffman (2001) have
ignored the possibility that the mass shell may have a nonzero initial peculiar
velocity, for $\Omega_{\rm m}(0) = 0.3$, $\Omega_{\rm \Lambda}(0) = 0.7$, and
initial time as $z = 0$, on average, the analytic estimation for $\delta_c$
based on the spherical tophat collapse model appears to provide a good
approximation to the actual threshold (see also Busha et al.\ 2003). Dunner et
al.\ (2006) showed that, on average, about $10\%$ of the mass enclosed by
$r_{\rm bound,max}$ is not bound, while about the same percentage is bound mass
that lays beyond this radius. Therefore, they claimed that this radius encloses
as much mass as will remain bound to the distant future (leaving about as many
bound galaxies outside as unbound ones inside).

\subsection{The bias in the $\zeta$ and $F$ values estimated using N-body
simulations due to the exclusion of baryons}

The Millennium simulations include only DM particles, while the halo mass
estimated from the data using galaxy dynamics is of the total matter. If
the baryon fraction, $f_{\rm b}$, is independent of the halo mass, mass ratio
estimations using only the DM are identical to the ones of total matter.
However, this is not the case, and the baryon fraction depends on halo mass
because in halos with lower mass baryons more easily escape the halo
gravitational potential well (Lin et al.\ 2003; Giodini et al.\ 2009; Dai et
al.\ 2010, though see also Gonzalez et al.\ 2007 who found that $f_{\rm b}$ is
independent on the halo mass) .

Here we roughly evaluate the bias in the $\zeta$ and $F$ values estimated using
N-body simulations (such as the Millennium simulations) due to the exclusion of
baryons. Generally, high mass halos have a higher baryon fraction. Therefore,
$\zeta$ and $F$ predictions based on N-body simulations are expected to
decrease (they are biased upwards) when taking baryons into consideration.

Dai et al.\ (2010) assumed that all the gas pressure is thermal and estimated
the baryon fraction dependence on the X-ray temperature, $T_{\rm X}$, as
\begin{equation}
 \log f_{\rm b} = (-1 \pm 0.02)+(0.2 \pm 0.03)\log T_{\rm X}
\label{f_b-T relation}
\end{equation} at $1\; keV \lesssim T_{\rm X} \lesssim 10\; keV$. In order to
convert the temperature into the halo mass, we use the mass-temperature
relation derived by Wojtak \& Lokas (2010),
\begin{equation}
M_{\rm vir} = (7.85 \pm 0.55)[T_{\rm X}/(5\; keV)]^{1.54 \pm 0.12}10^{14}
h_{0.7}^{-1} M_{\odot} \ .
\label{M-T relation}
\end{equation} This relation was derived from 23 nearby ($z < 0.1$) relaxed
galaxy clusters whose masses were estimated by kinematic data, and whose $T_{\rm
X}$ were taken from Horner (2001). The temperature range mentioned above is
equivalent to the $6.6\times 10^{13} \lesssim M_{\rm vir} \lesssim 2.3\times
10^{15}$ $h_{0.7}^{-1}$ $M_{\odot}$ mass range. Although not all of our
infalling and accreted halos are relaxed nor are their masses defined to be
the virial one, we use these expressions to roughly estimate the effect of
including baryons to the $\zeta$ and $F$ values estimated from the Millennium
simulations.

Since $M_{\rm tot} \equiv M_{\rm DM}+M_{\rm b}$, where $M_{\rm b}$ is the baryons 
mass, and $M_{\rm b} \equiv M_{\rm tot} f_{\rm b}(M_{\rm tot})$, we get
\begin{equation}
M_{\rm DM} = M_{\rm tot} (1-f_{\rm b}(M_{\rm tot})) \ .
\end{equation} The corrected mass ratio is therefore
\begin{equation}
\zeta \equiv \frac{M_{\rm tot,small}}{M_{\rm tot,mp}} =
\frac{M_{\rm DM,small}}{M_{\rm DM,mp}}\times \mathrm{f}_{\rm corr}(M_{\rm
DM,small},M_{\rm DM,mp}) \ , 
\end{equation} where the correction factor is
\begin{equation}
 \mathrm{f}_{\rm corr}(M_{\rm DM,small},M_{\rm DM,mp}) = \frac{1-f_{\rm
b}(M_{\rm tot,mp}(M_{\rm DM,mp}))}{1-f_{\rm b}(M_{\rm tot,small}(M_{\rm
DM,small}))} \ .
\label{mathrm f}
\end{equation}

For correcting $F$, we need to insert the denominator of the correction factor
into the integration over $\zeta$ in eq.~\ref{Estimated F}. However, a first
order correction for a narrow $\zeta$ range is to multiply $F$ by
$\mathrm{f}_{\rm corr}$. We estimate the correction for our lowest mass ratio
bin, $\zeta \sim 0.1$, where the correction is the largest. We take $M_{\rm
tot,mp}$ to be $M_{\rm caustics,vir}$, which on average is $8\times 10^{14}$
$h_{0.73}^{-1}$ $M_{\odot}$, so for  $\zeta \sim 0.1$ $M_{\rm tot,small} =
8\times 10^{13}$ $h_{0.73}^{-1}$ $M_{\odot}$. Then, we convert the masses to
$T_{\rm X}$ using eq.~\ref{M-T relation}, and convert the $T_{\rm X}$ to
$f_{\rm b}$ using eq.~\ref{f_b-T relation} to yield $\mathrm{f}_{\rm corr} =
0.96$. This 4\% correction is much smaller than our uncertainties.

\section{Results}
\label{Results}

In this section, we present our results for the first portion of CLASH clusters:
A963, A2261, A1423, RXJ2129, A611, MACSJ1206, and CL2130, which is
in the foreground of the cluster RXJ2129. Table~\ref{Clusters centers
and redshifts info table} gives information about the clusters' centers,
redshifts, and number of cluster members identified by the D99 method (for more
details see Appendix~\ref{D99 interlopers removal method}), which also includes 
infalling matter. The cluster centers are estimated in three different ways:
X-ray peak (taken from P12), BCG location (taken from P12, except for the
CL2130 BCG location, which is taken from Koester et al.\ 2007), and galaxy
surface density peak. For all clusters, the X-ray peaks and the BCG locations
are in agreement within $3$ arcsec (except for CL2130 where we do not have X-ray peak). 
The X-ray peaks and
the galaxy surface density peaks are also in agreement within $\sim 1 \sigma$ 
for all clusters, except MACSJ1206 where it is $\sim 2.5 \sigma$ and A2261 
where the galaxy surface density peak is at a halo fused below the cluster core (see our
definition for the cluster core at \textsection~\ref{Over density}). 

The clusters' redshifts are estimated once by taking the
median of cluster members (see \textsection~\ref{D99 interlopers removal
method}), and once by a Gaussian fit to the galaxies' velocity histogram (see
\textsection~\ref{First velocity cut}). We find both redshift estimations are
in agreement within $10^{-3}$ accuracy (see table~\ref{Clusters centers and
redshifts info table}). This agreement is when each cluster has $\sim$ a few
hundred cluster members. At lower numbers of galaxies, the Gaussian fit
uncertainty is higher, and the D99 method is not very reliable when there are
less than $\sim 100$ cluster members (Margaret Geller, private communication).
D99 mentioned that clusters may have multiple X-ray peaks. Thus, he suggested
defining the center of the cluster as the galaxy surface density peak. However,
all of the clusters in our sample have a single X-ray peak, and since its
uncertainty is smaller (and in agreement with the BCG location)
we choose it to be the cluster center.

\begin{turnpage}
\begin{table*}
\caption{Clusters' centers, redshifts, and number of members \label{Clusters
centers and redshifts info table}}
\begin{center}
\begin{tabular}{|l||c|c|c|c|c|c|c|c|c|}
\hline
               & \multicolumn{2}{|c|}{X-ray center (J2000)} &
\multicolumn{2}{|c|}{BCG location (J2000)} &
\multicolumn{2}{|c|}{Galaxy surface number density peak location}  & & &\\  
\hline
 Cluster name  &  RA   &  DEC  &   RA  &  DEC  & RA   & DEC
 & $z_{\rm c}$ & $z_{\rm c}$  & $N_{\rm gal}$   \\ 
&[hh:mm:ss (deg)]&[hh:mm:ss (deg)]&[dd:mm:ss (deg)]&[dd:mm:ss
(deg)]&[deg]&[deg]&(cluster members)&(Gaussian fit)&  \\
\hline
\hline

A963   & 10:17:03.74 (154.2656)& 39:02:49.2 (39.047)& 10:17:03.6 (154.265) &
39:02:49.42 (39.0471)& $154.2563 \pm 0.0117$ & $39.0500 \pm  0.0117$ & $0.204$ & 
$0.2039 \pm  3 \times 10^{-4}$ & $202^{+19}_{-49}$ \\
\hline

A2261  & 17:22:27.25 (260.6135)& 32:07:58.6 (32.1329)& 17:22:27.20 (260.6133) &
32:07:56.96 (32.1325)& $260.5820 \pm 0.0139$ & $32.0298 \pm 0.0139$ & $0.2251$ &
$0.225 \pm 3 \times 10^{-4}$ & $189^{+21}_{-30}$ \\

\hline

A1423  & 11:57:17.26 (179.3219)& 33:36:37.4 (33.6104)& 11:57:17.37 (179.3224)
& 33:36:39.5 (33.611)& $179.3047 \pm 0.0179$ & $33.6144 \pm 0.0179$ & $0.2141$ &    
$0.2142 \pm 2 \times 10^{-4}$ & $257^{+18}_{-29}$ \\

\hline

A611   & 08:00:56.83 (120.2368)& 36:03:24.1 (36.0567) & 08:00:56.82 (120.2368) &
36:03:23.58 (36.0565) & $120.2534 \pm 0.0155$ & $36.0763 \pm 0.0155$ & $0.2871$ & 
$0.287  \pm 2 \times 10^{-4}$ & $244^{+25}_{-39}$ \\

\hline

RXJ2129& 21:29:39.94 (322.4164)& 00:05:18.8 (0.0886)& 21:29:39.96 (322.4165)&
00:05:21.16 (0.0892) & $322.4170  \pm 0.022$ & $0.0898 \pm 0.0220 $ & $0.2339$ &
 $0.2341 \pm 2 \times 10^{-4}$ & $ 305^{+23}_{-27}$ \\

\hline

MACSJ1206 & 12:06:12.28 (181.5512)& -08:48:02.4 (-8.8007)& 12:06:12.15
(181.5506) & -08:48:03.32 (-8.8009) & $181.5380 \pm 0.0042$ & $-8.7953 \pm 0.0042$ & 
$0.4397$ & $0.4391 \pm 3 \times 10^{-4}$ & $519^{+43}_{-54}$ \\

\hline 

CL2130  & - & - & 21:30:27 (322.6123) & -00:00:24.48 (-0.0068) & $322.6034 \pm
0.0192$ & $-0.0090 \pm 0.0192$ & $0.1361$ & $0.1360  \pm 2 \times 10^{-4}$ & 
$337^{+18}_{-15}$ \\

\hline 

\end{tabular}
\end{center}
\tablecomments{The cluster centers are estimated in three different ways: X-ray
peak (taken from P12), BCG location (taken from P12, except for the CL2130 BCG
location which is taken from Koester et al.\ 2007), and galaxy surface density
peak. The clusters' redshifts are estimated once by taking the median of cluster
members (see Appendix~\ref{D99 interlopers removal method}), and once by a
Gaussian fit to the galaxies velocity histogram (see \textsection~\ref{First
velocity cut}). The rightmost column is for the number of cluster member and
infalling galaxies identified using the caustic method.}
\end{table*}
\end{turnpage}

\subsection{Clusters' mass profiles}
\label{Clusters mass profiles}

\begin{figure*}
\centering
\epsfig{file=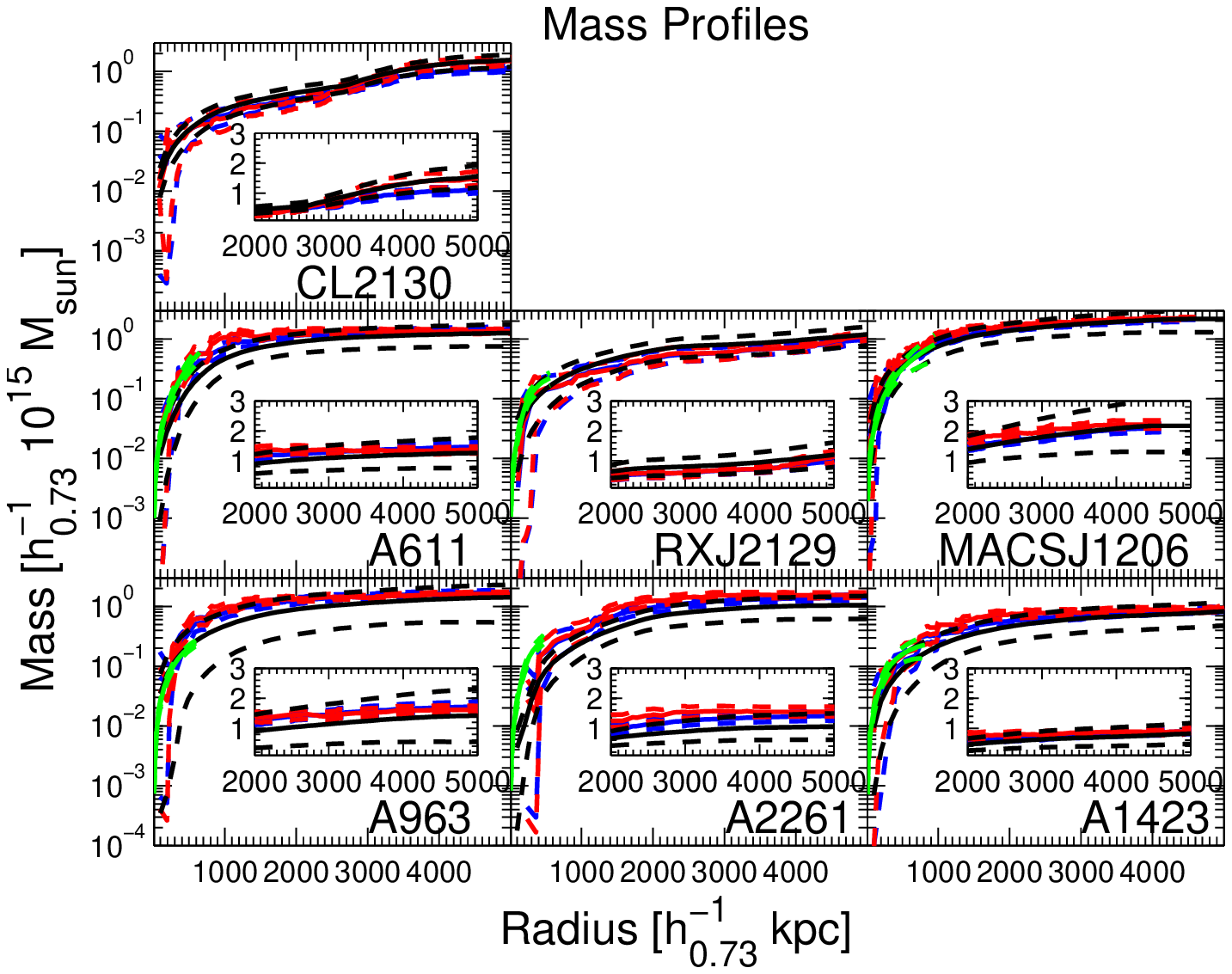, width=20cm, clip=}
\caption{Mass profiles. The caustic (black curves), virial (blue curves),
projected (red curves), and X-ray (green curves) mass profiles
are estimated. For all the mass estimators, solid curve represents the mean
value and the dashed curves the $\pm$ 1-$\sigma$. The caustic mass
profile is extrapolated to radii $\lesssim 200$ $h_{0.73}^{-1}$ kpc. In a few
cases and mainly at small radii, the projected and virial mass profiles decrease with
an increasing radius. The mass estimation at these radii is
based on a low number of galaxies (3-6), so the uncertainty is large. In any case, the
decrease is negligible comparing with the uncertainty at these radii. For
convenience, we zoom in the $2000-5000$ $h_{0.73}^{-1}$ kpc radius range of
the mass profiles of each cluster.
\label{mass profiles}}
\end{figure*}

The clusters' mass profiles are shown in figure~\ref{mass profiles}. For each
cluster, the caustic (black curves), virial (blue curves), projected (red
curves), and X-ray (green curves) mass profiles are estimated. For each of these
profiles, the solid curve represents the mean value and the dashed curves the
$\pm 1\sigma$ uncertainty. For convenience, we zoom in the
$2000-5000$ $h_{0.73}^{-1}$ kpc radius range of the mass profiles of each
cluster. For consistency, all dynamical mass profiles are
estimated after using the same interloper removal method. Since we estimate
the caustic mass profile, it is natural to use the caustic interloper
removal method. 

The virial radius is estimated to be the radius where $\overline{\rho}(<r) =
\Delta_c \rho_{\rm crit}$, where $\rho_{\rm crit}$ is the critical density of
the universe at the cluster redshift. The final overdensity to the critical
density at collapse is taken to be $\Delta_c = 18\pi^2+82x-39x^2$, where $x
\equiv \Omega_m(z)-1$ (Bryan \& Norman 1998). For estimating $r_{\rm 200}$,
which is the radius where the final overdensity to the critical density at
collapse is $200$, we take $\Delta_c = 200$. Then we estimate $M_{\rm vir}$ and
$M_{\rm 200}$ masses by $M_{\rm vir} = \frac{4\pi}{3} \Delta_c \rho_{\rm crit}
r_{\rm vir}^3$ and $M_{\rm 200} = \frac{4\pi}{3} 200 \rho_{\rm crit} r_{\rm
200}^3$, respectively. For estimating the $r_{\rm 200}$, $r_{\rm vir}$, $M_{\rm
200}$, and $M_{\rm vir}$ uncertainties, we make $10^4$ realizations for each of
the clusters' mass profiles. In each realization and for each mass bin, we
randomly take a value from a Gaussian distribution. The Gaussian mean and
standard deviation are taken to be the bin's mass and its uncertainty,
respectively. Finally, we estimate the $r_{\rm 200}$, $r_{\rm vir}$, $M_{\rm
200}$, and $M_{\rm vir}$ uncertainties as the standard deviations of all these
realizations. In table~\ref{Mvir rvir table}, we present our estimations for
these uncertainties for two of the dynamical mass profiles, caustics and
projected (the virial mass profiles are quite similar to the projected mass
profiles, see figure~\ref{mass profiles}). The last column in table~\ref{Mvir
rvir table} is the uncorrected virial mass estimated using the projected mass
estimator, $M_{\rm proj, vir}$, which is needed in \textsection~\ref{Accreted
halos}.

Note that $M_{\rm vir}$ is the cluster's mass within a sphere with an average
density of $\Delta_c$ times the critical one, not to be confused with the
virial mass profile, $M_{\rm v}(r)$ (see \textsection~\ref{Virial and projected
mass proxies}).

\begin{table*}
\caption{Estimations for the clusters' $r_{\rm 200}$, $r_{\rm vir}$, $M_{\rm
200}$, $M_{\rm vir}$ using two different mass estimators: caustics and
projected. 
\label{Mvir rvir table}}
\begin{center}
\begin{tabular}{|c||c|c|c|c|c|c|c|c|c|}
\hline
               & \multicolumn{4}{|c|}{Caustics} &
\multicolumn{5}{|c|}{Projected} \\ 
\hline
 Cluster name  & $r_{\rm caustics, 200}$ & $r_{\rm caustics, vir}$   &
$M_{\rm caustics, 200}$ & $M_{\rm caustics, vir}$ & $r_{\rm Cproj, 200}$ &
$r_{\rm Cproj, vir}$   & $M_{\rm Cproj, 200}$ & $M_{\rm Cproj, vir}$ & $M_{\rm
proj, vir}$\\ 
               &  [$h_{0.73}^{-1}$ kpc] & [$h_{0.73}^{-1}$ kpc] &    
[$10^{15}$ $h_{0.73}^{-1}$ $M_{\odot}$] &[$10^{15}$ $h_{0.73}^{-1}$ $M_{\odot}$]
 & [$h_{0.73}^{-1}$ kpc] & [$h_{0.73}^{-1}$ kpc]  &  [$10^{15}$
$h_{0.73}^{-1}$ $M_{\odot}$] &[$10^{15}$ $h_{0.73}^{-1}$ $M_{\odot}$]
&[$10^{15}$ $h_{0.73}^{-1}$ $M_{\odot}$]  \\
\hline
\hline

A963  & $1763 \pm 333$ & $2255 \pm 407$ & $0.82 \pm 0.48$ & $0.97 \pm 0.54$ & 
$2092 \pm  75$ & $2574 \pm  75$ & $1.36 \pm 0.13$ & $1.45 \pm 0.11$ & $1.76 \pm 0.13$ \\ 
\hline

A2261 & $1361 \pm 251$ & $1935 \pm 320$ & $0.38 \pm 0.23$ & $0.64 \pm 0.29$ & 
$2016 \pm 249$ & $2538 \pm 196$ & $1.25 \pm 0.34$ & $1.43 \pm 0.25$ & $2.01 \pm 0.19$ \\ 

\hline

A1423  & $1234 \pm 229$ & $1648 \pm 289$ & $0.28 \pm 0.16$ & $0.39 \pm 0.20$ & 
$1660 \pm 162$ & $2060 \pm 147$ & $0.69 \pm 0.16$ & $0.75 \pm 0.13$ & $0.94 \pm 0.10$ \\ 

\hline

A611   & $1701 \pm 259$ & $2120 \pm 298$ & $0.80 \pm 0.32$ & $0.93 \pm 0.35$ & 
$2013 \pm  91$ & $2396 \pm  95$ & $1.32 \pm 0.15$ & $1.34 \pm 0.14$ & $1.59 \pm 0.11$ \\ 

\hline

RXJ2129 & $1446 \pm 225$ & $1904 \pm 279$ & $0.46 \pm 0.21$ & $0.61 \pm 0.25$ & 
$1230 \pm 182$ & $1553 \pm 207$ & $0.29 \pm 0.12$ & $0.33 \pm 0.14$ & $1.02 \pm 0.16$ \\ 

\hline

MACSJ1206 & $1916 \pm 225$ & $2303 \pm 268$ & $1.34 \pm 0.45$ & $1.52 \pm 0.51$ & 
$2090 \pm  63$ & $2462 \pm  59$ & $1.74 \pm 0.15$ & $1.85 \pm 0.13$ & $2.31 \pm 0.13$ \\ 

\hline

CL2130 & $1282 \pm 149$ & $1716 \pm 182$ & $0.29 \pm 0.10$ & $0.38 \pm 0.11$ & 
$1144 \pm 157$ & $1518 \pm 203$ & $0.21 \pm 0.08$ & $0.26 \pm 0.09$ & $0.71 \pm 0.10$ \\ 

\hline
\end{tabular}
\end{center}
\end{table*}

\subsection{Clusters' dynamical states}
\label{Substructure level}

In this section, we estimate the clusters' dynamical states using the proxies
described in \textsection~\ref{Substructure and relaxation tests}. In
table~\ref{Clusters substructure level table}, we present our substructure
level estimations using the DS test for two different values of $N_{\rm local}$,
$10$ and $\sqrt{N_{\rm gal}}$, and after cleaning the interlopers with the two
different procedures, HK96 and D99. The DS test is estimated considering only
galaxies within the caustic virial radius. We also estimate the clusters'
relaxation state using cluster center displacements (see
\textsection~\ref{Relaxation tests}) after cleaning interlopers using D99
method.
\begin{table*}%[h]
\caption{Clusters' substructure levels and relaxation degrees \label{Clusters
substructure level table}}
\begin{center}
\begin{tabular}{|c||c|c|c|c|c|c|c|}
\hline
               & \multicolumn{4}{|c|}{D99} & \multicolumn{2}{|c|}{HK96} \\ 
\hline
 Cluster name  &  DS($N_{\rm local}=10$)   &  DS($N_{\rm local}=\sqrt{N_{\rm
gal}}$) & $r_{\rm sp}$ & $r_{\rm cp}$ &   DS($N_{\rm local}=10$)   & 
DS($N_{\rm local}=\sqrt{N_{\rm
gal}}$)  \\ 
               &   $P(\Delta_{\rm s}>\Delta_{\rm obs})$ [\%]                    
   &       $P(\Delta_{\rm s}>\Delta_{\rm obs})$ [\%]      &              
        &     &    $P(\Delta_{\rm s}>\Delta_{\rm obs})$ [\%]                    
   &       $P(\Delta_{\rm s}>\Delta_{\rm obs})$ [\%]       \\
\hline
\hline

A963   & $1.95$ &  $1.33$ &  $0.06 \pm 0.06$ & $0.30$ & $0.01$ &  $0.03$ \\ 

\hline

A2261  & $21.20$ &  $14.32$ &  $0.1 \pm 0.09$ & $0.63$ & $1.17$ &  $1.36$ \\ 

\hline

A1423  & $18.90$ &  $19.30$ &  $0.07^{+0.13}_{-0.07}$ & $0.31$ & $7.43$ &  $7.16$ \\ 

\hline

A611   & $6.10$ &  $6.94$ &  $0.19 \pm 0.11$ & $0.39$ & $8.05$ &  $1.58$ \\ 

\hline

RXJ2129 & $35.20$ &  $20.17$ &  $0^{+0.15}$ & $0.48$ & $13.99$ &  $14.64$ \\ 

\hline

MACSJ1206 & $27.88$ &  $15.06$ &  $0.07 \pm 0.04$ & $0.15$ & $0.69$ &  $0.16$ \\ 

\hline

CL2130 & $0.20$ &  $0.22$ &  $0.06^{+0.09}_{-0.06}$ & $0.81$ & $0.20$ &  $0.17$ \\ 

\hline

\end{tabular}
\end{center}
\tablecomments{Columns under D99 and HK96 are estimations made after removing
interlopers using the D99 and HK96 procedures, respectively. Columns 2 and 6 are
substructure levels using the DS test when $N_{\rm local}=10$, and columns 3 and
7 are the same but for $N_{\rm local}=\sqrt{N_{\rm gal}}$. Columns 4 and 5 are
relaxation estimations, after cleaning interlopers using the D99 method, using the 
$r_{\rm sp}$ and $r_{\rm cp}$ proxies, respectively.}
\end{table*}

In most cases, taking $N_{\rm local} = \sqrt{N_{\rm gal}}$ gives about
the same or lower DS values in both interloper removal methods. This
reinforces Bird's (1994) claim that using a non-constant value for the number of
nearest neighbors increases the sensitivity of this test to significant
structure. Another outcome is that removing interlopers with HK96 gives, in
most cases, a larger substructure level than with D99. Since the D99 method, on
average, gives a lower substructure level, we use it to clean interlopers before
we define accreted and infalling satellite halos.

\subsection{The contribution of different mass ratios to cluster growth}
\label{Accreted halos}

In this section, we first show the 2D spatial distribution of the cluster members 
and infalling galaxies when the galaxies are identified using the D99 method.
Then, out of these bound
galaxies, we identify accreted and infalling satellite halos and estimate their
dynamical properties. Finally, we estimate the differential fraction of the
cluster mass accreted and compare it to our simulations-based
predictions (see \textsection~\ref{Expected fraction of cluster mass
accretion}).

\begin{figure*}
\centering
\includegraphics[width=7.5cm,angle=0]{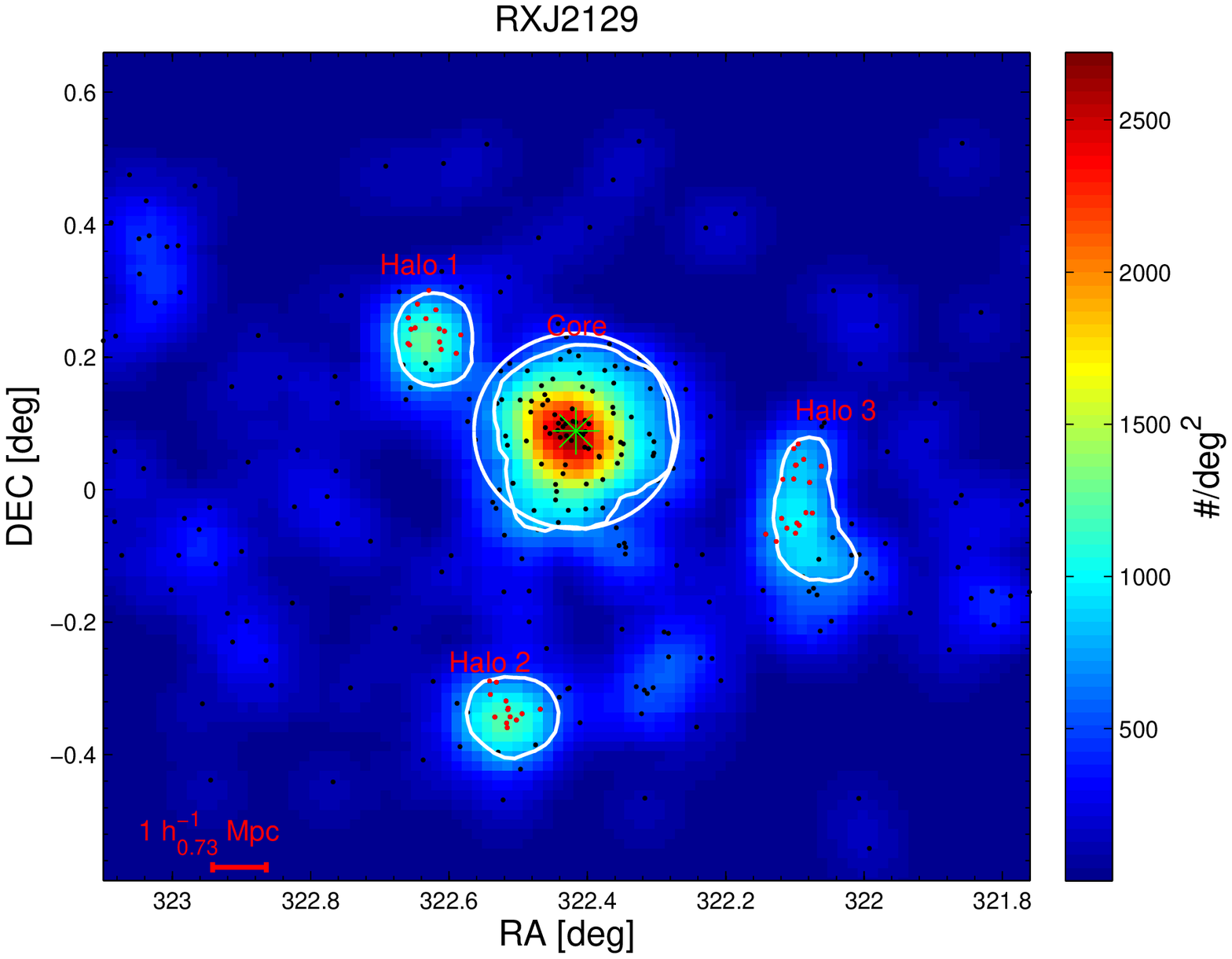}
\hspace{0.5cm}
\includegraphics[width=7.5cm,angle=0]{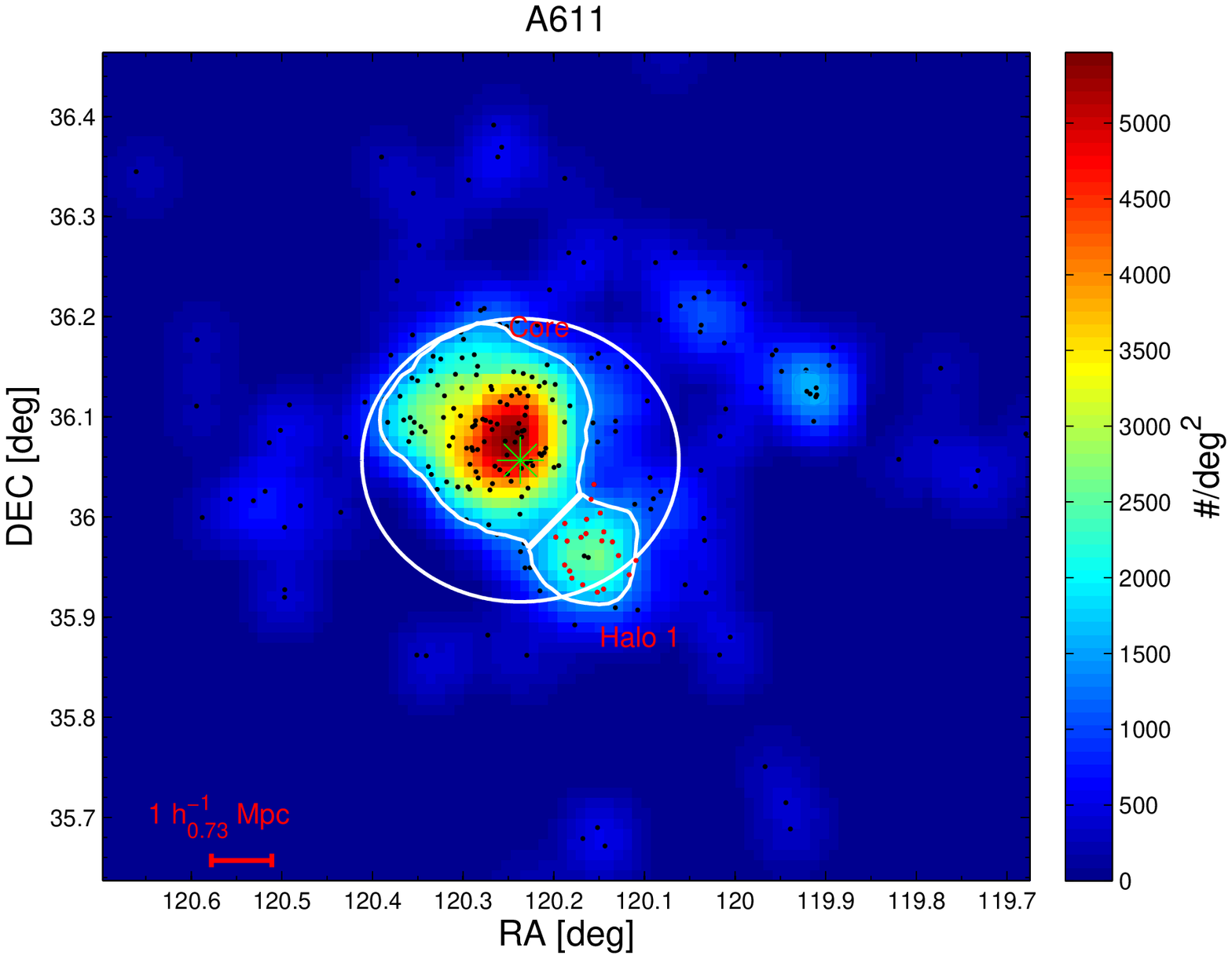}
\vskip 0.15cm
\includegraphics[width=7.5cm,angle=0]{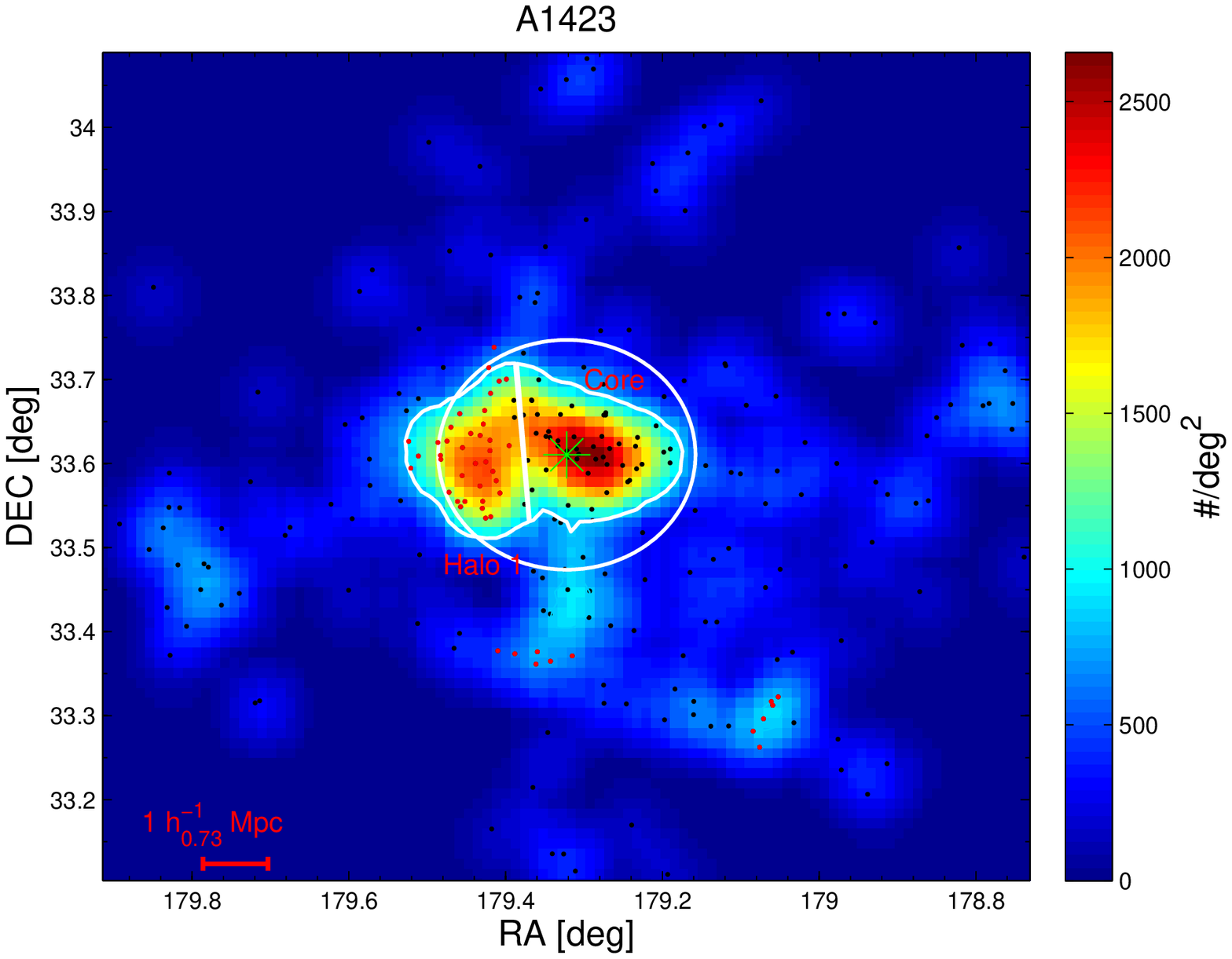}
\hspace{0.5cm}
\includegraphics[width=7.5cm,angle=0]{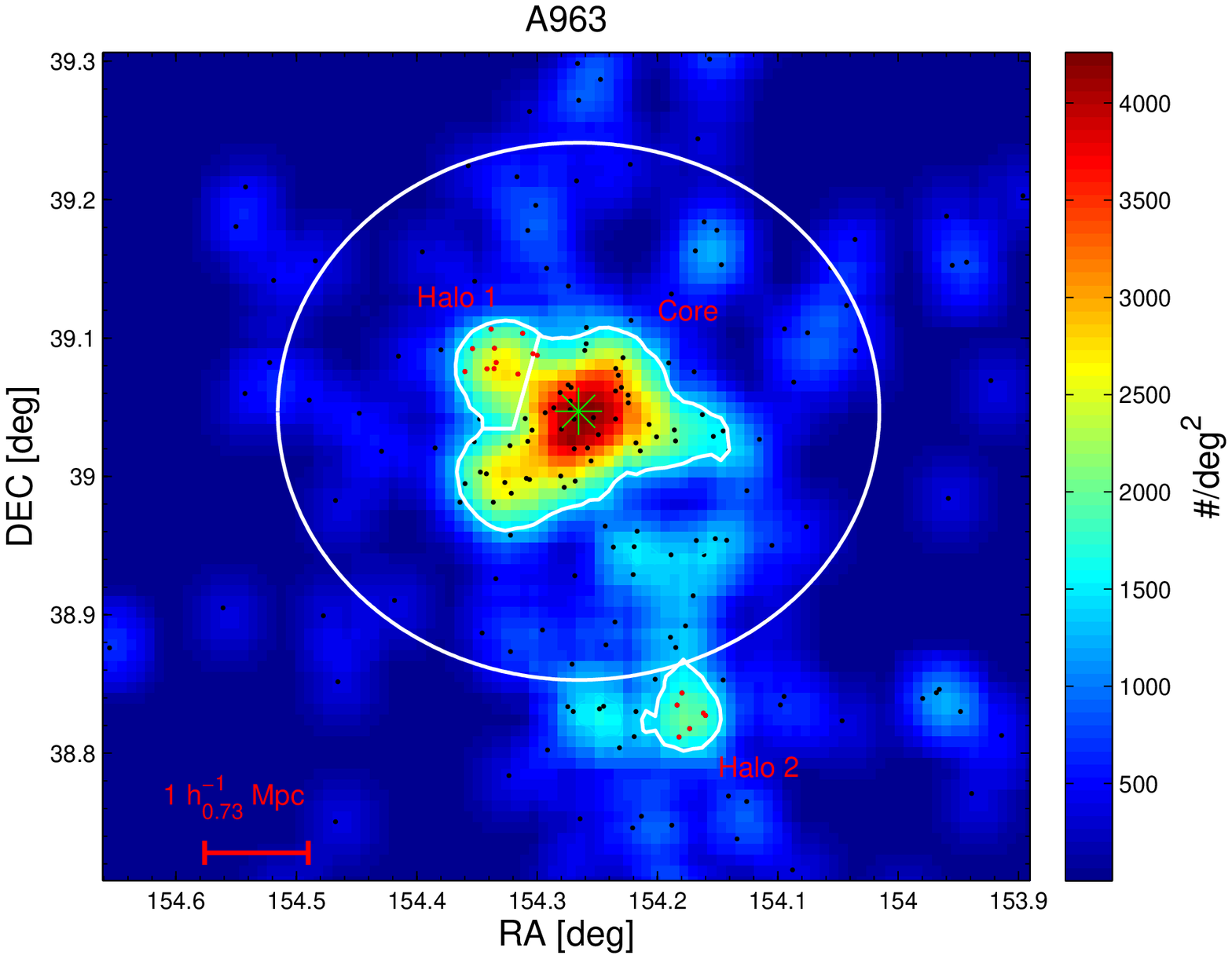}
\vskip 0.15cm
\includegraphics[width=7.5cm,angle=0]{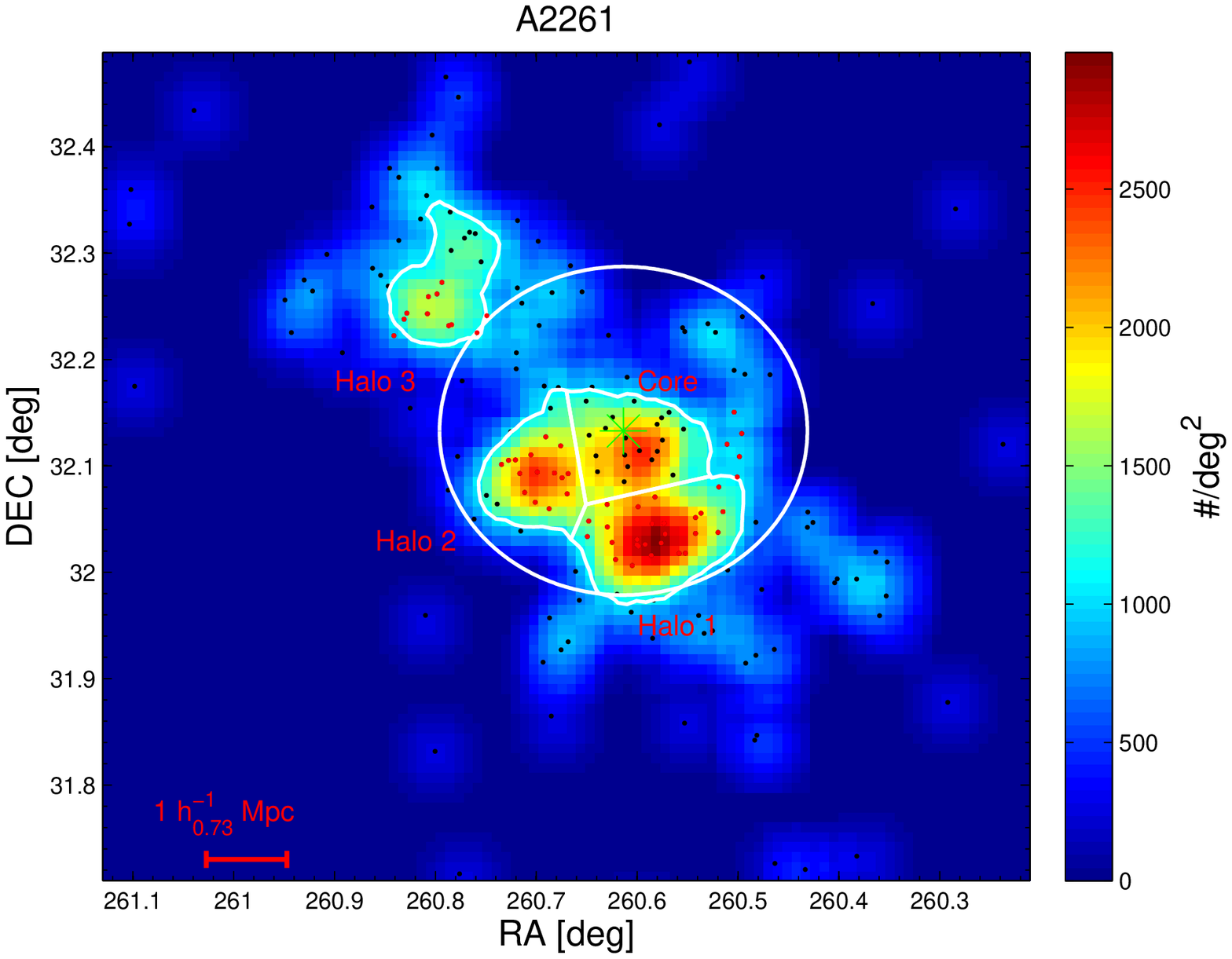}
\hspace{0.5cm}
\includegraphics[width=7.5cm,angle=0]{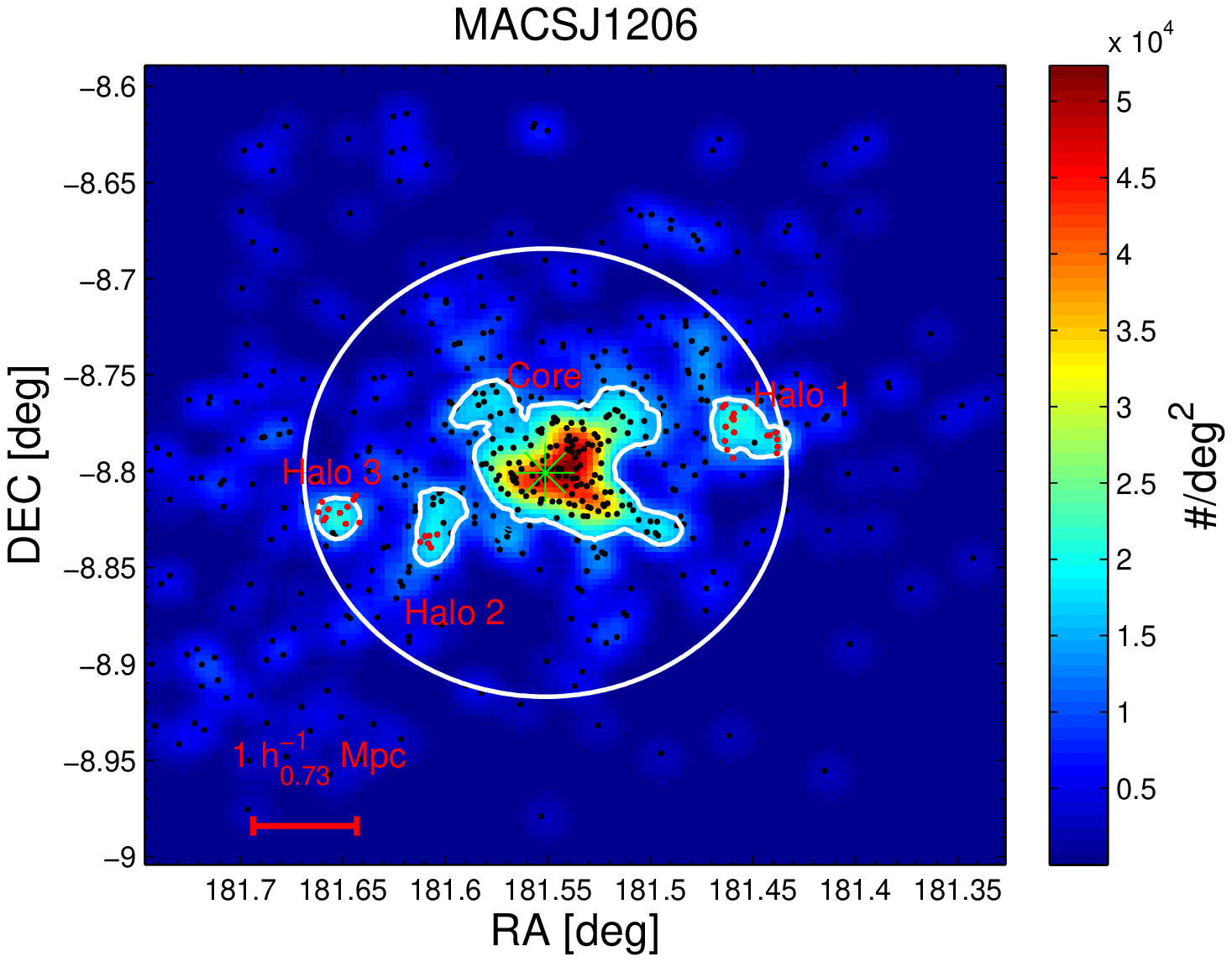}
\vskip 0.15cm
\includegraphics[width=7.5cm,angle=0]{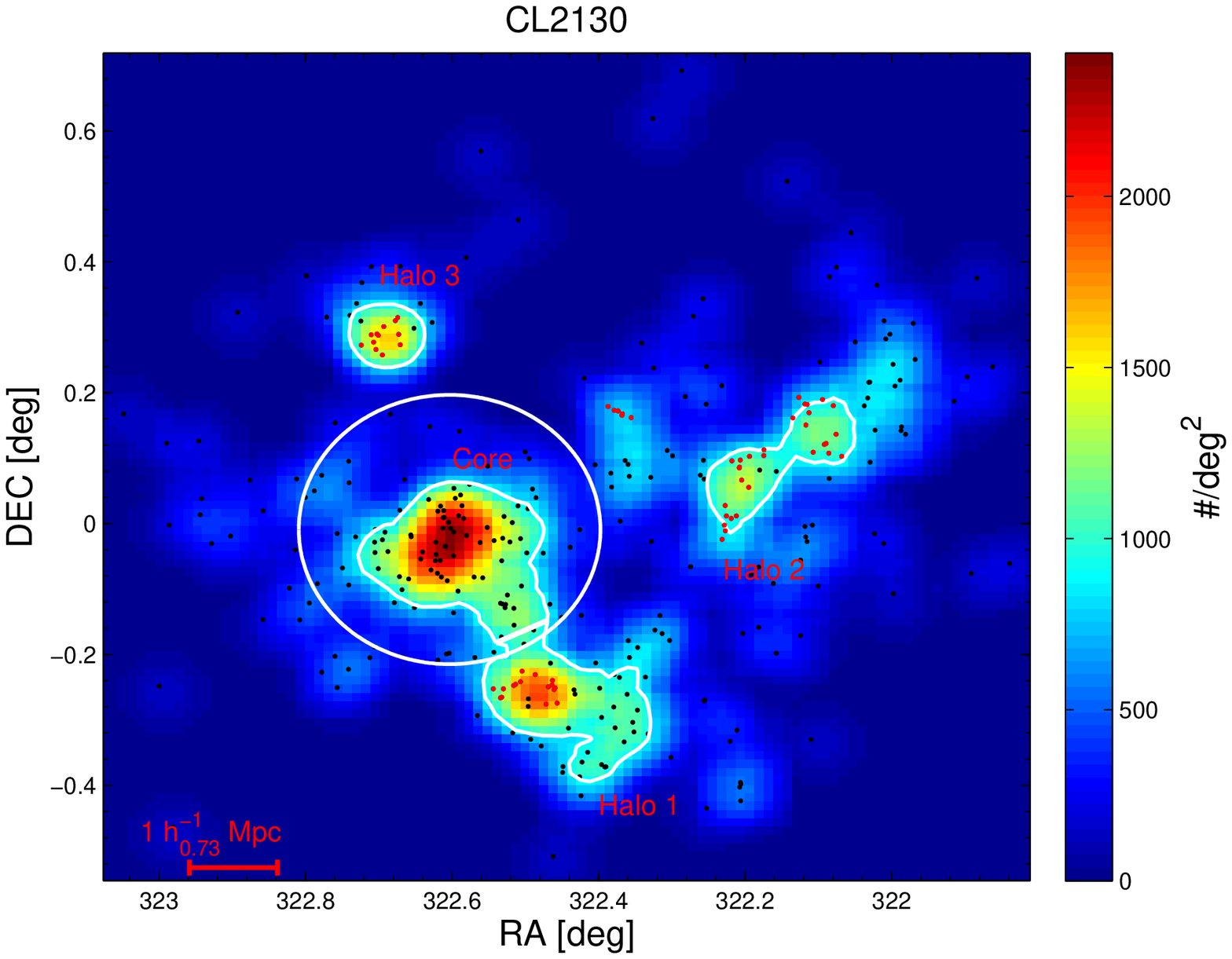}
\hspace{0.5cm}
\caption{Bound galaxies surface density and identified halos. Clusters
members and infalling galaxies are identified using the D99 method. The white
contours are overdensities found by using the overdensity HF (see
\textsection~\ref{Over
density}). The galaxies surface densities are smoothed and galaxy surface
density peaks with $\sim 3 \sigma$ above the average are identified. White
contours surround galaxy surface densities corresponding to a significance of
 $\sim 2 \sigma$ above the average. Red galaxies are groups identified using the
FoF HF (see \textsection~\ref{FoF}) with $b_{\rm frac} = 0.54$ and $N_{\rm
gal,min} = 6$. Green asterisk mark the X-ray peaks (when exists). White circles
mark the virial radius estimated using the caustic mass profile. 
\label{Clusters_n_close_halos_DEC_vs_RA}}
\end{figure*}

In figure~\ref{Clusters_n_close_halos_DEC_vs_RA}, we show for each cluster its
cluster members and infalling galaxies (black and red dots) and their smoothed 
surface density. We identify halos
using the two HFs described in \textsection~\ref{Halos identification}. The
white contours are overdensities found by using the overdensity HF (see
\textsection~\ref{Over density}). Red dots are galaxy groups identified using
the FoF HF (see \textsection~\ref{FoF}). White circles mark the virial radius
estimated using the caustic mass profile.

In table~\ref{Halos info}, we present the properties of all the 2$\sigma$
regions (including the clusters' cores) identified by the overdensity HF. This
includes the galaxy surface density peaks, virial and projected masses,
$P(\Delta_{\rm s}>\Delta_{\rm obs})$, LOS velocities (after removing the
cluster mean velocity), mean redshifts, and number of galaxies. Here we
use the projected and virial mass estimations and not the caustics ones, since
in these identified regions there are low numbers of galaxies, $\lesssim 100$,
and the caustic mass estimation is less reliable in such cases (Margaret
Geller, private communication). 

\begin{turnpage}
\begin{table*}

\caption{Information about the 2$\sigma$ regions identified by the overdensity
HF \label{Halos info}}
\begin{center}
\begin{tabular}{|l||l|c|c|c|c|c|c|c|c|c|c|c|}
\hline
 Cluster Field & Halo name  &  RA  & DEC & $M_{\rm vir}$  & $M_{\rm proj}$
 & $P(\Delta_{\rm s}>\Delta_{\rm obs})$  & $r_{\rm sp}$ & $r_{cp}$ &
$v_{\rm los}$ & $z$ & $\sigma$\footnote{The velocity dispersion uncertainty here
is only due to measurements uncertainties.}
& $N_{\rm gal}$ \\
              &            & [deg] & [deg] & [$10^{14}$ $h_{0.73}^{-1}$
$M_{\odot}$] & [$10^{14}$ $h_{0.73}^{-1}$ $M_{\odot}$] & [\%] & & & [km sec$^{-1}$] & &
[km sec$^{-1}$]  &  \\
\hline
\hline

A963   & Core  & $154.2563 \pm 0.0117$ &  $39.0500 \pm 0.0117$ &  $8.93 \pm 
1.35$ &  $11.17 \pm 1.95$ &  $60.22$ &  $0.32 \pm 0.30$ &  $0.44$ &  $-80 
\pm 4$ & $0.2037$ & $946 \pm 2$ &  $52$ \\ 

A963   & Halo 1 & $154.3264 \pm 0.0117$ &  $39.0733 \pm 0.0117$ &  $6.63 \pm
 2.84$ &  $7.23 \pm 1.68$ &  $41.79$ &  $0.57^{+0.67}_{-0.57}$ &  $0.35$ &  $117 \pm 
11$ & $0.2045$ & $1218 \pm 8$ &  $10$ \\ 

A963   & Halo 2 & $154.1785 \pm 0.0117$ &  $38.8242 \pm 0.0117$ &  $0.26 \pm 
0.30$ &  $0.25 \pm 0.13$ &  $75.18$ &  $0.84 \pm 0.81$ &  $0.60$ &  $-172 \pm 
12$ & $0.2033$ & $354 \pm 5$ &  $6$ \\

\hline

A2261  & Core & $260.6006 \pm 0.0139$ &  $32.1134 \pm 0.0139$ &  $3.21 \pm 0.96$ &  
$3.57 \pm 0.89$ &  $52.19$ &  $0^{+0.45}$ &  $0.34$ &  $-101 \pm 7$ & $0.2246$ & 
$649 \pm 5$ &  $20$ \\ 

A2261  & Halo 1 & $260.5820 \pm 0.0139$ &  $32.0298 \pm 0.0139$ &  $5.01 \pm 1.33$ &  
$4.61 \pm 0.96$ &  $18.27$ &  $0.45 \pm 0.40$ &  $0.43$ &  $244 \pm 7$ & $0.2260$ & 
$743 \pm 28$ &  $27$ \\ 

A2261  & Halo 2 & $260.7027 \pm 0.0139$ &  $32.0948 \pm 0.0139$ &  $3.03 \pm 0.99$ &  
$3.02 \pm 0.93$ &  $16.90$ &  $0.64 \pm 0.49$ &  $0.67$ &  $141 \pm 8$ & $0.2256$ & 
$615 \pm 11$ &  $18$ \\ 

A2261  & Halo 3 & $260.8049 \pm 0.0139$ &  $32.2527 \pm 0.0139$ &  $2.62 \pm 0.64$ &  
$2.68 \pm 0.58$ &  $10.27$ &  $2.55 \pm 0.49$ &  $1.69$ &  $-355 \pm 9$ & $0.2236$ & 
$562 \pm 13$ &  $15$ \\ 

\hline

A1423  & Core & $179.3047 \pm 0.0179$ &  $33.6144 \pm 0.0179$ &  $3.85 \pm 0.78$ &  
$4.60 \pm 1.05$ &  $3.10$ &  $0.21^{+0.37}_{-0.21}$ &  $0.37$ &  $104 \pm 9$ & $0.2144$ & 
$590 \pm 45$ &  $49$ \\ 

A1423  & Halo 1 & $179.4242 \pm 0.0179$ &  $33.6024 \pm 0.0179$ &  $4.89 \pm 0.92$ &  
$5.13 \pm 0.96$ &  $10.52$ &  $0.85 \pm 0.41$ &  $0.69$ &  $-128 \pm 6$ & $0.2135$ & 
$642 \pm 3$ &  $38$ \\

\hline

A611   & Core & $120.2534 \pm 0.0155$ &  $36.0763 \pm 0.0155$ &  $12.55 \pm 1.48$ &  
$14.78 \pm 1.70$ &  $56.32$ &  $0.49 \pm 0.29$ &  $0.15$ &  $116 \pm 3$ & $0.2875$ & 
$883 \pm 17$ &  $104$ \\ 

A611   & Halo 1 & $120.1500 \pm 0.0155$ &  $35.9626 \pm 0.0155$ &  $2.61 \pm 0.85$ &  
$2.55 \pm 0.80$ &  $93.55$ &  $0.96 \pm 0.56$ &  $0.72$ &  $-219 \pm 7$ & $0.2861$ & 
$568 \pm 3$ &  $20$ \\

\hline

RXJ2129 & Core & $322.4170 \pm 0.0220$ &  $0.0898 \pm 0.0220$ &  $6.86 \pm 1.28$ &  
$7.95 \pm 1.35$ &  $9.86$ &  $0^{+0.29}$ &  $0.02$ &  $-46 \pm 4$ & $0.2338$ & 
$631 \pm 15$ &  $79$ \\ 

RXJ2129 & Halo 1 & $322.0791 \pm 0.0220$ &  $-0.0424 \pm 0.0220$ &  $2.98 \pm 1.75$ &  
$3.84 \pm 1.44$ &  $37.11$ &  $0.49^{+0.52}_{-0.49}$ &  $0.88$ &  $-129 \pm 8$ & $0.2335$ & 
$516 \pm 3$ &  $20$ \\ 

RXJ2129 & Halo 2 & $322.6227 \pm 0.0220$ &  $0.2367 \pm 0.0220$ &  $1.78 \pm 0.47$ &  
$1.68 \pm 0.43$ &  $1.12$ &  $0.82 \pm 0.61$ &  $0.62$ &  $-47 \pm 8$ & $0.2338$ & 
$420 \pm 4$ &  $17$ \\ 

RXJ2129 & Halo 3 & $322.5199 \pm 0.0220$ &  $-0.3362 \pm 0.0220$ &  $1.68 \pm 0.58$ &  
$1.70 \pm 0.42$ &  $23.54$ &  $0^{+0.61}$ &  $0.07$ &  $586 \pm 9$ & $0.2364$ & 
$484 \pm 2$ &  $16$ \\

\hline

MACSJ1206 & Core & $181.5380 \pm 0.0042$ &  $-8.7953 \pm 0.0042$ &  $10.74 \pm 1.12$ &  
$13.29 \pm 1.56$ &  $74.87$ &  $0.34 \pm 0.18$ &  $0.24$ &  $41 \pm 13$ & $0.4402$ & 
$1041 \pm 88$ &  $149$ \\ 

MACSJ1206 & Halo 1 & $181.4591 \pm 0.0042$ &  $-8.7727 \pm 0.0042$ &  $2.45 \pm 0.88$ &  
$2.50 \pm 0.68$ &  $53.46$ &  $0^{+0.48}$ &  $0.91$ &  $243 \pm 32$ & $0.4412$ & 
$878 \pm 37$ &  $18$ \\ 

MACSJ1206 & Halo 2 & $181.6057 \pm 0.0042$ &  $-8.8375 \pm 0.0042$ &  $3.98 \pm 1.38$ &  
$7.34 \pm 3.41$ &  $79.84$ &  $0.48^{+0.51}_{-0.48}$ &  $1.11$ &  $119 \pm 43$ & $0.4406$ & 
$1006 \pm 43$ &  $13$ \\ 

MACSJ1206 & Halo 3 & $181.6536 \pm 0.0042$ &  $-8.8263 \pm 0.0042$ &  $3.07 \pm 1.52$ &  
$3.22 \pm 1.38$ &  $79.34$ &  $1.06 \pm 0.72$ &  $0.78$ &  $-153 \pm 44$ & $0.4393$ & 
$1022 \pm 12$ &  $9$ \\

\hline

CL2130  & Core & $322.6034 \pm 0.0192$ &  $-0.0090 \pm 0.0192$ &  $4.17 \pm 0.53$ &  
$4.99 \pm 0.71$ &  $1.29$ &  $0.20^{+0.30}_{-0.20}$ &  $0.68$ &  $-67 \pm 4$ & $0.1357$ & 
$607 \pm 10$ &  $65$ \\ 

CL2130  & Halo 1 & $322.4756 \pm 0.0192$ &  $-0.2646 \pm 0.0192$ &  $1.73 \pm 0.53$ &  
$3.53 \pm 1.20$ &  $9.61$ &  $0.33^{+0.35}_{-0.33}$ &  $0.60$ &  $-320 \pm 6$ & $0.1348$ & 
$417 \pm 5$ &  $42$ \\ 

CL2130  & Halo 2 & $322.2072 \pm 0.0192$ &  $0.0549 \pm 0.0192$ &  $2.04 \pm 0.83$ &  
$3.31 \pm 1.02$ &  $0.15$ &  $0.94 \pm 0.47$ &  $1.16$ &  $718 \pm 7$ & $0.1387$ & 
$528 \pm 6$ &  $24$ \\ 

CL2130  & Halo 3 & $322.6801 \pm 0.0192$ &  $0.2849 \pm 0.0192$ &  $0.33 \pm 0.11$ &  
$0.40 \pm 0.10$ &  $11.49$ &  $0.90 \pm 0.67$ &  $0.44$ &  $-834 \pm 9$ & $0.1328$ & 
$271 \pm 6$ &  $14$ \\

\hline

\end{tabular}
\end{center}
\tablecomments{Columns 3-13 are: galaxy surface density peak location, virial and
projected mass, $P(\Delta_{\rm s}>\Delta_{\rm obs})$, relaxation levels for both proxies, 
LOS velocity (after removing the cluster redshift), mean redshift, velocity dispersion, 
and number of galaxies in the 2$\sigma$ regions identified by the overdensity HF.}
\end{table*}
\end{turnpage}

We take each identified satellite halo and estimate its mass ratio, $\zeta
\equiv M_{\rm small}/M_{\rm mp}$, assuming $M_{\rm mp} \approx M_{\rm cluster}$
(for more details, see \textsection~\ref{Expected fraction of cluster mass
accretion}). For $M_{\rm small}$ and $M_{\rm mp}$, we use the uncorrected
projected mass estimator, since many of the satellite halos do not have enough
galaxies for the correction. Thus, the accreted and infalling satellite halos'
masses are $M_{\rm small} = M_{\rm proj}$ (taken from table~\ref{Halos info}).
For consistency, in estimating $\zeta$, we use the same mass estimator to
estimate the cluster mass, which is taken to be the mass inside the virial
radius, i.e. $M_{\rm cluster} = M_{\rm proj, vir}$ (the last column in
table~\ref{Mvir rvir table}). We divide the mass ratio range to a low number of
bins, $3$, because we have a low number of satellites. For each bin, we
estimate the differential fraction of the cluster mass accreted, $F(\zeta)$. To estimate
$F(\zeta)$, we calculate for each cluster the
sum of the masses of all identified satellite halos in the relevant $\zeta$
range. Then to be consistent with our predictions, we divide it by the cluster
mass at the time when all identified matter falls into it, i.e. $z_1$ (for more
details see \textsection~\ref{Estimating z_f}), using eq.\ref{M(z)}. Finally, we
average out $F(\zeta)$ over all the clusters.

For estimating the $F(\zeta)$ uncertainties, we make $10^2$ realizations where
in each one and for each cluster we randomly take a value from a Gaussian
distribution for both the infall or accreted satellite and cluster masses.
The Gaussian mean and standard deviation are taken to be the mass estimation
and its uncertainty, respectively. For each realization and for each cluster, we
estimate $F(\zeta)$ as is explained above. Then, we average out $F(\zeta)$
over all clusters. Finally, we estimate the $F(\zeta)$ uncertainties as the
standard deviation of all the realized $F(\zeta)$.

\begin{figure*}
\centering
\epsfig{file=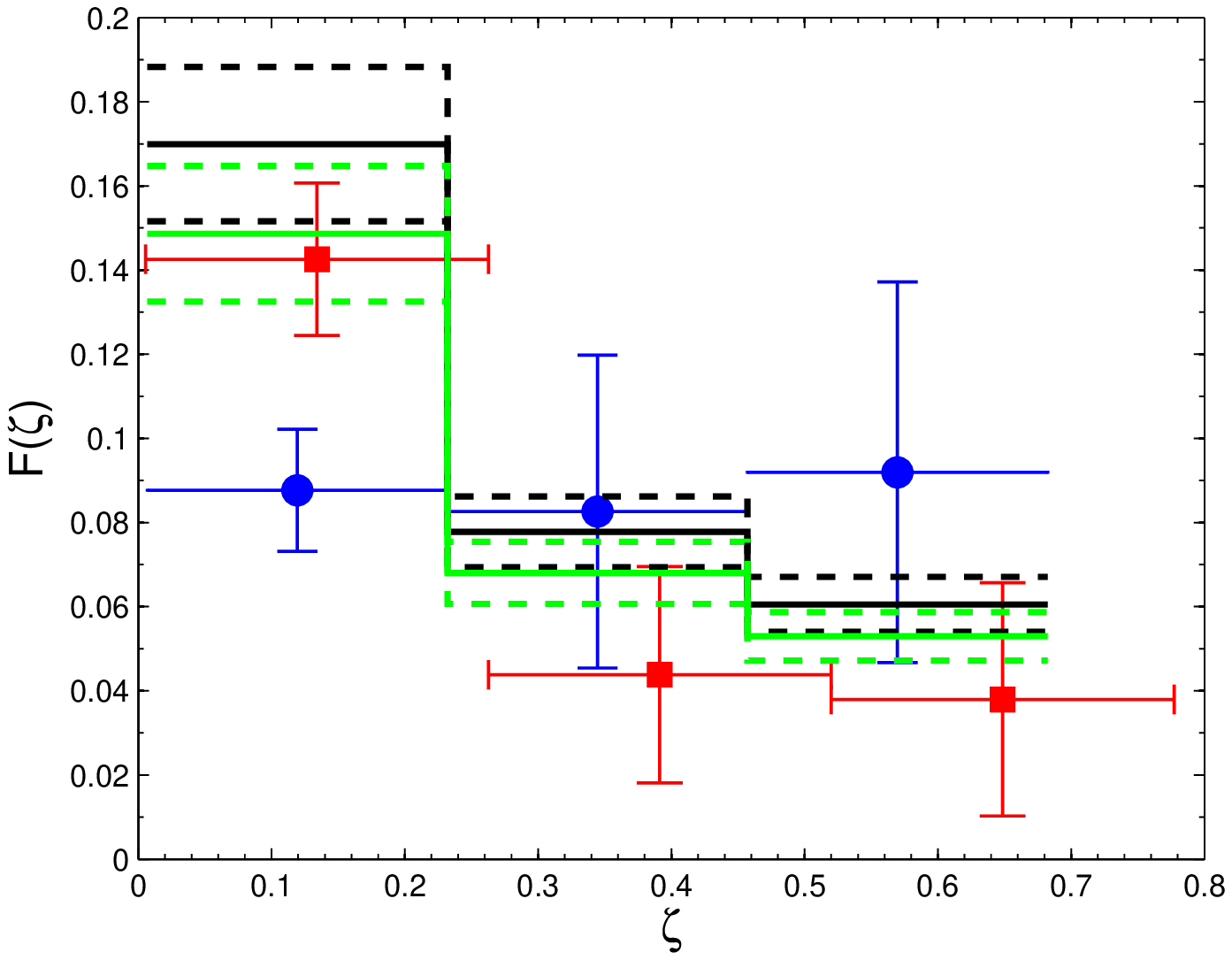, width=8cm,clip=}
\epsfig{file=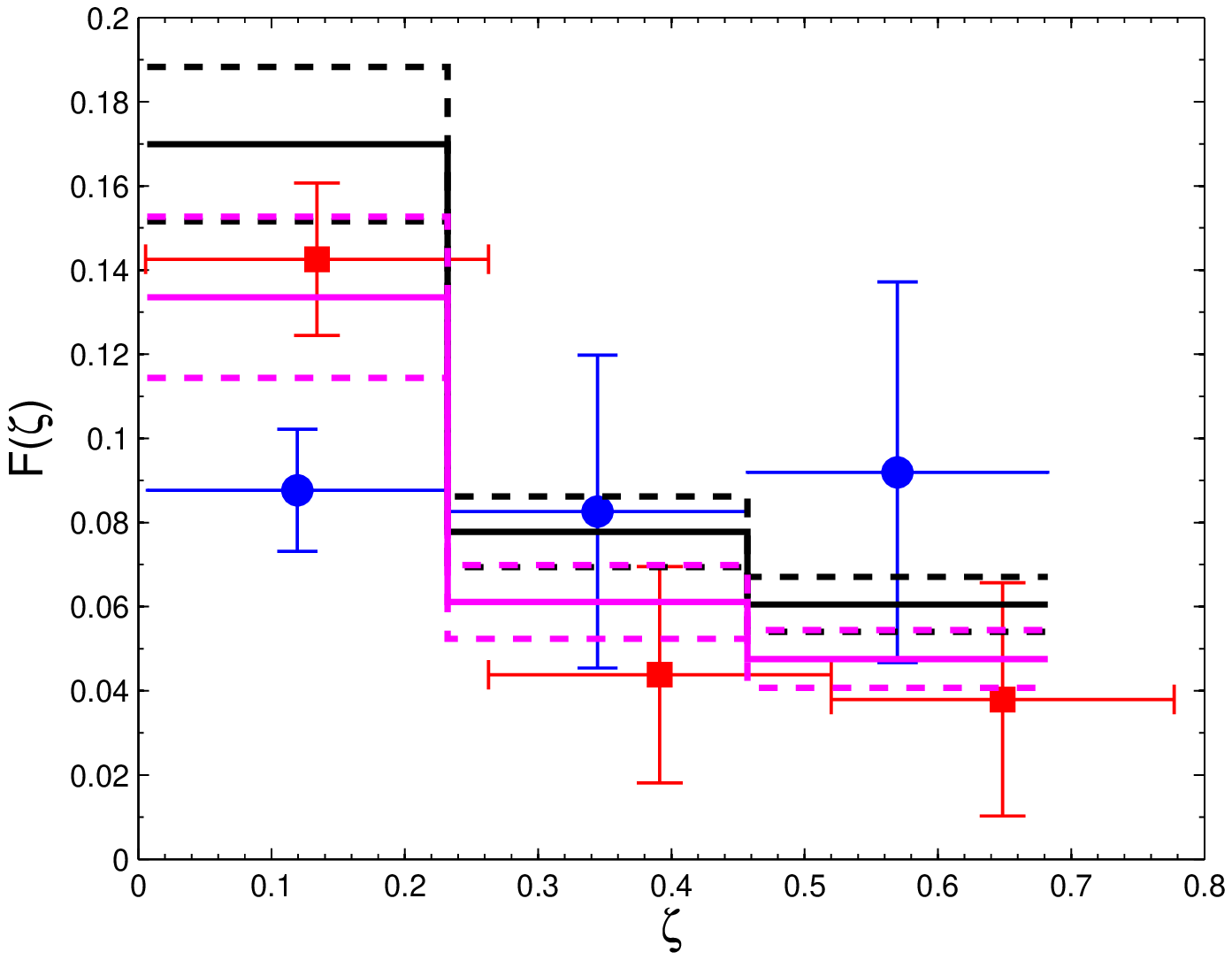, width=8cm,clip=}
\caption{Differential fraction of the cluster mass accreted, $F(\zeta)$, vs the
halo mass ratio, $\zeta$. Blue circles and red squares are for satellites halos
identified by the overdensity and FoF HFs, respectively. Left panel: black (green) 
curves are our simulations-based predictions (for more details, see
\textsection~\ref{Expected fraction of cluster mass accretion}) using WMAP7
cosmology and $\alpha=0.15$ ($\alpha=0.133$). Right panel: black (purple) curves are our
simulations-based predictions using WMAP7 (WMAP1) cosmology. Scaling by $\sigma_8$, 
the theoretical expectation using WMAP9 is in between the expectations from WMAP1 and WMAP7 and closer to the latter.
In both panels, solid and dashed curves represent the simulations-based predictions mean 
and 1$\sigma$ uncertainty, respectively.
\label{Fractional contribution to halo growth}}
\end{figure*}

In figure~\ref{Fractional contribution to halo growth}, we show the
simulations-based theoretical expectation of $F(\zeta)$ that
we derive following the procedure described in \textsection~\ref{Expected
fraction of cluster mass accretion}. For $M_{\rm obs}$, we insert the clusters
$M_{\rm proj, vir}$ (for consistency with observational based $\zeta$ values).
We first estimate the expected $F(\zeta)$ for each cluster by inserting the cluster's 
$M_{\rm obs}$, $z_{\rm c}$, and $z_{\rm f}$ values to eq.~\ref{Estimated F}. Then 
we take the final theoretical expectation of $F(\zeta)$ (solid curves) and its uncertainty (dashed curve) 
to be the mean and standard deviation, respectively, of the expected $F(\zeta)$ of all clusters. 

In the left panel, we
present the theoretical expectations using the two different $\alpha$ values
mentioned in \textsection~\ref{Expected fraction of cluster mass accretion},
$0.15$ and $0.133$ (black and green curves, respectively). As is mentioned in
\textsection~\ref{Correcting sigma_8}, Angulo et al.\ (2012) scaled the
Millennium simulations, which are based on the values found by analyzing WMAP1,
to the ones based on the values found by analyzing WMAP7. In the right panel, we
present the theoretical expectations using two different cosmologies WMAP7 and
WMAP1 (black and purple curves, respectively). Scaling by $\sigma_8$, the theoretical 
expectations using WMAP9 are in between the
expectations from WMAP1 and WMAP7 and closer to the latter. On top of the
predicted $F(\zeta)$, we plot our derived values using the two different HFs,
i.e. overdensity (blue circles) and FoF (red squares). The theoretical
expectation $F(\zeta)$ bins are the same as the ones of the estimated $F(\zeta)$
using the overdensity HF. 

\begin{figure*}
\centering
\epsfig{file=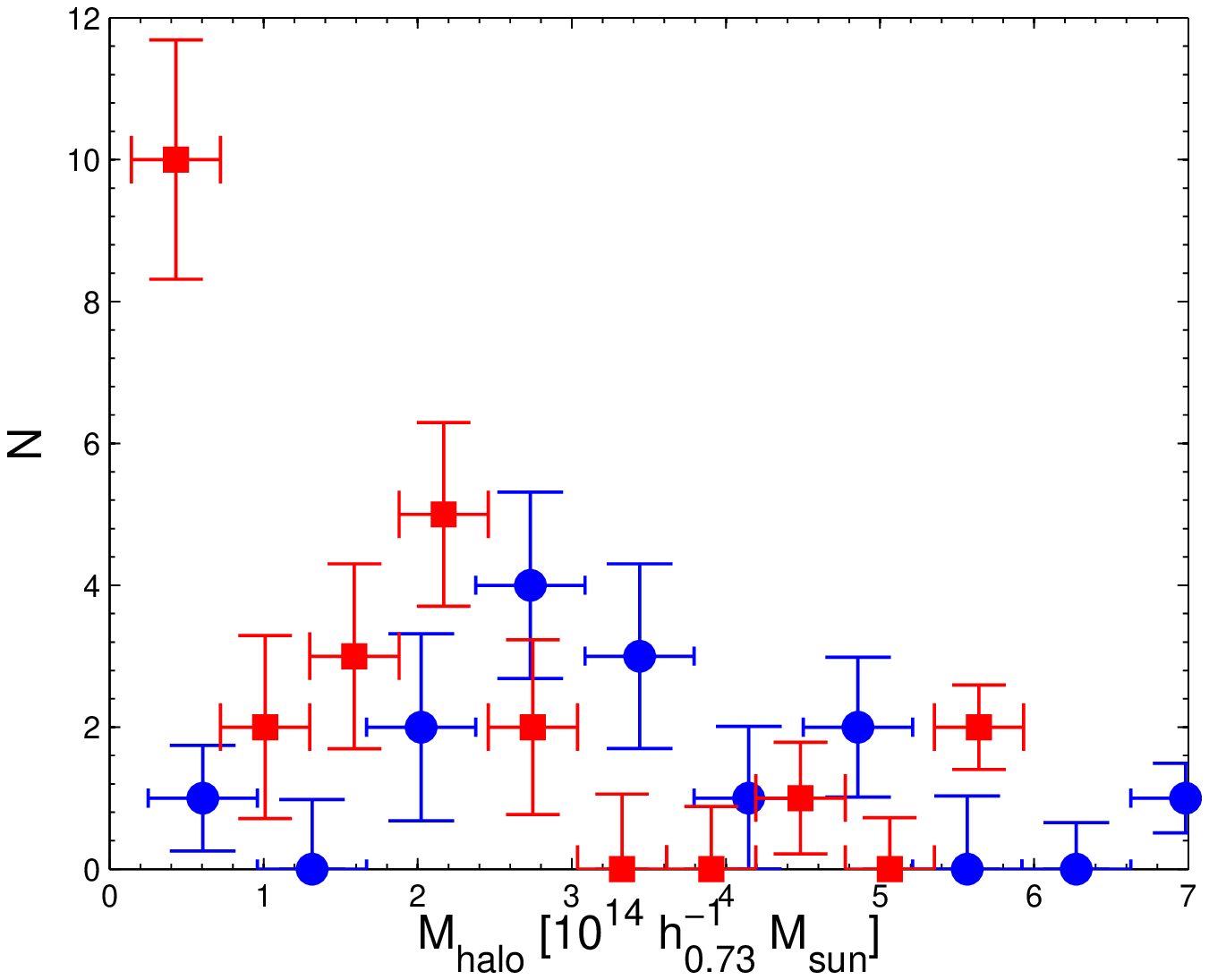,width=8cm,clip=}
\caption{Histogram of the infall and accreted satellites' masses. Blue circles
and red squares are for halos identified by the overdensity and FoF HFs,
respectively.
\label{N_vs_Mhalo}}
\end{figure*}

In figure~\ref{N_vs_Mhalo}, we plot the infalling and accreted satellite
mass histogram, where the satellite halos
are identified using the two different HFs, i.e. overdensity (blue circles) and
FoF (red squares). In both halo definitions, at masses lower than about $2
\times 10^{14}$ $h_{0.73}^{-1}$ M$_{\odot}$, the efficiency of detecting halos
decreases. Dividing this number by the averaged cluster mass, which again is
taken to be $M_{\rm proj, vir}$, we get $\zeta \sim 0.14$. Thus, at the
$\zeta$ bin with the lowest value, the number of identified halos is
underestimated. 

The $\zeta$ bin with the highest value may be underestimated as well, if
mergers with high $\zeta$ are missed. These mergers may be missed for
two reasons. First, all the clusters in our sample (except for CL2130) were
chosen to be quite X-ray relaxed (P12). Therefore, this sample is biased
against large $\zeta$ values. Second, mergers with large $\zeta$ are more
rare. Taking eq.~\ref{N_m} (with our clusters' masses and redshifts and
integrating over the $z_{\rm f} \leqslant z \leqslant z_{\rm c}$ redshift range)
we expect the averaged (over all clusters) number of mergers per cluster to be
$ \approx  0.24 \pm 0.05$ and $ \approx 0.21 \pm 0.04$ for $N_{\rm m} (\zeta \gtrsim 
0.7,\alpha = 0.15)$ and $N_{\rm m} (\zeta \gtrsim 0.7,\alpha = 0.133)$,
respectively. This gives a total number of $\sim 1$ merger for our 7 clusters
sample. Thus, these high $\zeta$ mergers may not be seen due to our sample size.

\subsection{Substructure-Relaxation connection}
\label{Substructure-Relaxation connection}

\begin{figure*}
\centering
\epsfig{file=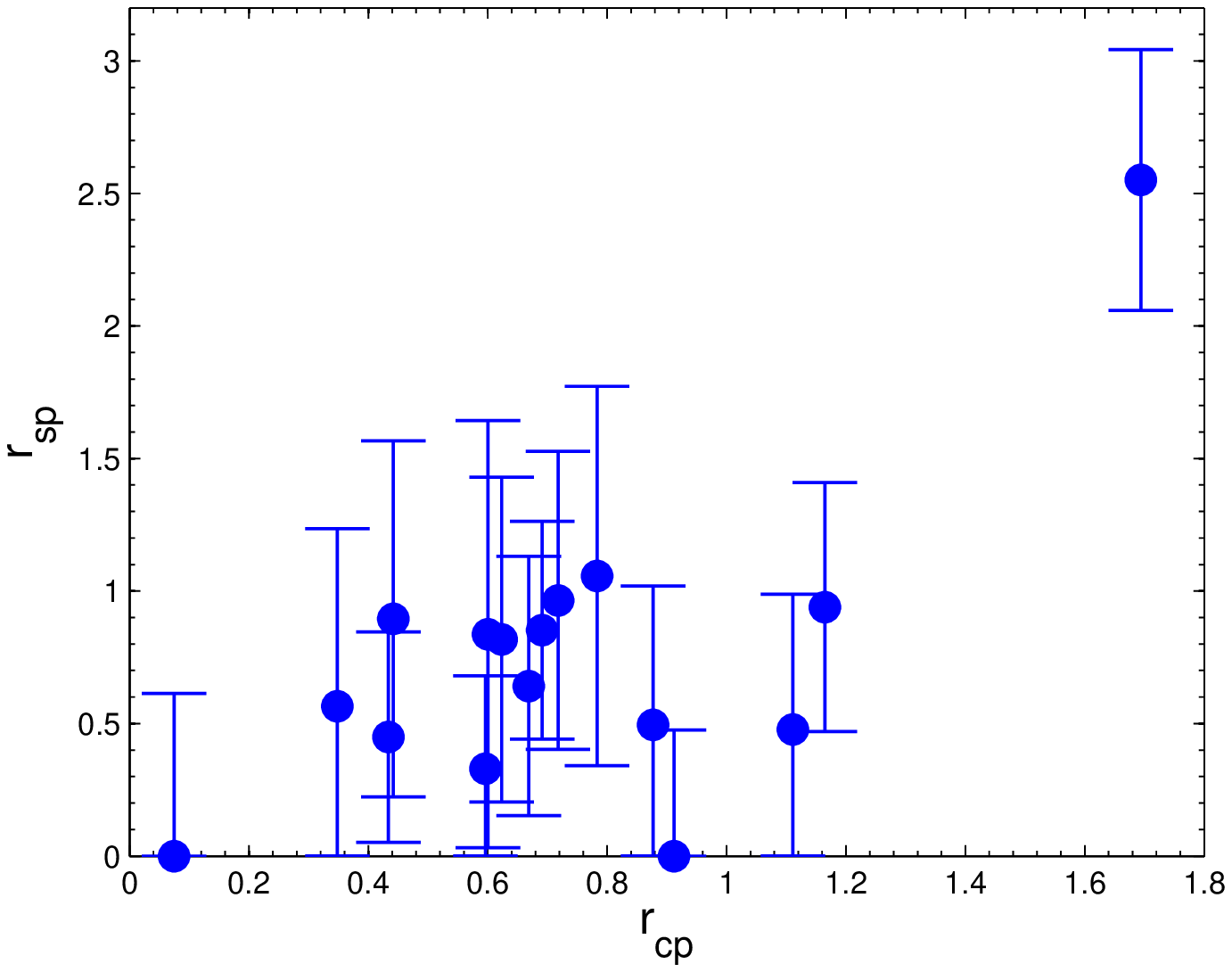,
width=8cm,clip=}
\epsfig{file=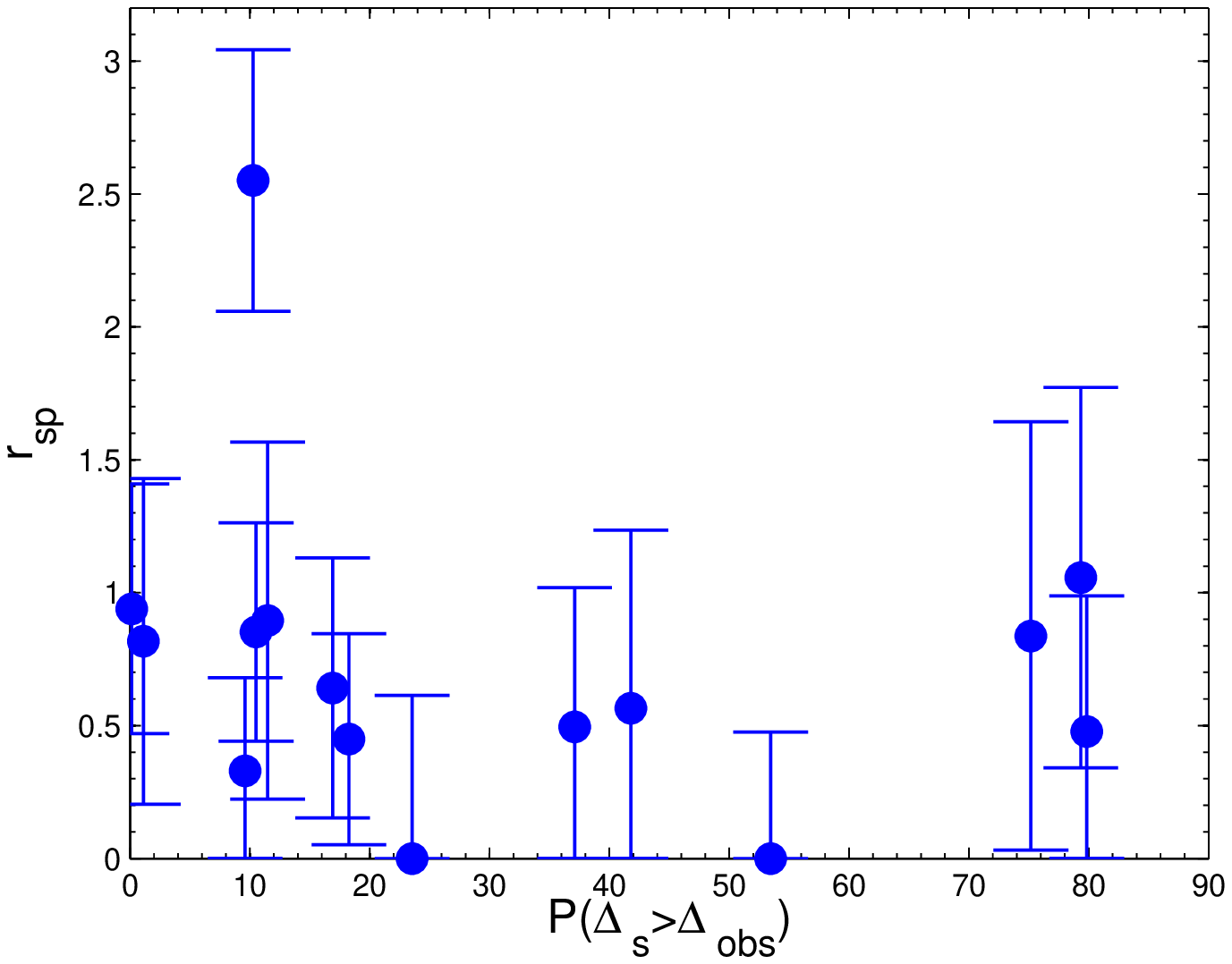,
width=8cm,clip=}
\caption{Left panel: estimated relaxation levels using the two proxies discussed
in \textsection~\ref{Relaxation tests}, $r_{\rm sp}$ and $r_{\rm cp}$, for the
infalling and accreted satellite halos. Right panel: estimated substructure
levels (using the DS test) vs the relaxation levels (using the $r_{\rm sp}$
proxy) for the same halos. 
\label{relaxation proxies and relaxation vs substructure}}
\end{figure*}

In this section, we test if the correlation between the substructure and
relaxation levels of the infalling and accreted satellite halos is strong. In
figure~\ref{relaxation proxies and relaxation vs substructure} (left panel), we
show the measured relaxation levels using the two proxies described in
\textsection~\ref{Relaxation tests}, $r_{\rm sp}$ and $r_{\rm cp}$. In
figure~\ref{relaxation proxies and relaxation vs substructure} (right panel), we
show the relaxation levels using the $r_{\rm sp}$ proxy vs substructure levels
using the DS test. 
The substructure levels do not correlate strongly with our estimated dynamical 
relaxation levels (using the $r_{\rm sp}$ proxy), while the correlation
between the two relaxation proxies is stronger.

\section{Discussion}
\label{Discussion}

In this paper, we estimate the mass and dynamical properties of seven
clusters and of the halos falling (and accreted) into them. Then we estimate the
contribution of different mass ratios to cluster growth and compare these
estimations to our expectations based on simulations.

Before estimating the clusters' dynamical properties, it is important to remove 
interlopers from them. We test two interloper removal methods, HK96 and D99.
We use the fact that the DS test is sensitive to outliers in order to
indicate which interloper removal method is most effective. Removing interlopers 
with HK96 gives a larger substructure level than with
D99 (see table~\ref{Clusters substructure level table}). Therefore, the D99
method is better in dealing with the presence of structures that are along
the LOS and have a small projected distance to the cluster center. Thus, we
use the D99 method to clean interlopers. 

Our analysis is heavily based on estimating the masses of clusters and accreted
and infalling halos. Thus, we examine four possible mass profile biases.
Namely, we check the effect of the interloper removal method, the mass
estimator used, the number of identified galaxies, and uneven sampling of the
sky's plane. The HK96 method yields wider caustics and, therefore, a higher
estimated mass (see Appendix~\ref{Mass profile biases}). The masses estimated, 
at large radii (i.e. about the virial radius), using the HK96 method were about 
$\sim 50\%-100\%$ higher than those estimated using D99 (when the higher values 
are for clusters with many massive 
infalling satellites, e.g. RXJ2129 and CL2130). Although, in about 2 $\sigma$ confidence 
level, the two methods yield virial masses which are in agreement.
This uncertainty is even greater than the one due to the projection effect, which,
for the caustic mass, is about 30\% (for about $100$ cluster members within 
the virial radius, Gifford et al.\ 2013; S11; Saro et al.\  2012; Gifford \& Miller 2013). 

Overall, we find a 1$\sigma$ agreement between the caustic, virial, and projected 
mass profiles. Although the
projected mass profile is in $\sim 1$ $\sigma$ agreement with the caustic mass
profile, it is usually higher. Therefore, the derived $r_{\rm 200}$, $r_{\rm
vir}$, $M_{\rm 200}$, and $M_{\rm vir}$ values for most clusters are higher when
using the projected mass profile than the caustics one (see table~\ref{Mvir rvir
table}). This is in agreement with Rines \& Diaferio (2008), who also found that
at large radii the projected mass profile tends to be higher than the caustic one.
At large radii, the caustic mass uncertainty is larger than the projected 
(and virial) mass uncertainty (see figure~\ref{mass profiles} and
table~\ref{Mvir rvir table}) because it also takes into consideration 
uncertainty due to the projection effect (though not with exactly 1$\sigma$ 
confidence level, see the notes in \textsection~\ref{Mass profile from velocity caustics} 
and \ref{Building the redshit-phase diagram} for more details).

As expected, when the number of galaxies is smaller, the statistic 
uncertainty is larger
(see Appendix~\ref{Mass profile biases}). Estimating the virial mass profile, taking about 15\% (when these 15\%
are $41$ and $63$ galaxies for A963 and RXJ2129, respectively) and 75\% of the
clusters' spectroscopically identified galaxies gave 30\%-50\% and 5\%-10\%
scatter over the mean, respectively. Uneven sampling of the sky's plane does not
insert a large uncertainty (see Appendix~\ref{Mass profile biases}). 
We test it by excluding a sector and estimating the
mass profile. Generally, excluding a sector increases both the statistical and
systematic (downwards) uncertainties. However, even when excluding a sector
with a large opening angle, $\pi/2$, at the virial radius, the mass is systematically 
lower by $\lesssim 7.5\%$, and the statistical uncertainty is $\lesssim 7.5\%$. If the
obscuration covers only part of the sector, the effect will be smaller, of
course. The closer the obscured part is to the cluster center, the larger the
bias because there are more galaxies at the center. 

Since among the two interloper removal methods we test we take the one that removes 
interlopers more efficiently, the three dynamical mass profiles we test are in 1 $\sigma$ 
agreement, and the uncertainty due to uneven sampling on the plane of the sky is not 
substantial, we conclude that these three mass biases are not substantial in 
our case. However, a very low number of galaxies increases substantially the statistical 
uncertainty. Indeed, the uncertainties in the halo mass estimations (see table~\ref{Halos info}) 
are larger than the ones of the clusters (see the last column in table~\ref{Mvir rvir table}).

We find good agreement between the dynamical mass profiles and the X-ray
derived ones at their overlapping radii range, except for A2261 where at 
small radii, i.e. $r < 418 $ $h_{0.73}^{-1}$ kpc, the X-ray profile is $\sim 2-3 \sigma$ 
above the dynamical mass profile. At these radii, the dynamical mass profile 
is estimated using spectroscopic data of less than 9 galaxies. In addition,
in this cluster, the separation between the X-ray and the galaxy surface number density 
peaks is the largest. The X-ray and dynamical mass profiles are combined together
and give a mass profile range from $\sim 10$ $h_{0.73}^{-1}$ kpc to beyond the
virial radius. 

Biviano et al.\ (2013) analyzed MACSJ1206 using the same spectroscopic data. 
They used different procedures, including the use of two other techniques to remove interlopers, 
and added photometric data in five bands which were derived from Subaru Suprime-Cam, to derive the cluster mass. 
Still, their cluster mass estimation, $M_{\rm 200} = 1.4 \pm 0.2$ $h_{0.7}^{-1}$ M$_{\odot}$, is in excellent 
agreement with our results.
Umetsu et al.\ (2012) used a combined WL distortion, magnification, and SL analysis 
of wide-field 5-band Subaru imaging and 16-band HST observations and found that for MACSJ1206 $M_{\rm vir} = 1.1 
\pm 0.2 \pm 0.1$ $10^{15}$ $h^{-1}$ M$_{\odot}$ (at the virial overdensity of 132), corresponding to $M_{\rm 200} = 
0.98 \pm 0.19 \pm 0.10$ $10^{15}$ $h^{-1}$ M$_{\odot}$. This is in excellent agreement with our (and Biviano et al.\ 2013) results.
Coe et al.\ (2012), who used strong- and weak-lensing (hereafter SL and WL, 
respectively), estimated that
$M_{\rm vir} = 2.2 \pm 0.2$ $10^{15}$ $h_{0.7}^{-1}$ M$_{\odot}$ for A2261
when assuming a spherical halo. This value is higher than our dynamical based
mass estimations (see table~\ref{Mvir rvir table}). Okabe et al.\ (2010), who
used high-quality Subaru/Suprime-Cam imaging data to conduct a detailed
WL study, found that $M_{\rm vir}$ is $6.96^{+2.17}_{-1.59}$,
$6.65^{+1.75}_{-1.42}$, and $6.71^{+2.73}_{-1.96}$ $10^{14}$ $h^{-1}$ M$_{\odot}$, for
A963, A611, and RXJ2129, respectively. The values they estimated for all 
clusters are in agreement with our results. Rodriguez-Gonzalvez et al.\ (2012) estimated the mass of A611 and
A1423 using Sunyaev-Zeldovich (hereafter SZ) measurements and found $M_{\rm 200} = 4^{+0.7}_{-0.8}$ $10^{14}$
$h^{-1}$ M$_{\odot}$ and $M_{\rm 200} =  2.2 \pm 0.8$ $10^{14}$ $h^{-1}$
M$_{\odot}$, respectively. Both are in $\sim 1\sigma$ agreement with our
results. 
Newman et al.\ (2013a) used SL and WL and found that
$M_{\rm 200}$ is $4.07_{-1.19}^{+1.17}$ $10^{14}$ $h_{0.7}^{-1}$ M$_{\odot}$ and
$3.63_{-0.81}^{+1.27}$ $10^{14}$ $h_{0.7}^{-1}$ M$_{\odot}$ for A963 and A611,
respectively. All these estimations are also in agreement with ours. Overall,
there is general good agreement between our dynamical based mass estimations and
the ones derived by lensing and SZ. In one case (A2261), the mass
estimation based on lensing is higher than our dynamical one. Our results are in
general agreement with Geller et al.\ (2013), who compared the mass profiles
based on the caustic technique with WL measurements taken from the
literature for 19 clusters. They found that at $3r_{\rm 200}$ the WL
overestimates the caustic profile by about $20\% - 30\%$, probably due to the
impact of superposed large-scale structures. However, at $r_{\rm vir}$, they
found the WL and caustic mass profiles in very good agreement. 

Turning to halo accretion, accreted and soon to be accreted halos are
identified by first using D99 methodology to identify cluster members and
infalling galaxies. Then we define halos using two different HFs, i.e.
overdensities of the smoothed galaxy surface density (for more details, see
\textsection~\ref{Over density}) and FoF (for more details, see
\textsection~\ref{FoF}). The use of two different HFs increases the
reliability of our results and shows differences arising from the halo
identification scheme. The number of identified halos and their sizes depend on
the HF (see figure~\ref{Clusters_n_close_halos_DEC_vs_RA}) and the values
inserted into their free parameters. Generally, here the FoF HF tends to break
halos, that are identified by the overdensity HF, into a few smaller halos. 
In addition, the FoF HF identifies more low mass halos with a filament-like 
appearance (see the fields of A1423 and CL2130 in figure~\ref{Clusters_n_close_halos_DEC_vs_RA}).
As a result of these two differences between the two HFs, the estimated 
F($0.2 \lesssim \zeta \lesssim 0.7$) and F($0.01 \lesssim \zeta \lesssim 0.2$) 
are higher and lower when using the overdensity HF. 
Nevertheless, a general agreement between the HFs is found both in the detection of halos (see
figure~\ref{Clusters_n_close_halos_DEC_vs_RA} and figure~\ref{N_vs_Mhalo}) and
the estimated $F(\zeta)$ values at the $0.2 \lesssim \zeta \lesssim 0.7 $ range. 

In the $0.2 \lesssim \zeta \lesssim 0.7$ range, our observational based
estimation for the $F(\zeta)$ profile is in $\sim 1 \sigma$ agreement with our
theoretically predicted one. At low mass ratios, $\zeta \lesssim 0.2$, the estimated 
$F(\zeta)$ using the overdensity HF is underestimated since the detection efficiency 
decreases at low masses, $\sim 2 \times 10^{14}$ $h_{0.73}^{-1}$ M$_{\odot}$. 
The estimated $F(\zeta)$ using the FoF HF is not overestimated but only because generally
here the  
FoF HF tends to break halos, that are identified by the overdensity HF, into a few smaller 
halos, and it identifies more low mass halos with a filament-like appearance. As a result, too many
low mass halos were identified (see figure~\ref{N_vs_Mhalo}).
At large mass ratios, $\zeta \gtrsim 0.7$, we do not detect halos probably because
all the clusters in this sample (except for CL2130) were chosen to be quite
X-ray relaxed (P12). Therefore, this sample is biased against large $\zeta$
values. Another explanation is that these mergers are rare and may be missed due
to our sample size. In addition, in cases where a satellite is partly or fully 
within the cluster, our cluster mass estimation includes part (or all) of the 
satellite's mass. Therefore, in these cases, $\zeta$ is underestimated. Thus, at 
large $\zeta$ values, the estimated $F(\zeta)$ is also underestimated.

Using the overdensity HF, we estimate these halos' centers, redshifts, LOS 
velocities, masses, relaxation states, and substructure levels. Except for the
four mass biases we mention above, the mass estimators we use to measure the
identified satellites are based on the assumption that the system is in
dynamical equilibrium. The mass estimation of unrelaxed halos due to LOS mergers
can be overestimated (Takizawa et al.\ 2010). As we see in
figure~\ref{relaxation proxies and relaxation vs substructure} (left panel), the
halos relaxation spectrum is quite wide and some of them may not be relaxed,
though our displacement relaxation proxies are more sensitive when the merger
axis is in the plane of the sky. Therefore, some of the halos masses may be
overestimated. Regarding the substructure levels, the mass merger ratio
limitation, for which there is a marginal detection, depends on the sample size
(P96). Because in these halos there is a low number of spectroscopically
identified galaxies, these substructure level estimations are sensitive for a
mass merger ratio limitation of about 1/2 (P96). This narrow mass merger ratio
range may explain why there is no significant correlation between the
substructure and our estimated dynamical relaxation levels.

\section{Summary}
\label{Summary}

For the first time, we test a key outcome in the $\Lambda$CDM cosmological model: the
contributions of mergers with different mass ratios to cluster-sized
halo growth. At the $0.2 \lesssim \zeta \lesssim 0.7$ range, we find a 
$\sim 1 \sigma$ agreement with the cosmological
model. 

Our other main conclusions are:

\begin{enumerate}

\item There is good agreement between the caustic, virial, and
projected mass profiles. 

\item As in Diaferio et al.\ (2005) and Geller et al.\ (2013), who explored 
the consistency between mass profiles derived from the caustic method and 
those derived using weak lensing measurements, we also find that our caustic 
mass estimations at $r_{\rm 200}$ and $r_{\rm vir}$ are in agreement with their 
corresponding WL and SZ estimations in the literature.

\item Different interloper removal methods affect the substructure level
estimation dramatically, or in other words, the DS test is very sensitive to
the method used for removing interlopers.

\item The D99 interloper removal method is more adequate to deal with the
presence of close and LOS substructure than HK96.

\item Mass uncertainties can be substantial due to the different cluster members
identification methodology ($\sim 50\%-100\%$ at the virial radius, although, 
in about 2 $\sigma$ confidence level, the two methodologies yield virial masses which 
are in agreement) and the very
low number of galaxies ($30\%-50\%$ and $5\%-10\%$ when taking $15\%$ and $75\%$
of the clusters spectroscopically identified galaxies, respectively). The
uncertainty is less substantial in the case of uneven sampling of the sky's
plane (when excluding a sector with a large opening angle, $\pi/2$, at the 
virial radius, the mass is
systematically lower by $\lesssim 7.5\%$ and the systematic uncertainty is
$\lesssim 7.5\%$). All the dynamical mass estimators we used were in agreement
within 1$\sigma$, but in most cases the projected mass profiles were higher than
the caustic mass profiles.

\end{enumerate}

\acknowledgements
We thank Margaret J. Geller, Kenneth Rines, Michael Kurtz, and Antonaldo Diaferio for 
enabling this work by generously providing their redshift data for A611 and CL2130, and 
especially for providing the redshift data for A963, A2261, A1423, and RXJ2129 in 
advance of publication. We also thank them for many helpful discussions. We thank Susan 
Tokarz for performing the extensive data reduction on the Hectospec redshift observations 
and Perry Berlind and Mike Calkins for providing their expertise in operating the Hectospec 
instrument. In addition, we acknowledge very useful discussions with Ana Laura Serra, Dan 
Gifford, Mark Neyrinck, Yuval Birnboim, Eyal Neistein, and Maxim Markevitch. We thank the 
anonymous referee for useful comments. DL thanks Eran Ofek for his publicly available Matlab 
scripts. This research is supported in part by NASA grant HST-GO-12065.01-A. MM acknowledges 
support from PRIN INAF 2009 and ASI (agreement Euclid phase B2/C). AZ is supported by contract 
research ``Internationale Spitzenforschung II/2-6'' of the Baden W\"urttemberg Stiftung.
Facilities: MMT (Hectospec), VLT (VIMOS), Magellan (IMACS), Chandra X-ray Observatory, SDSS,
CFHT (MegaCam).

\appendix

\section{Survey completeness}
\label{Survey completeness}

\begin{figure*}
\centering
\epsfig{file=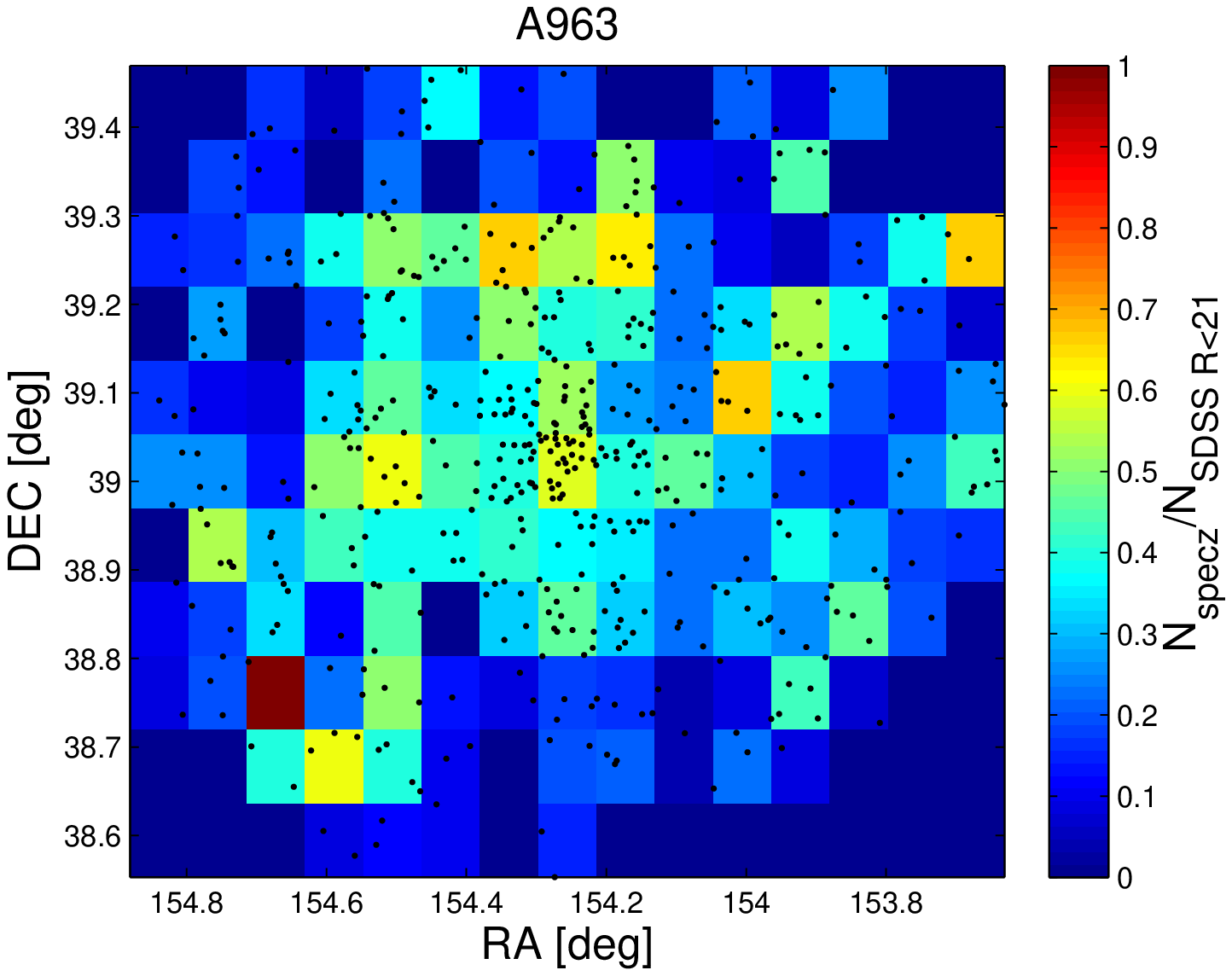, width=8cm,
clip=}
\epsfig{file=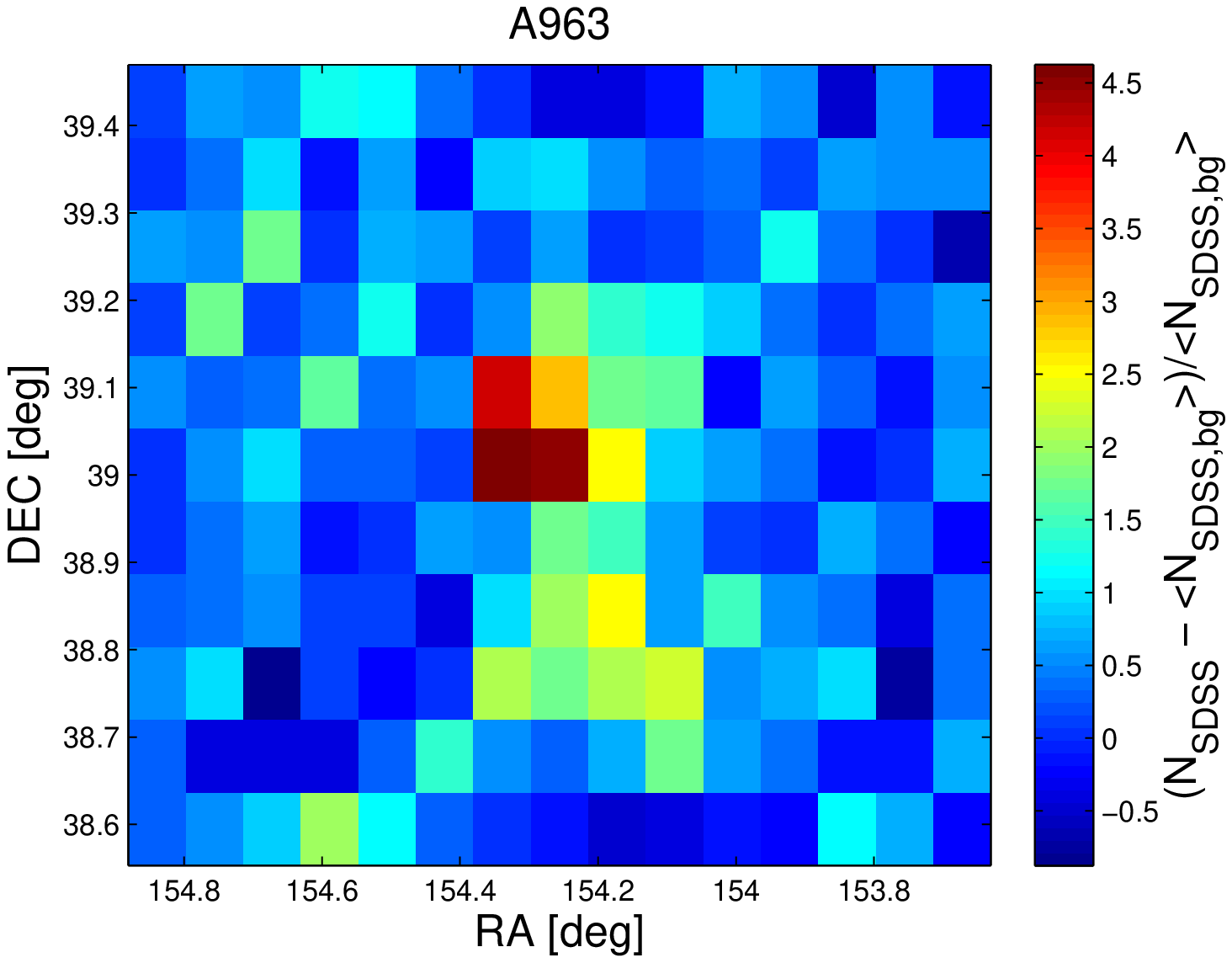, width=8cm,
clip=}
\caption{A963 targets selection (as described in
\textsection~\ref{Spectroscopy})
plane of the sky completeness. Left panel: the plane of the sky is divided to
5 x 5 arcmin regions. Then we calculate in each region the ratio between
Hectospec (spectroscopic) sources and SDSS photometric sources which are above
the Hectospec limiting R magnitude, i.e. $\sim 21$. Black dots indicate the
Hectospec spectroscopically identified galaxies. Right panel: testing real plane
of the sky differences in the galaxies distribution by plotting
$(N_{\rm SDSS}-N_{\rm SDSS,bg})/N_{\rm SDSS,bg}$, where $N_{\rm SDSS,bg}$ is the
$(N_{\rm SDSS}$ background which is estimated as the mean of the four 5 x 5
arcmin corner regions. 
\label{Survey completeness A963}}
\end{figure*}

\begin{figure*}
\centering
\epsfig{file=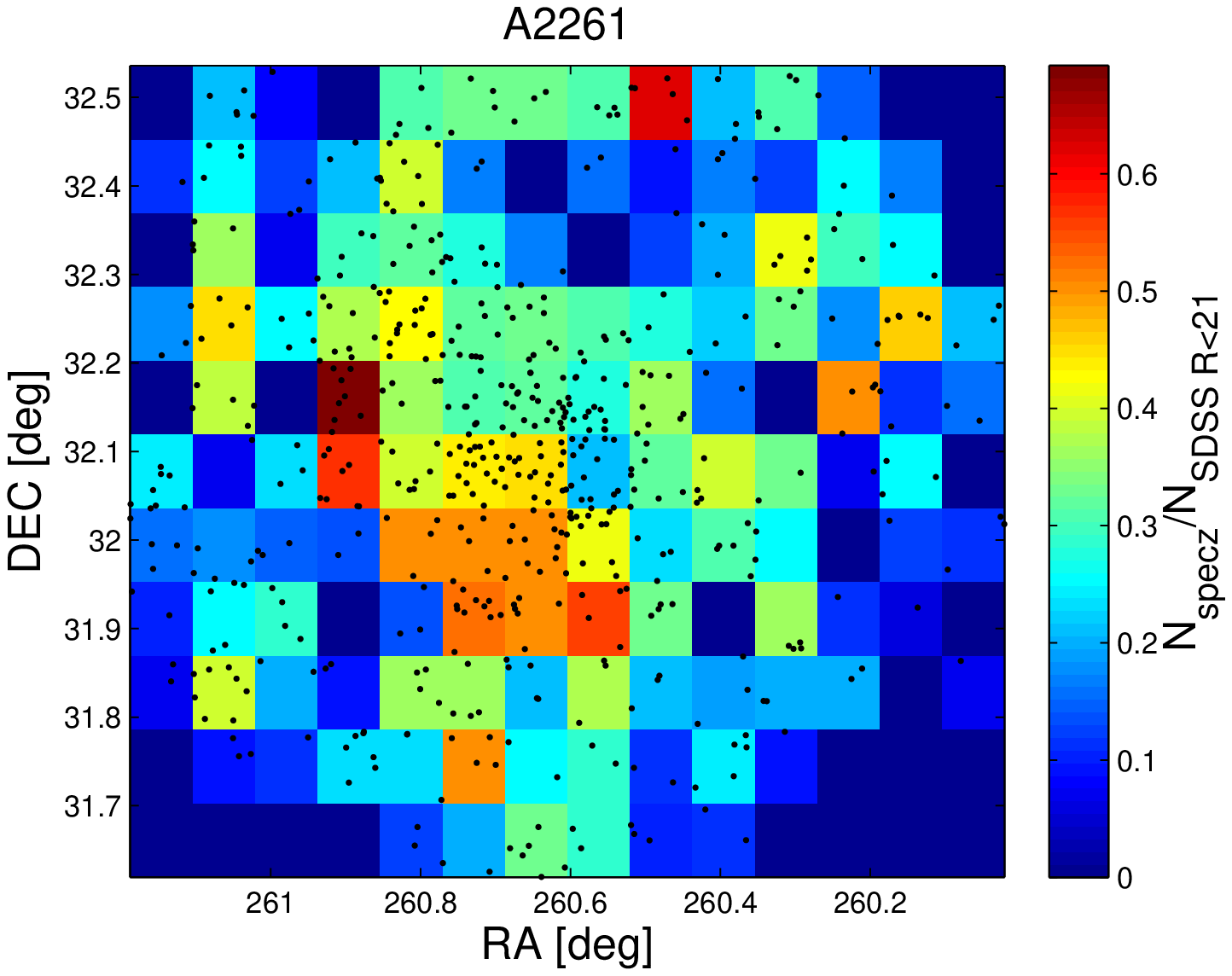, width=8cm,
clip=}
\epsfig{file=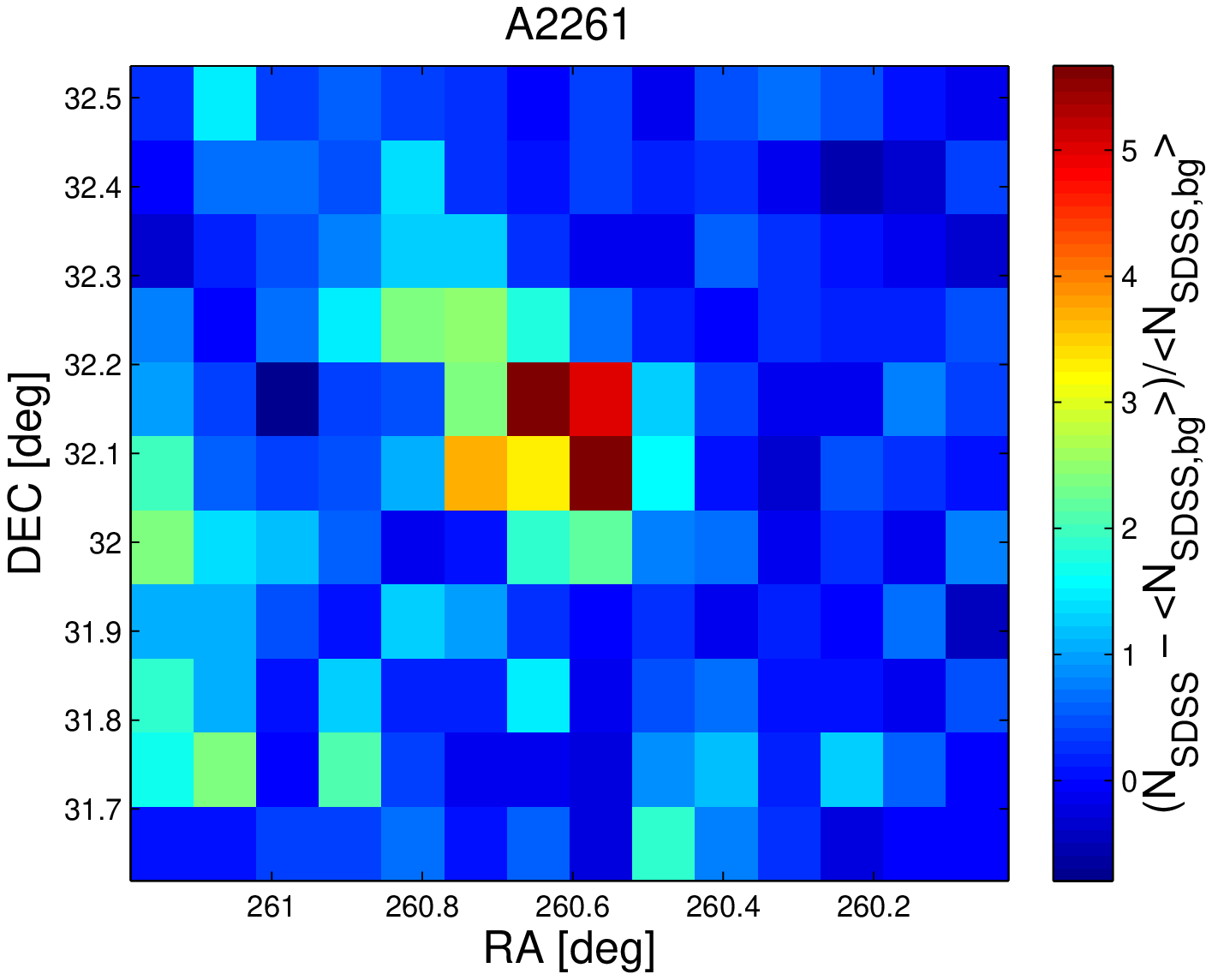, width=8cm,
clip=}
\caption{The same as figure~\ref{Survey completeness A963} but for A2261.
\label{Survey completeness A2261}}
\end{figure*}

\begin{figure*}
\centering
\epsfig{file=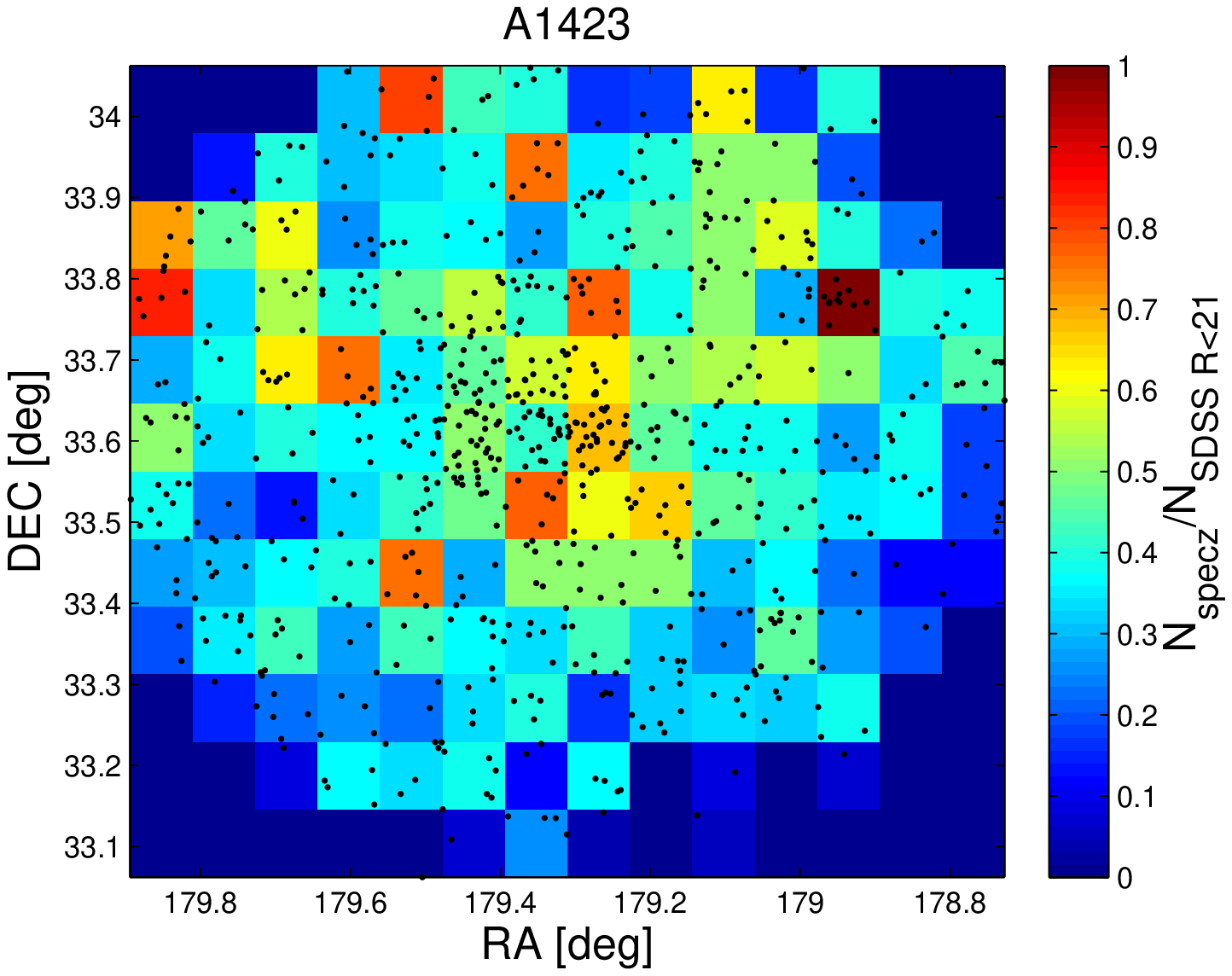, width=8cm,
clip=}
\epsfig{file=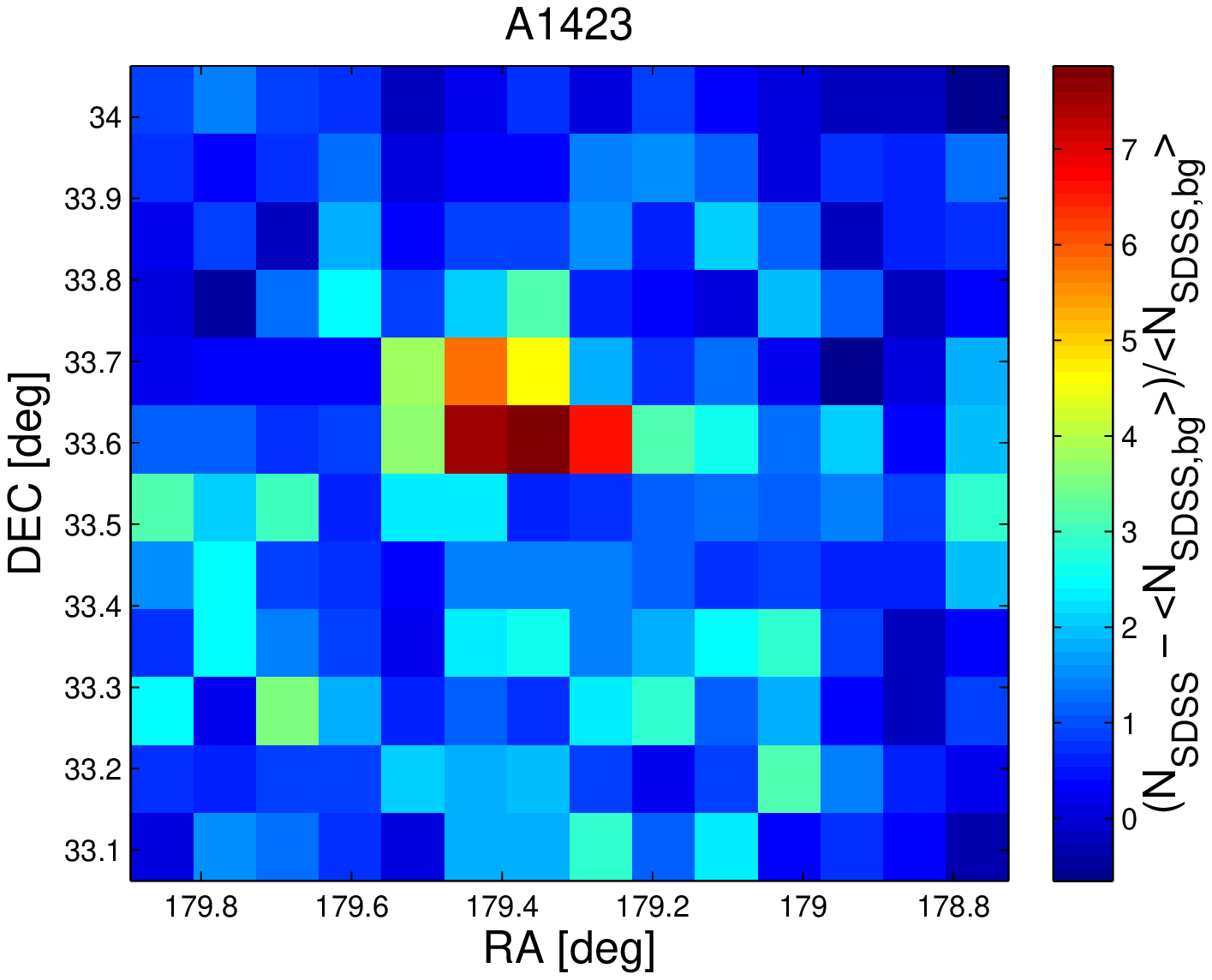, width=8cm,
clip=}
\caption{The same as figure~\ref{Survey completeness A963} but for A1423.
\label{Survey completeness A1423}}
\end{figure*}

\begin{figure*}
\centering
\epsfig{file=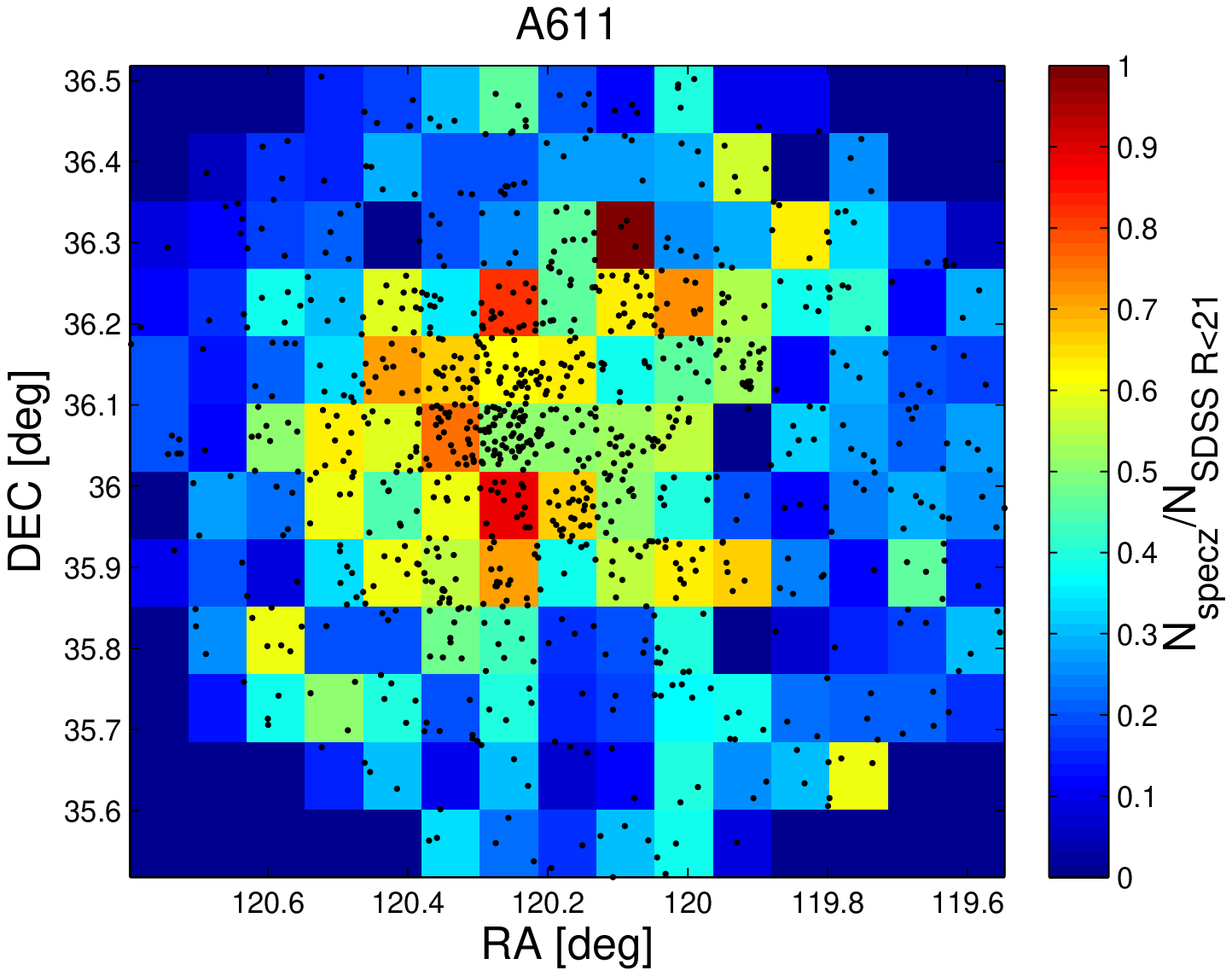, width=8cm, clip=}
\epsfig{file=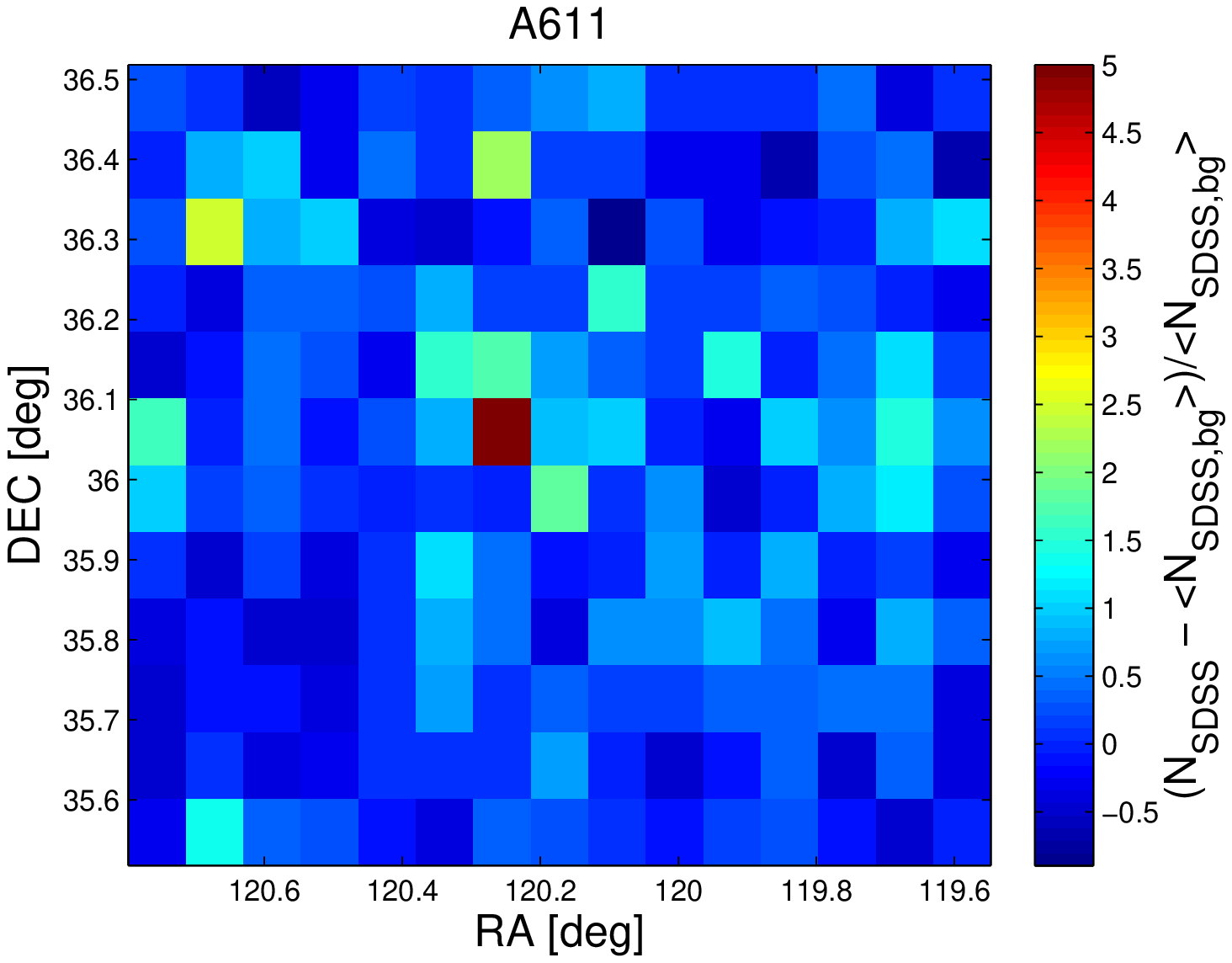, width=8cm,
clip=}
\caption{The same as figure~\ref{Survey completeness A963} but for A611.
\label{Survey completeness A611}}
\end{figure*}

\begin{figure*}
\centering
\epsfig{file=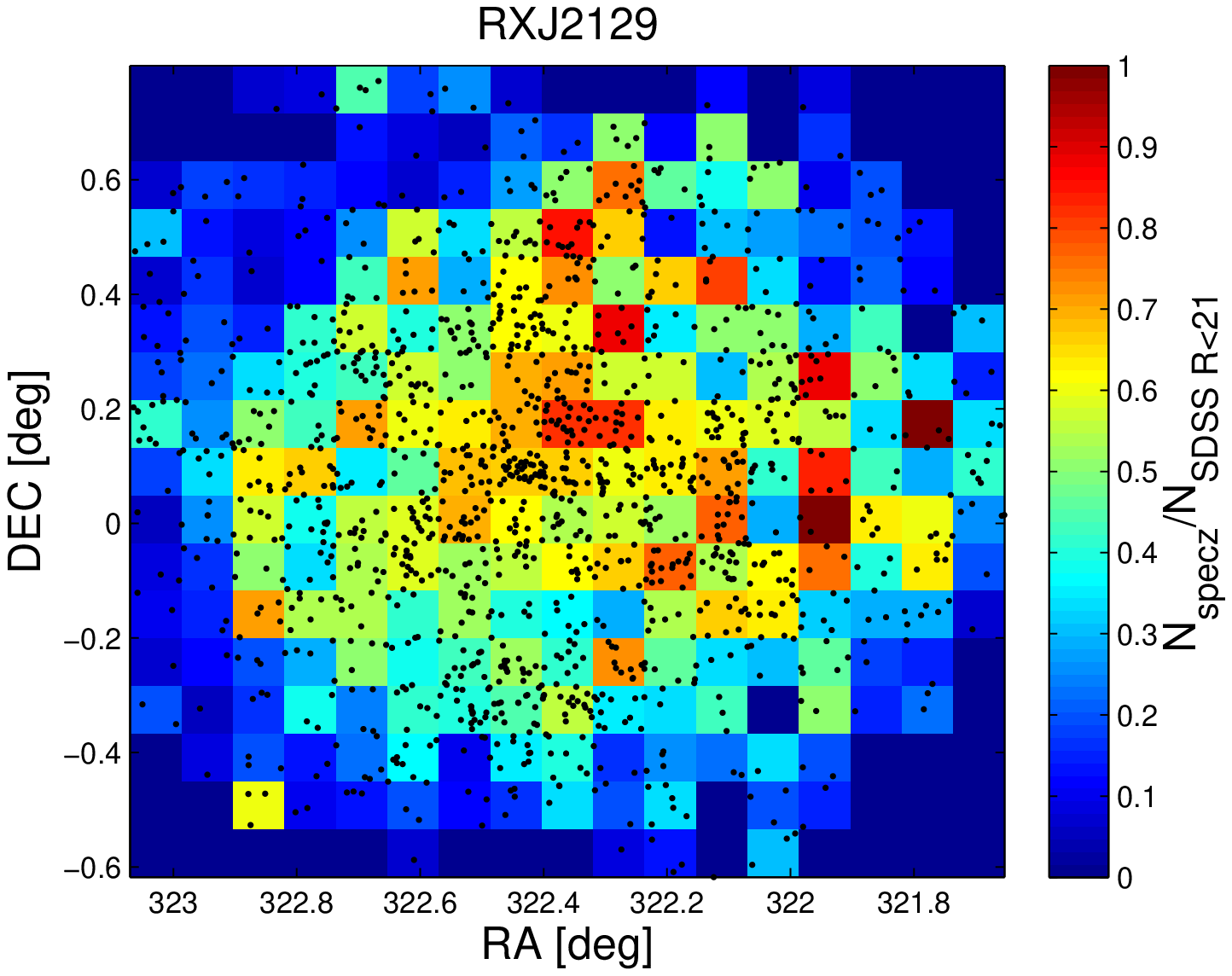, width=8cm,
clip=}
\epsfig{file=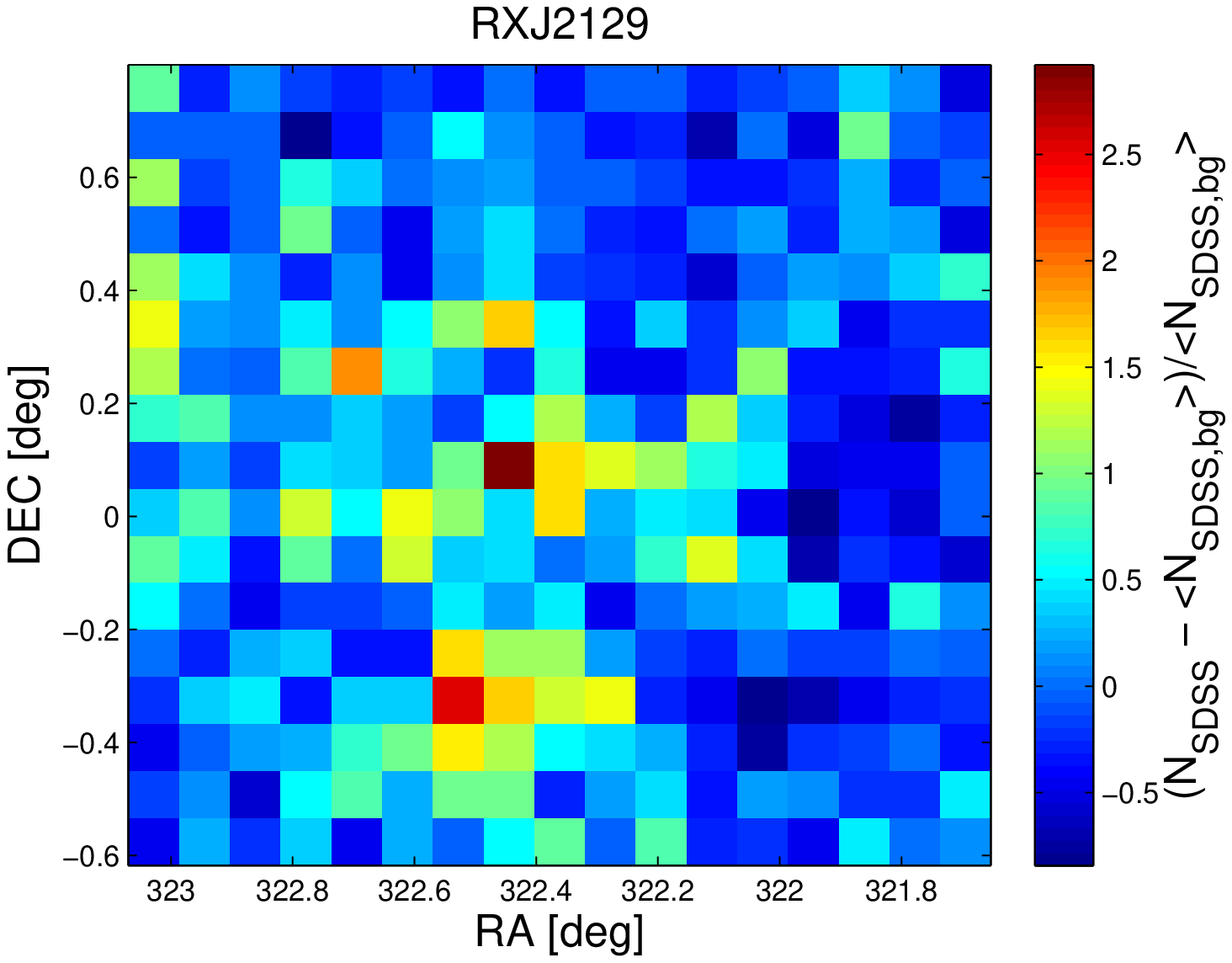, width=8cm,
clip=}
\caption{The same as figure~\ref{Survey completeness A963} but for RXJ2129.
\label{Survey completeness RXJ2129}}
\end{figure*}

Target selection (as described in \textsection~\ref{Spectroscopy}) plane of the
sky completeness is checked by comparing those targets to more uniformly
selected sources. Hectospec and VIMOS sources are compared to SDSS
and Canada France Hawaii Telescope (CFHT) sources, respectively. In
figures~\ref{Survey completeness A963}-\ref{Survey completeness RXJ2129}, for
each of the clusters observed using Hectospec we divide its plane of the sky to
5 x 5 arcmin regions. We then calculate in each region the ratio between the
Hectospec (spectroscopic) sources, $N_{\rm specz}$, and the SDSS photometric
galaxies, which are above the Hectospec limiting R magnitude, i.e. $\sim 21$,
$N_{\rm SDSS}$. This procedure allows us to see if there are regions with large
coherent patterns of higher than average incompleteness, which can happen due to
bright stars, for example. Regions may look incoherent not due to incompleteness
but rather due to real plane of the sky differences in the galaxies'
distribution. Therefore, in order to check for the latter case we, also plotted
the ratio $(N_{\rm SDSS}-N_{\rm SDSS,bg})/N_{\rm SDSS,bg}$, where $N_{\rm
SDSS,bg}$ is the $N_{\rm SDSS}$ background, which is estimated as the mean of
the four 5 x 5 arcmin corner regions. 

\begin{figure*}
\centering
\epsfig{file=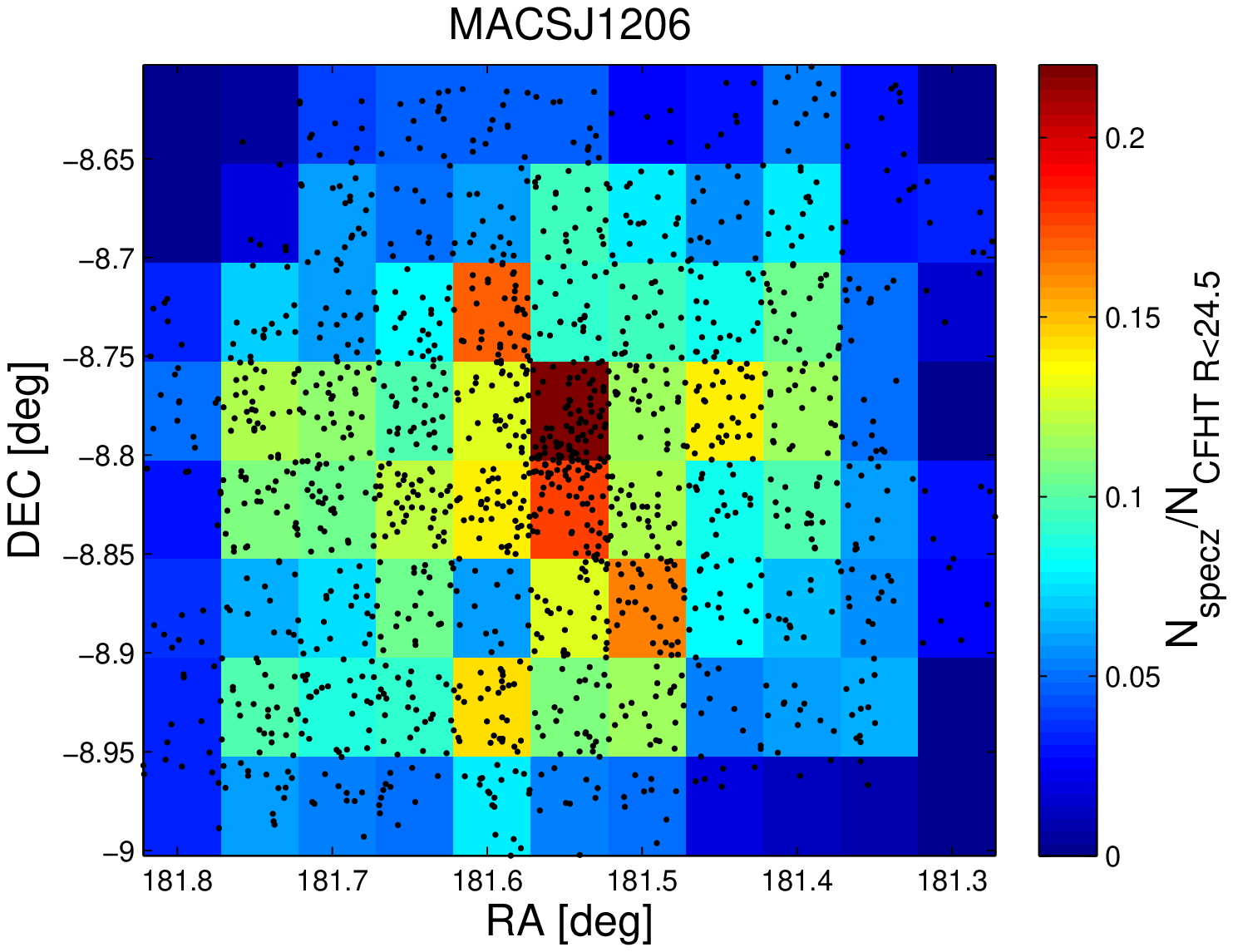,
width=8cm,clip=}
\epsfig{file=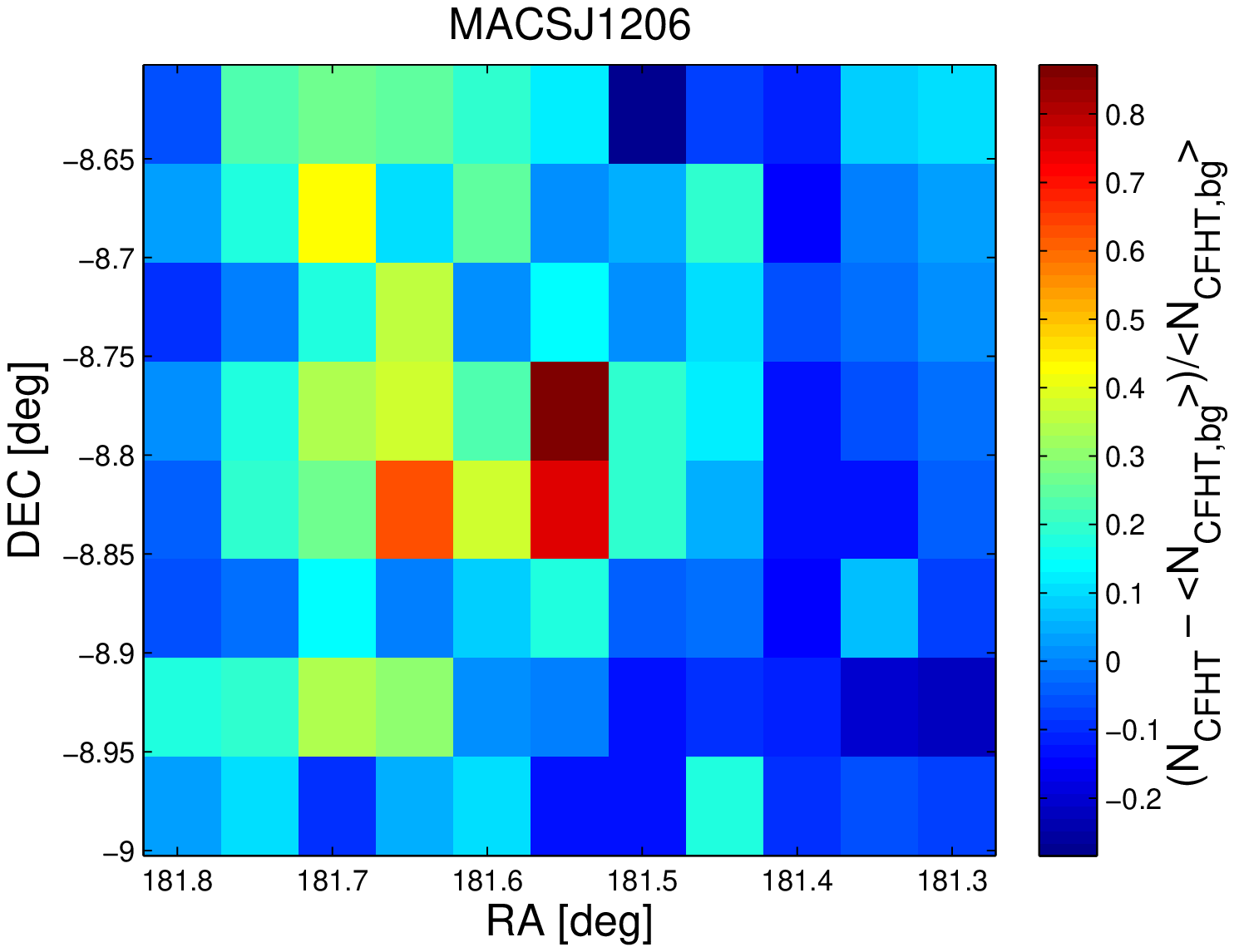,
width=8cm,clip=}
\caption{MACSJ1206 targets selection (as described in
\textsection~\ref{Spectroscopy})
plane of the sky completeness. Left panel: the plane of the sky is divided to
3 x 3 arcmin regions. Then we calculate in each region the ratio between
VIMOS (spectroscopic) sources and CFHT photometric sources with $R \le
24.5$. Black dots indicate the VIMOS spectroscopically identified galaxies.
Right panel: testing real plane of the sky differences in the galaxies
distribution by plotting $(N_{\rm CFHT}-N_{\rm CFHT,bg})/N_{\rm CFHT,bg}$, where
$N_{\rm CFHT,bg}$ is the $(N_{\rm CFHT}$ background which is estimated as the
mean of the four 3 x 3 arcmin corner regions. 
\label{Survey completeness MACSJ1206 24.5 mag}}
\end{figure*}

In figures~\ref{Survey completeness MACSJ1206 24.5 mag}, we divide MACSJ1206
plane of the sky to 3 x 3 arcmin regions. Then we calculate $N_{\rm
specz}/N_{\rm CFHT}$ and $(N_{\rm CFHT}-N_{\rm CFHT,bg})/N_{\rm CFHT,bg}$ for
sources with $R \le 24.5$.

Note, in Appendix~\ref{Mass profile biases} we estimate the uncertainty in the 
virial mass profile due to plane of the sky incompleteness sampling.

\section{Techniques for removal of non-cluster galaxies} 
\label{Removal of non-cluster galaxies - techniques description}

In the following two subsections, we briefly describe the two methods we use to
remove interlopers.

\subsection{den Hartog \& Katgert interloper removal method}
\label{HK96 interlopers removal method}

The HK96 procedure is an iterative method, which uses the estimated mass profile
(and assuming spherical symmetry) to infer the maximum LOS velocity profile that
a cluster member galaxy can have. We make an initial cut in the velocities as
described in \textsection~\ref{First velocity cut}. Then in each iteration a
mass profile is estimated and the maximum LOS velocity profile is inferred. Two
options for dynamically based mass profiles are the virial and projected, see
\textsection~\ref{Virial and projected mass proxies}. Here we choose to use the
projected mass profile (choosing the virial mass profile yields similar
results). Galaxies with LOS velocities exceeding the maximum LOS
velocity at their projected radius are considered to be interlopers and are
removed. Then we estimate a new mass profile and so on. The method converges in
1 to 10 iterations.

The maximum LOS velocity profile is calculated as follows. Since the
velocity anisotropy is not known, galaxies are considered to have circular
(tangential) or infalling (radial) orbits. The circular velocity is $v_{\rm
cir}(r) = \sqrt{GM(r)/r}$, and if all orbits are circular with this velocity
the cluster is in virial equilibrium, since $T_{\rm cir} = \frac{1}{2} \int
\rho(r) v_{\rm cir}^2(r) dr = -\frac{1}{2} U$. For bound galaxies, the upper
limit for the infall velocity is $T_{\rm infall}=|U|$, so $v_{\rm infall}(r) =
\sqrt{2}v_{\rm cir}(r)$.

We ignore the presence of the virialized central region, and assume that each
galaxy is either on a radial orbit towards the center of the cluster, or on
a purely circular orbit. Under the first assumption, we calculate the upper
limit to the LOS velocities, $V_{\rm max}(R)$, at projected distance
R, as the maximum value of the LOS component of the infall velocity
for all positions r on the line of site within the turnaround radius, i.e.
$v_{\rm infall}(r_{\rm max})\cos(\theta)$. Under the second assumption, one uses
the maximum LOS component of the circular velocity, i.e. $v_{\rm
cir}(r_{\rm max})\sin(\theta)$. Combining both assumptions yields
\begin{equation}
 V_{\rm max}(R) = \max \{v_{\rm infall}[r_{\rm max}(R)]\cos[\theta(R)],v_{\rm
cir}[r_{\rm max}(R)]\sin[\theta(R)]\} \ ,
\end{equation} where $\theta(R)$ is the angle between the radial vector $r$ and
the LOS at $R$, and $r_{\rm max}(R)$ is the position (along the LOS at $R$)
where the maximum occurs.

\subsection{Caustics interloper removal method}
\label{D99 interlopers removal method}

The caustic method has a few steps. First we identify the cluster members
candidates using a binary tree. These candidates determine the cluster's
velocity dispersion and size. Then these two parameters together with all the 
galaxies after the velocity cut (see \textsection~\ref{First velocity cut}) 
are used to determine the velocity space density threshold $\kappa$, which 
determines the caustics and the final cluster members. Below we describe each 
of these steps in more details.

\subsubsection{Binary tree}
\label{Binary tree}

A binary tree is a method to estimate the similarity between the different
galaxies. In each step, the two galaxies with the highest similarity are
combined into one group. The process continues until all galaxies are joined to
one group, which is called the root of the binary tree. Here we use this method
to find cluster members candidates, which are needed for the D99 procedure. 

As Serna \& Gerbal (1996) suggested, we use the galaxy pairwise binding energy
as a measure of similarity:
\begin{equation}
 E_{i,j} = -G\frac{m_i m_j}{|{\bf r}_i-{\bf r}_j|}+\frac{1}{2}\frac{m_i m_j}{m_i +
m_j}({{\bf v}_i}-{{\bf v}_j})^2 \ , \label{pairwise binding energy} 
\end{equation} where $G$ is the gravitational constant, $m_i$, $m_j$, ${\bf r}_i$,
${\bf r}_j$, ${{\bf v}_i}$, ${{\bf v}_j}$ are the masses, positions, and velocities of
the two galaxies. To calculate eq.~\ref{pairwise binding energy}, one needs the
full velocity space information (6D), though in our case only three coordinates
are available. However, Serna \& Gerbal (1996) found that the projection
instability between 6D and 3D is quite low, which means that the ``observed''
structures are reasonably similar to the ``intrinsic'' 6D ones. Therefore, we
used eq.~\ref{pairwise binding energy} as our similarity measurement when the
2D spatial position, ${\bf R}$, and the velocities are the LOS ones,
$v$, so 
\begin{equation}
 E_{i,j} = -G\frac{m_i m_j}{|{\bf R}_i-{\bf R}_j|}+\frac{1}{2}\frac{m_i m_j}{m_i +
m_j}(v_i-v_j)^2 \ . 
\end{equation}

To build the binary tree, we proceed as follows:
(i) Each galaxy is a group $G_{\nu}$. (ii) We compute the similarity between
each two groups $G_{\rm \mu}$, $G_{\rm \nu}$ with the single linkage method:
$E_{\mu \nu} = \min \{E_{\rm ij} \}$ where $E_{\rm ij}$ is the similarity
between the member $i \in G_{\rm \mu}$ and the member $j \in G_{\rm \nu}$.(iii)
We replace the two groups with the largest similarity (smallest binding energy
$E_{\mu \nu}$) with a group $G_{\rm k}$. The number of independent groups
is decreased by one. (iv) The procedure is repeated from (ii) until we are left
with only one independent group.

\subsubsection{Identifying the cluster member candidates}
\label{Estimation of sigma_pl}

After building the binary tree (as is described above and taking $m_i = m_j =
10^{12}$ $h_{0.73}^{-1}$ M$_{\odot}$ to be
consistent with S11), we need to identify the cluster member
candidates. D99 suggested doing so by first identifying the main branch of the
binary tree. The binary tree starts at the root when there is only one group,
which is actually the end product of the binary tree procedure. We walk through
the tree along the main branch. At each step, node, the main branch splits into
two branches. The branch with the larger number of galaxies is the continuation
of the main branch. The velocity dispersion, $\sigma_x$, of the galaxies hanging
from a given node, $x$, shows a characteristic behavior when walking towards the
leaves along the main branch. Initially (at the root of the tree where the
galaxy pairwise binding energy is the highest) it decreases rapidly, then it
reaches a plateau and again drops rapidly towards the end of the walk. D99 claimed 
that the plateau, $\sigma_{pl}$, is a clear indication of the presence of the nearly
isothermal cluster: at the beginning of the walk, $\sigma_x$ is large because we
include galaxies outside the main halo; at the end of the walk $\sigma_x$ is low
since we consider the very central galaxies of a given subclump. 

The two nodes $x_1$ and $x_2$ that limit the $\sigma_{pl}$ plateau are good
candidates for the substructure and cluster identification, respectively.
Locating $x_1$ and $x_2$ and estimating sigma plateau, $\sigma_{pl}$, is not a
trivial procedure. Their values depend on the estimation procedure, and
on the definition of similarity used to build the binary tree (which may change
a little the $\sigma_x$ - $x$ diagram). D99, who applied the method to
dissipationless cosmological N-body simulations where galaxies form and evolve
according to semi-analytic modeling, used a hyperbolic function to fit the
plateau and to chose the substructure and cluster thresholds as the inflection
points of that fit. While S11, who applied the caustic technique
to clusters extracted from a cosmological hydrodynamic simulation of a
$\Lambda$CDM universe, used a different technique (see their \textsection 4.2).
S11 claimed that their algorithm turns out to be more accurate
in determining the thresholds since it improves the $\sigma_{\rm caus}$ /
$\sigma_{\rm true}$ relation (see their figure 9).

The picture, however, can be more complicated since a few levels of plateaus
with different lengths can be seen in data due to substructure. Here we use a
similar approach to S11, but taking into consideration that a few
plateaus can be present in the $\sigma_x$ vs $x$ diagram.
The $\sigma_{pl}$ node region location is determined by the minimum
ratio between the standard deviation and average of the $\sigma_x$ over a range
of at least $N_{\rm node,min}$ number of (main branch) nodes, i.e.
\begin{equation}
 \delta_{\rm min} = \min \left \{\frac{\Delta
\sigma_{x_1,x_2}}{<\sigma_{x_1,x_2}>} \right \}_{\rm
N \geqslant N_{\rm node,min}} \ ,
\end{equation} where we take $N_{\rm node,min}=7$ to have a minimal number of
galaxies for the standard deviation estimation. The cluster identifying node,
i.e. $x_2$, is determined by the largest node that is bellow $\sigma_{pl}(1+\#
\delta_{\rm min})$, where $\#$ was taken to be $2$.

\subsubsection{Building the redshift-phase diagram}
\label{Building the redshit-phase diagram}

The galaxies hanging from node $x_2$ and constitute the main group of the 
binary tree determine the velocity dispersion and mean projected distance of the 
members. These two are then used in the D99 procedure, which locates the
caustics using all the galaxies and determines the radial dependence of their 
amplitude (in units of velocity). Galaxies that are inside the caustics are 
considered to be
cluster members. Of course, some of these galaxies might still be interlopers,
but their number is typically a few percent and has little effect on dynamical
analyses (Serra \& Diaferio 2013; Antonaldo Diaferio, private communication).

In more details, a redshift-space diagram is constructed from all the galaxies 
after the velocity cut (see \textsection~\ref{First velocity cut}). Then we 
determine the threshold $\kappa$ that defines the
caustic location through $f(R,v)=\kappa$. Here $f(R,v)$ is the galaxy
density distribution in the redshift-space diagram, smoothed with a
multidimensional adaptive kernel (Silverman 1986; Pisani 1993; Pisani 1996)\footnote{
Gifford et al.\ (2013), who used simulations, claim that a standard fixed kernel 
also recovers the cluster mass estimates with low scatter and bias.}. For
the calculation of the optimal smoothing length (see D99 eq. 20),
the two coordinates $R$ and $v$ must have the same units. Therefore, we divide
them by their maximum value. In addition, Silverman (1986, page 77) suggests rescaling
these coordinates to avoid extreme differences of spread in the various
coordinate directions, so we further divide them by their dispersion. The 
maximum and dispersion of both $R$ and $v$ are determined from the cluster
member candidates\footnote{Previously in the caustic method one of the major 
input parameter was the rescaling parameter, $q$, which
sets the scaling between the quantities $R$ and $v$ within the smoothing
procedure (D99). There is no simple a priori choice for this parameter value.
Usually it was chosen to be $q=25$ (e.g. D99; Rines et al.\ 2003), which was
obtained by the ratio of the two coordinates uncertainties (D99). It was claimed
that different values of $q$ in the $10-50$ range have little effect on the
results (D99; Rines et al.\ 2002). However, even in this $q$ range, Rines et
al.\ (2002) tested the effect of different $q$ values on the A2199 supercluster
estimated mass profile, and found that although the mass mean does not depend
strongly on $q$ the mass uncertainty does. In addition, Reisenegger et al.\
(2000) found that $q$ values out of the $10-50$ range are more suitable in the
case of the Shapley supercluster. In this paper, we rescale $R$ and $v$ in an 
automatic procedure instead of choosing the scaling by hand.}.

The parameter $\kappa$ is chosen by minimizing the quantity $S(\kappa,\langle
R\rangle)=|\langle v_{esc}^2\rangle_{\kappa,\langle R\rangle}-4 \langle
v^2\rangle|^2$, where $\langle v_{esc}^2\rangle_{\kappa,\langle R\rangle}=
\int_0^{\langle R\rangle}A^2(R)\phi(R)dR/\int_0^{\langle R\rangle}\phi(R)dR$ is
the mean value of the square of the caustic amplitude $A(R)$ within $\langle R
\rangle$ (the mean projected radius of the cluster members), $\phi(R)=\int
f(R,v)dv$, and $\langle v^2\rangle ^{1/2}$ is the one-dimensional velocity
dispersion of the cluster members. The caustics are the value of $v(R)$ at the
projected radius $R$ at the point where $f(R,v)= \kappa$. 

In general, $f(R,v)$ is not symmetric around the cluster redshift, and since we
assume spherical symmetry, we take $A(R) = \min\{|v_u(R)|,|v_d(R)|\}$, when
$v_u(R)$ and $v_d(R)$ are the caustics above and below the clusters' redshift,
respectively. Taking the minimum of the upper and lower caustics is more robust
than $A(R) = (v_u(R)-v_d(R))/2 $ against interloper contamination and presence
of massive substructure. Finally, since for any realistic system $d\ln A / d
\ln R \lesssim 1/4 $, D99 claimed that to control the contamination by
background and foreground galaxies efficiently a further step is needed. We
follow D99 and accept only values of $A(R)$ that yield $\frac{d\ln A}{d
\ln R} \leqslant \zeta_{\rm caustics}$, when $\zeta_{\rm caustics} = 1/4$.
Otherwise we impose new values of $A(R)$ that yield $d\ln A / d \ln R =
1/4$. Note that D99 and S11 took $\zeta_{\rm caustics} = 1$ and
$\zeta_{\rm caustics} = 2$, respectively.

In table~\ref{Clusters centers and redshifts info table}, we also present the
clusters' redshift estimations as suggested in D99. The cluster center in redshift 
space is the median of the cluster member velocity distribution.

\paragraph{Caustic uncertainty}
\label{Caustics uncertainties}

A larger $\kappa$ indicates a smaller number of galaxies within the caustics,
therefore a larger uncertainty. Similarly, a poorly sampled cluster yields a
small $\max \left\{f(R,v)\right\}$, therefore, again, a large uncertainty. If
the area surrounding the caustics is poorly populated and the area within the
caustics is well populated, $\kappa$ is low and $\max \left\{f(R,v)\right\}$ is
large and the uncertainty is small. Therefore, the caustic uncertainty is
taken to be $d A(R)/A(R) = \kappa/ \max \left\{f(R,v)\right\}$, where the
maximum is found along the $v$-axis at each $R$ (D99; and Antonaldo Diaferio,
private communication).\footnote{By analyzing a sample of 3000 simulated clusters 
with masses of $M_{200}>10^{14}$ $h_{0.73}^{-1}$ M$_{\odot}$, S11 tested the deviation 
of this recipe from the 1 $\sigma$ confidence level. They found that on average at $r_{200}$ 
the upper caustic uncertainty is underestimated by about 24\% and the lower caustic 
uncertainty is overestimated by about a few per cent (see left panel in their figure 16; 
Ana Laura Serra, private communication).}

\subsubsection{The value of $F_{\beta}$}
\label{F_beta}

In order to calculate the caustic mass (see eq.~\ref{M_from_A}), one has to set
a value for $F_{\beta}$. This factor, which absorbs in it the galaxy
velocity anisotropy profile ($\beta$),\footnote{The $\beta$, in this subsection, is 
not to be confused with the $\beta$ parameter in eq.~\ref{N_m}} varies slowly with the radius and,
therefore, taken to be constant (D99). D99 followed Diaferio \& Geller 1997 and
set $F_{\beta}$ to be $0.5$, finding that the resulting caustic method recovers
the true cluster mass within a factor of 2 at large radii, $r \sim (0.3 - 6)\
r_{200}$, at least with the low concentration parameters of his simulated
clusters. D99's hyperbolic fit to the $\sigma_x$ - $x$ diagram makes the
thresholds ($x_1$ and $x_2$) to be below (in the case of substructure threshold)
and above (in the case of the cluster threshold) the plateau in most cases. An increase 
in $x_2$ increases the cluster galaxies
velocity dispersion (for more details, see Appendix~\ref{D99 interlopers removal
method}). As a result, the caustics are wider. The S11 algorithm
places $x_1$ and $x_2$ on the plateau, which means that the caustics are slightly
narrower (Ana Laura Serra, private communication). To compensate for that, they
adopted a higher value for $F_{\beta}$, $0.7$, instead of $0.5$.\footnote{Lemze et al.\ (2009), 
who analyzed A1689 using galaxy dynamics, derived the cluster's $F_{\beta}$ 
profile (see their figure 9), using the dynamical information and the mass density profile 
derived in Lemze et al.\ (2008). However, they did not attempt to constrain the cluster's 
$F_{\beta}$ mean value because, at that time, the only value suggested in previous works was $0.5$.
Even if the mean $F_{\beta}$ was constrained for this cluster, a large scatter is expected between
the $F_{\beta}$ profiles of different clusters (D99; S11). Thus, the mean $F_{\beta}$
of A1689 can be different than the averaged over many clusters.} 
This large difference in the value of $F_{\beta}$ is because the caustics are sensitive to
the $<v^2>$ value (D99), which depends on the binary tree cutting method. Since our thresholds 
finding algorithm is closer to the one used in S11 (see Appendix~\ref{D99 interlopers removal method}), 
we adopted $F_{\rm \beta} = 0.7$.

In an independent work from S11, Gifford et al.\ (2013) used simulations and also found a high
value for the caustic mass scaling factor, $F_{\beta} = 0.65$. They also gave another explanation for 
the $0.5-0.7$ scatter in the $F_{\rm \beta}$ values (see their \textsection 4.3).

\subsection{Velocity-space diagram for MACSJ1206} 
\label{Velocity-space diagram for MACSJ1206}

\begin{figure*}
\centering
\epsfig{file=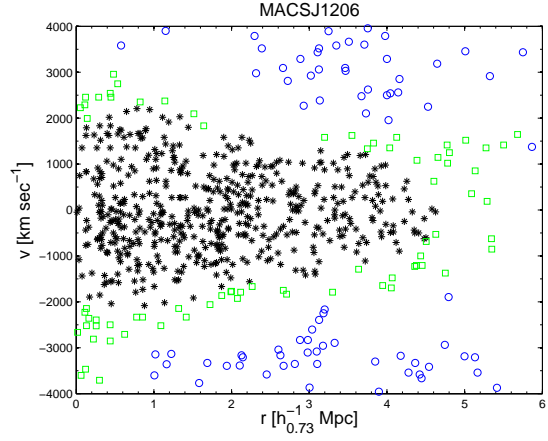, width=8cm, clip=}
\caption{Phase-space diagram of MACSJ1206. Galaxies, which are left after cleaning interlopers using the D99 method, are marked by black asterisk.
Galaxies, which are left after cleaning interlopers using the HK96 method but are considered interlopers using the D99 method, are marked by green squares.
Galaxies, which are considered to be interlopers by both the D99 and HK96 methods, are marked by blue circles. 
\label{MACSJ1206 removing interlopers}}
\end{figure*}

In this section, we show the velocity-space diagram (see textsection~\ref{Removal of non-cluster galaxies}) 
of MACSJ1206, which has the highest number of bound galaxies in our sample. In the velocity-space diagram, we 
clean interlopers using the two different methods: HK96 and D99 (see also Biviano et al.\ (2013), who used two 
other techniques to remove interlopers from this cluster). In figure~\ref{MACSJ1206 removing interlopers}, 
galaxies, which are left after cleaning interlopers using the D99 method, are marked by black asterisk. Galaxies, which are 
left after cleaning interlopers using the HK96 method but are considered interlopers using D99 method, are marked by green squares.
Galaxies, which are considered to be interlopers by both the D99 and HK96 methods, are marked by blue circles. 

\section{Mass profile biases}
\label{Mass profile biases}

\begin{figure*}
\centering
\epsfig{file=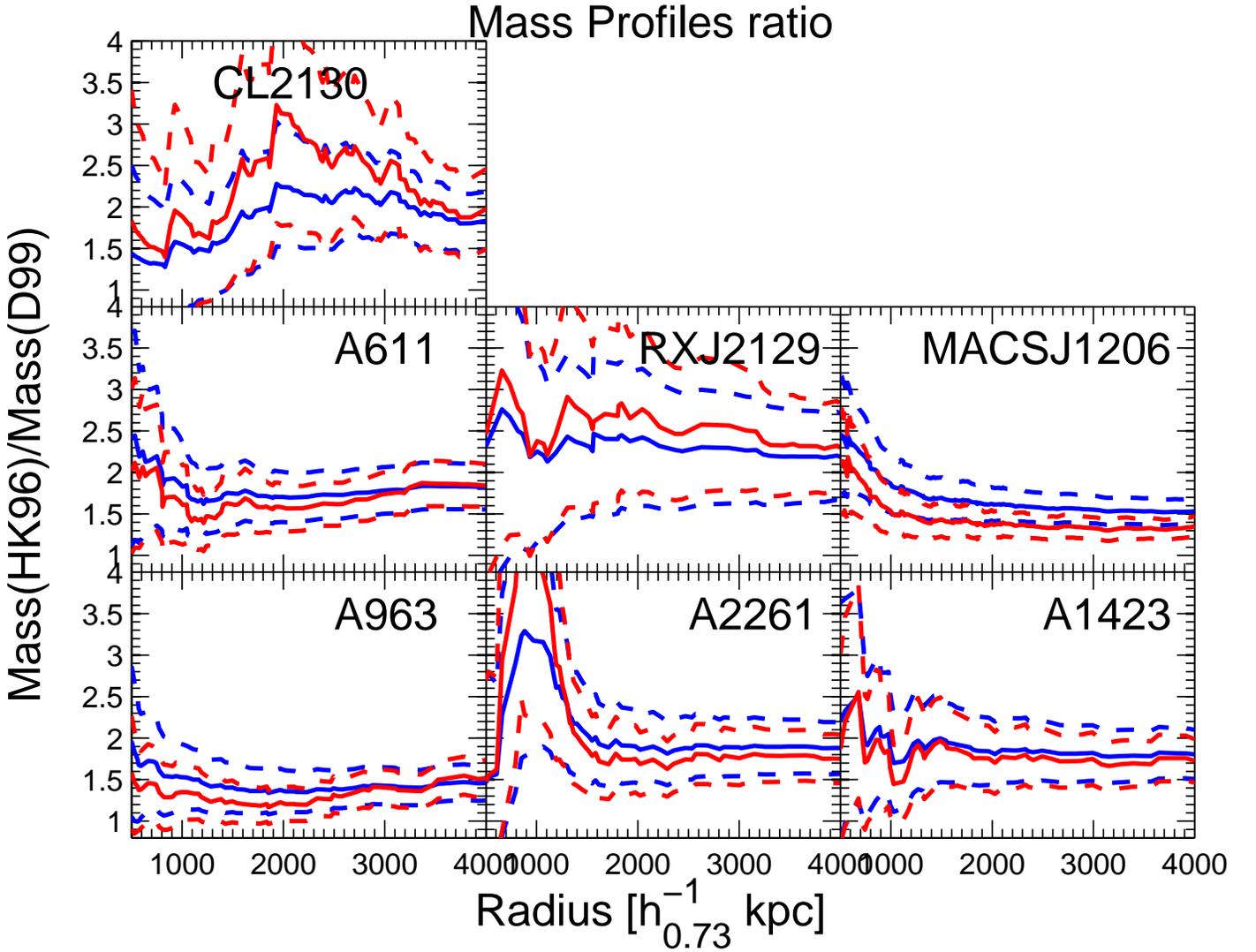, width=20cm, clip=}
\caption{Mass profiles ratio inferred by using different interlopers removal
methods, HK96 and D99. Blue and red curves are for the corrected virial and
projected mass profiles, respectively. The upper and lower dash curves represent
the $\pm 1\sigma$ width of the distribution uncertainty. 
\label{mass profiles ratio}}
\end{figure*}

In this section, we examine various sources that may affect the mass profile
(see also Biviano et al.\ 2006), irrespective of the mass estimator used. The 
first one is the dependence of the mass profile on the
interloper removal method. In figure~\ref{mass profiles ratio}, we plotted the
ratio between the mass profiles when using HK96 and D99 interloper removal
methods (for more details, see \textsection~\ref{Removal of non-cluster
galaxies}). We show this ratio for both the corrected virial (blue curve) and
projected (red curve) mass profiles. 

\begin{figure*}
\centering
\epsfig{file=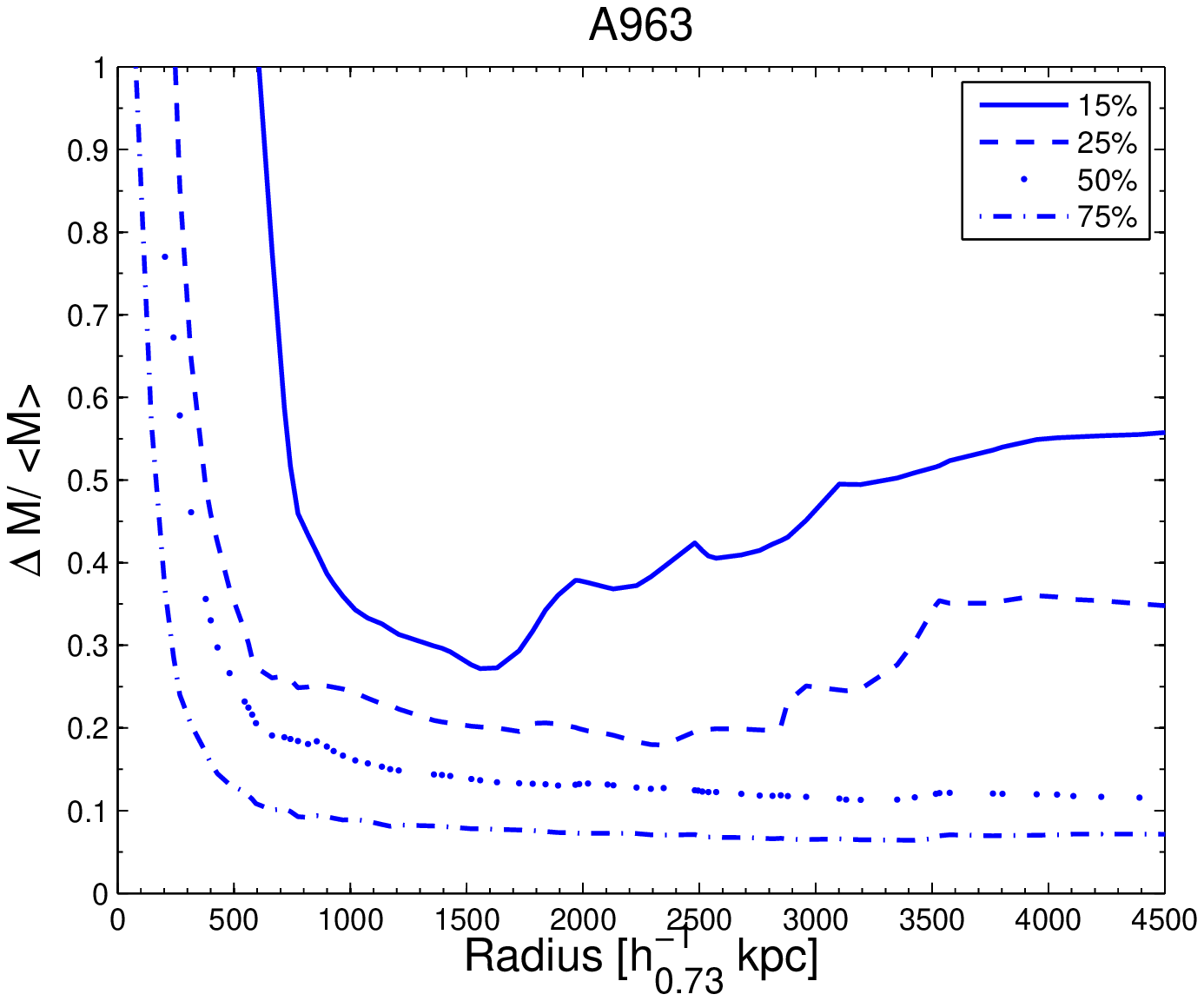, width=8cm, clip=}
\epsfig{file=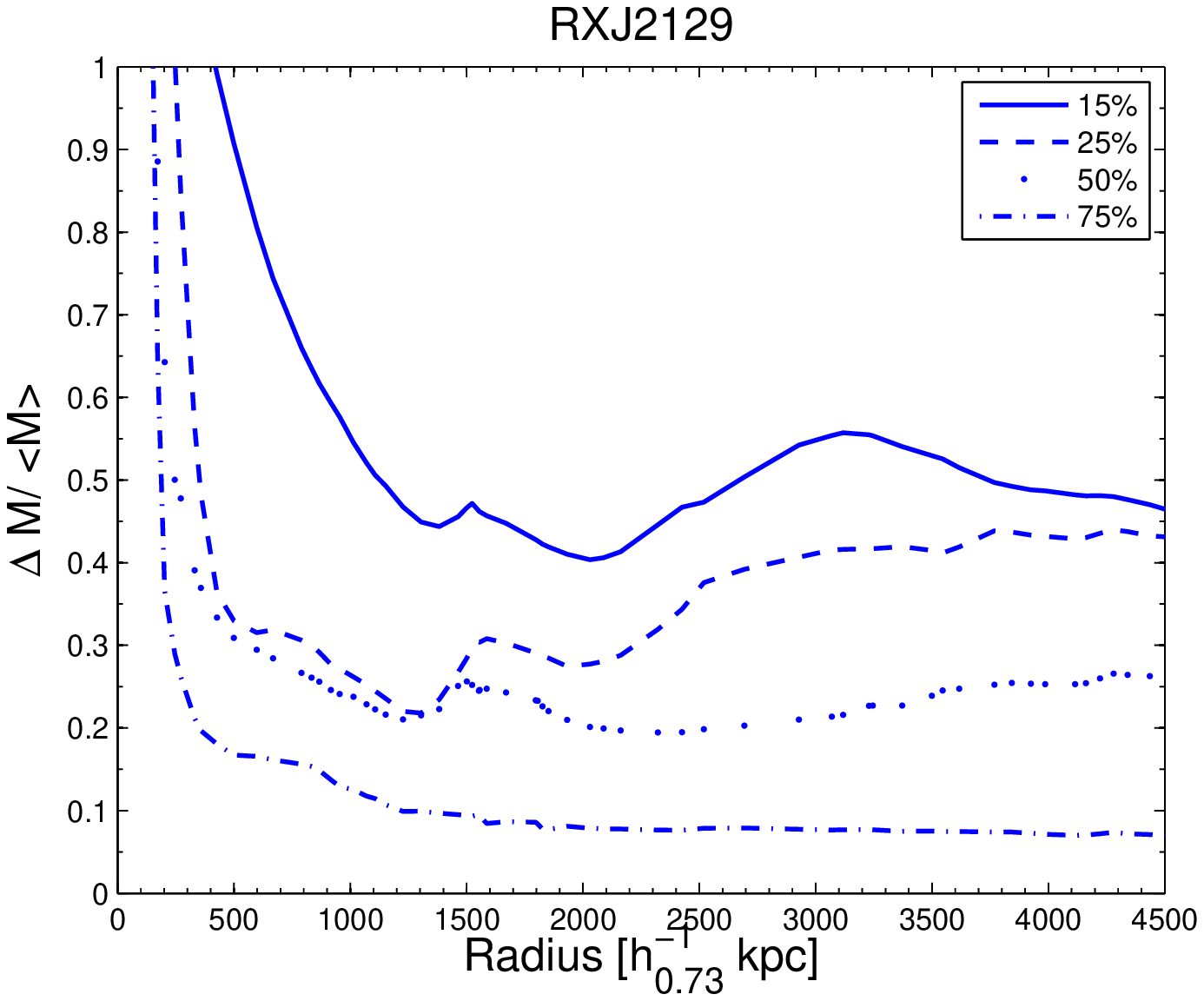, width=8cm, clip=}
\caption{Statistic uncertainty due to the number of galaxies. The statistical
uncertainty is estimated by randomly taking a fraction of the galaxy number 
after the initial velocity cut (see \textsection~\ref{First velocity cut}), removing
interlopers using the HK96 method, and estimating the corrected virial mass profile. 
This procedure is repeated 50 times for each fraction, and the statistical uncertainty 
is estimated by dividing the scatter by the mean of all the repetitions. Left panel, taking different galaxy fractions
from A963 galaxies. Solid, dashed, dotted, dotted-dashed curves are for taking
15\%, 35\%, 50\%, and 75\% of the galaxies, respectively. Right panel, the same
as the left panel but for RXJ2129. 
\label{Mvir_Mrand_std_Mrand_mean}}
\end{figure*}

A second test we make is to estimate the statistical uncertainty due to the
number of spectroscopic identified galaxies. We randomly take a fraction of
the galaxy number after the initial velocity cut (see \textsection~\ref{First
velocity cut}). Then we remove interlopers using HK96 method and estimate 
the corrected virial mass profile. This procedure is
done 50 times, and the statistical uncertainty due to the number of galaxies is
estimated as the scatter divided by the mean. In
figure~\ref{Mvir_Mrand_std_Mrand_mean}, we show the results of this test for
A963 and RXJ2129, and for a few different fractions of the spectroscopic
identified galaxy number. Solid, dashed, dotted, and dotted-dashed curves are
for taking 15\% (when these 15\% are $41$ and $63$ galaxies for A963 and
RXJ2129, respectively), 35\%, 50\%, and 75\% of the galaxies, respectively.   

\begin{figure*}
\centering
\epsfig{file=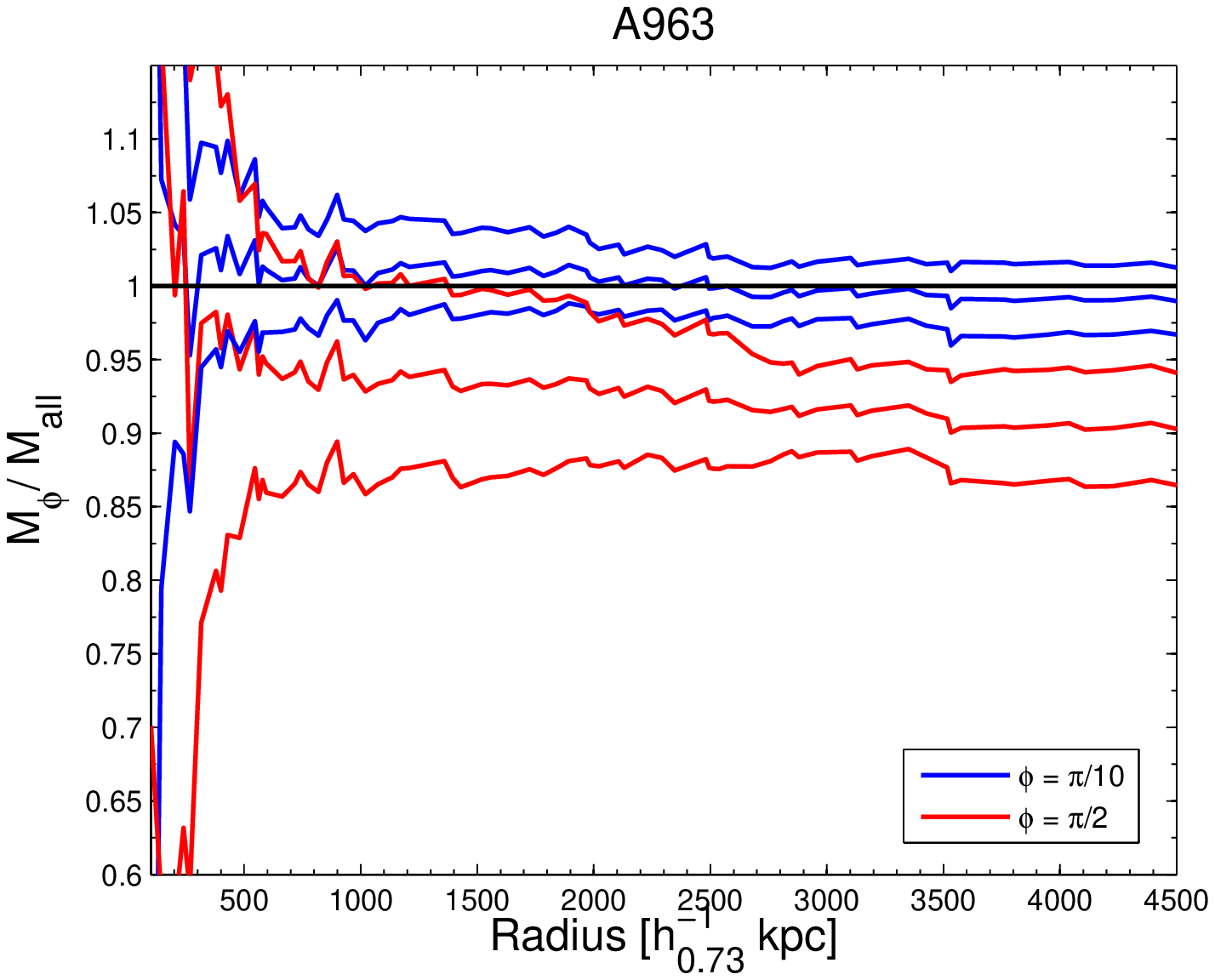, width=8cm, clip=}
\epsfig{file=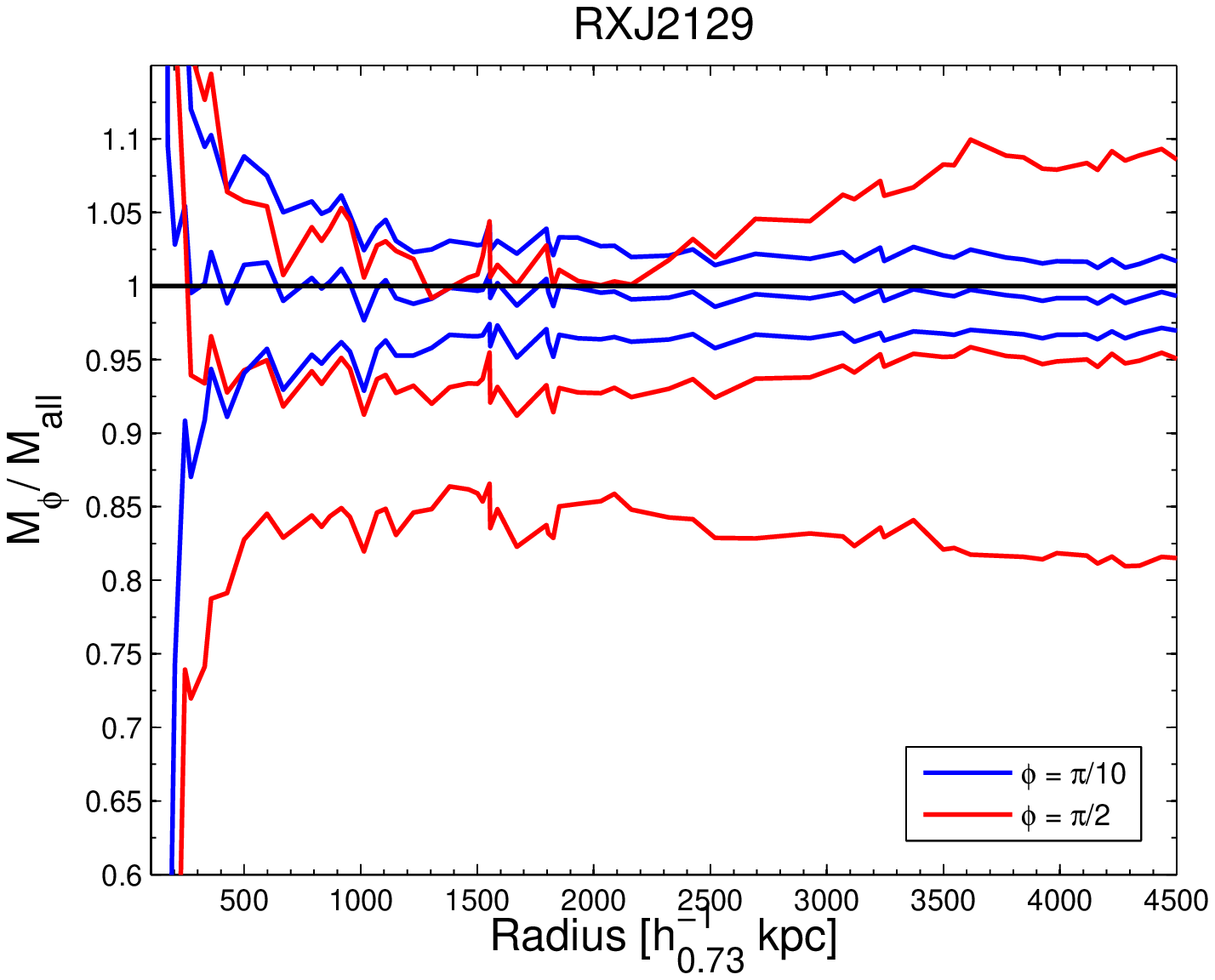, width=8cm, clip=}
\caption{Mass profile bias due to uneven sampling of the plane of the sky. We
estimate the virial mass profile after excluding galaxies in a sector with
opening angles of $\phi = \pi/10$ (blue curves) and $\pi/2$ (red curves) and removing
interlopers using the HK96 method. For each $\phi$, the
mass profile, $M(\phi)$, is estimated $N_{\rm \theta}=50$ times, where in each
time the sector is rotated by an angle $\theta$, so $\theta \times N_{\rm
\theta}$ gives $2 \pi$.  
Left panel is for A963, and right panel is for RXJ2129. In both figures, the middle
solid and the upper and lower solid curves represent the mean and 1$\sigma$ scatter,
respectively, of all rotations. 
\label{M_phi_M_all}}
\end{figure*}

Lastly, we estimate the uncertainty in the virial mass profile due to plane of
the sky incompleteness sampling. We take our spectroscopic data, exclude part
of them, cleaning them from interlopers using HK96 procedure, and estimate the
virial mass profile. The excluded part is taken to be a sector centered at the 
cluster center and with an opening angle $\phi$. Assuming a relaxed and spherical cluster, mass profiles estimated
from different sectors should be the same. Clusters, on the other hand, may not
be completely relaxed nor spherical. However, the level of relaxation can be
quantified (as we show in \textsection~\ref{Relaxation tests}), and the
ellipticity effect can be averaged out when taking a sample of clusters, each
with a different ellipticity. In figure~\ref{M_phi_M_all}, we plot the virial
mass profiles after excluding galaxies in sectors with opening angles of $\phi
= \pi/10$ (blue curves) and $\pi/2$ (red curves). 
If we exclude only an annulus within the sector, the bias is obviously smaller than excluding 
the whole sector. If we exclude an inner annuli the bias is larger than excluding an outer annuli 
(not shown) since the galaxies' density is the higher closer to the cluster center.
For each $\phi$, the
mass profile, $M(\phi)$, is estimated $N_{\rm \theta}=50$ times, where in each
time the sector is rotated by an angle $\theta$, so $\theta \times N_{\rm
\theta}$ gives $2 \pi$. The mass uncertainty from each sector is taken to be
the standard deviation over all $\theta$. 

\section{Estimating $\beta_{\rm MAH}(M_0)$ and $\gamma_{\rm MAH}(M_0)$ }
\label{beta_MAH and gamma_MAH}

\begin{figure*}
\centering
\epsfig{file=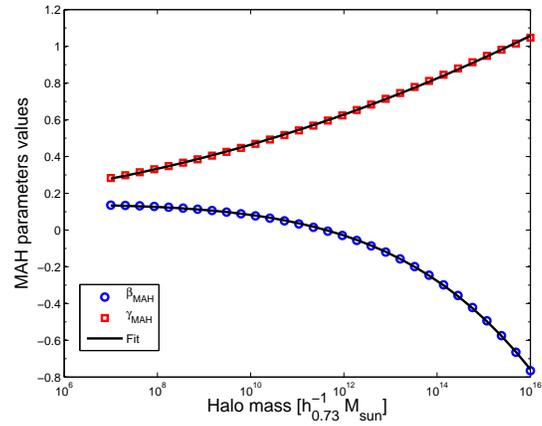, width=8cm, clip=}
\caption{MAH parameters fit. Blue circles and red squares are for $\beta_{\rm
MAH}$ and $\gamma_{\rm MAH}$, respectively. These are obtained by fitting
eq.~\ref{M(z)} to the MAH derived using eq.~\ref{mean mass growth rate}. Black
solid lines are the fit to $\beta_{\rm MAH}$ and $\gamma_{\rm MAH}$using
eq.~\ref{beta_MAH and gamma_MAH fitting expressions}.
\label{MAH parameters fit}}
\end{figure*}

Here we explicitly express $\beta_{\rm MAH}(M_0)$ and $\gamma_{\rm
MAH}(M_0)$ as functions of $M_0$. This allows us together with eq.~\ref{M(z)}
to calculate the averaged halos' MAH for any $M_0$. In order to do so, we
first derive $\beta_{\rm MAH}$ and $\gamma_{\rm MAH}$ for different $M_0$ values. 
For each $M_0$, we fit the expression given in eq.~\ref{M(z)} to the mass histories
inferred from the mean mass growth rates of halos, i.e. eq.~\ref{mean mass
growth rate}. Then, we fit the following phenomenological functions to the 
$\beta_{\rm MAH}$ and $\gamma_{\rm MAH}$ obtained for the different $M_0$ values:
\begin{eqnarray}
\begin{array}{ll}
 \beta_{MAH}(M_0) = C_1(1+(\log(M_0)/C_2)^{C_3})^{-1}+C_4 \\
 \gamma_{MAH}(M_0) = C_1(1+(\log(M_0)/C_2)^{C_3})+C_4 \ ,
\end{array}
\label{beta_MAH and gamma_MAH fitting expressions}
\end{eqnarray} where $C_1$, $C_2$, $C_3$, $C_4$, are the fitted expressions'
free parameters. The best-fit values are $C_1=(28.2684,6.3862)$,
$C_2=(28.6379,39.5429)$, $C_3=(5.8674,2.1171)$, and $C_4=(-28.1271,-6.2693)$,
when the left and right values are for $\beta_{\rm MAH}$ and $\gamma_{\rm MAH}$,
respectively. In figure~\ref{MAH parameters fit}, we plot $\beta_{\rm MAH}$
and $\gamma_{\rm MAH}$ as blue circles and red squares, respectively. On top, we
plot the fitted expression as a black solid line.

\end{document}